\documentclass[prl,twocolumn,amsmath,amssymb,10pt,aps,longbibliography,superscriptaddress,citeautoscript,bibnotes,floatfix]{revtex4-2}

\usepackage{graphicx} 
\usepackage{dcolumn}
\usepackage{bm}
\usepackage{graphicx,color,xcolor}
\usepackage{caption}
\usepackage{physics}
\usepackage{comment}
\usepackage{ulem}
\graphicspath{ {./Figures/} }
\usepackage{textcomp}
\usepackage[utf8]{inputenc}
\usepackage[T1]{fontenc}
\usepackage{amsmath}
\usepackage{amssymb}
\usepackage{amsthm}
\usepackage{siunitx}
\usepackage{physics}
\usepackage{graphicx}
\usepackage{subfigure}
\usepackage{tabularx}
\usepackage{booktabs}
\usepackage{multirow}
\usepackage[colorlinks=true,linkcolor=blue,citecolor=blue]{hyperref} 

\usepackage{xcolor}

\newcommand{\CEA}[1]{{\color{purple} #1}}

\usepackage{soul}

\begin{document}

\title{Emergent Quantum Phenomena of {Noncentrosymmetric Charge-Density Wave} in 1T-Transition Metal Dichalcogenides}
\author{Cheong-Eung Ahn}
\thanks{These authors contributed equally.}
\affiliation{Department of Physics, Pohang University of Science and Technology, Pohang, 37673, Republic of Korea}
\affiliation{Center for Artificial Low Dimensional Electronic Systems, Institute for Basic Science, Pohang 37673, Korea}

\author{Kyung-Hwan Jin}
\thanks{These authors contributed equally.}
\affiliation{Department of Physics, Jeonbuk National University, Jeonju, 54896, Republic of Korea}
\affiliation{Center for Artificial Low Dimensional Electronic Systems, Institute for Basic Science, Pohang 37673, Korea}

\author{Young-Jae Choi}
\affiliation{Center for Artificial Low Dimensional Electronic Systems, Institute for Basic Science, Pohang 37673, Korea}

\author{Jae Whan Park}
\affiliation{Center for Artificial Low Dimensional Electronic Systems, Institute for Basic Science, Pohang 37673, Korea}

\author{Han Woong Yeom}
\affiliation{Department of Physics, Pohang University of Science and Technology, Pohang, 37673, Republic of Korea}
\affiliation{Center for Artificial Low Dimensional Electronic Systems, Institute for Basic Science, Pohang 37673, Korea}

\author{Ara Go}
\email{arago@jnu.ac.kr}
\affiliation{Department of Physics, Chonnam National University, Gwangju 61186, Korea} 

\author{Yong Baek Kim}
\email{yongbaek.kim@utoronto.ca}
\affiliation{Department of Physics, University of Toronto, Toronto, Ontario M5S 1A7, Canada} 

\author{Gil Young Cho}
\email{gilyoungcho@postech.ac.kr}
\affiliation{Department of Physics, Pohang University of Science and Technology, Pohang, 37673, Republic of Korea}
\affiliation{Center for Artificial Low Dimensional Electronic Systems, Institute for Basic Science, Pohang 37673, Korea}
\affiliation{Asia-Pacific Center for Theoretical Physics, Pohang, Gyeongbuk, 37673, Korea}

\date{\today} 

\begin{abstract}
1T-transition metal dichalcogenides (TMD) have been an exciting platform for exploring the intertwinement of charge density waves and strong correlation phenomena. While the David star structure has been conventionally considered as the underlying charge order in the literature, recent scanning tunneling probe experiments on several monolayer 1T-TMD materials have motivated a new, alternative structure, namely the {anion-centered} David star structure. In this paper, we show that this novel {anion-centered} David star structure manifestly breaks inversion symmetry, resulting in flat bands with pronounced Rashba spin-orbit couplings. These distinctive features unlock novel possibilities and functionalities for 1T-TMDs, including the giant spin Hall effect, the emergence of Chern bands, and spin liquid that spontaneously breaks crystalline rotational symmetry. Our findings establish promising avenues for exploring emerging quantum phenomena of monolayer 1T-TMDs with this novel noncentrosymmetric structure.
\end{abstract}

\maketitle

1T-transition metal dichalcogenides (TMDs) have exhibited a surprisingly diverse set of emergent quantum phenomena, including ultra-fast manipulations of electronic states \cite{mak2012valley,xu2014tmd}, superconductivity~\cite{lu2015isingsc, yu2015gate, park2019emergent, lee2020stable, liu2016nature}, Mott insulators~\cite{sipos2008mott, Ross2011tmdmott, ligo2016tas2mott, naka2016nbse2mott, qiao2017mottness, liu2021direct, jung2023tas2mott}, topological insulators~\cite{li2014toptmd, lee2020stable, wata2018topexp}, and spin liquids~\cite{law2017tmdspliq, he2018spinon, chen2020tase2,ruan2021tase2,naka2021tanb}. At the heart of all these remarkable phenomena, forming a very particular charge density wave (CDW), known as the David star (DS), is crucial, as it serves as the origin of strong electronic correlation and geometric frustration in these materials~\cite{wil1975cdw,faze1980tas2,sipos2008mott, Ross2011tmdmott, ligo2016tas2mott, naka2016nbse2mott, qiao2017mottness, stoj2014tas2,rits2015tas2arpes}. Indeed, several recent experiments on samples with the DS CDW have shown the opening of Mott gap and potential spin liquid behaviors~\cite{cho2016tas2,chen2020tase2,ruan2021tase2,naka2021tanb}. These findings highlight the importance of the underlying charge order in understanding their strong correlation physics.

\begin{figure}
\includegraphics{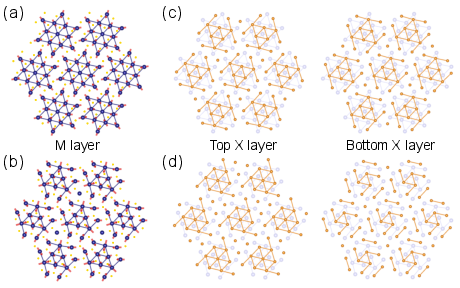}
\caption{\textbf{ACDS vs. DS Structures.} \textbf{(a)} and \textbf{(b)} represent the bonding networks of M atoms in the CDW states. Here, \textbf{(a)} is for the DS, and \textbf{(b)} is for the ACDS. The blue circles represent the M atoms. The arrows indicate the atomic displacement under the formation of the corresponding CDW patterns. Similarly, \textbf{(c)} and \textbf{(d)} represent the bonding networks of the top X layer (left) and the bottom X layer (right). Here, \textbf{(c)} is for the DS and \textbf{(d)} for the ACDS, respectively. The ACDS structure has distinct bonding networks of X atoms between the top and bottom X layers, leading to the broken inversion symmetry.}
\label{fig:MFig1}
\end{figure}

Scanning tunneling microscopes (STM) can probe the detailed structures of the monolayer TMD materials. In the DS CDW~[Fig.\ref{fig:MFig1}(a)], three bright protrusions around the DS center in the top chalcogenide layer are characteristically observed \cite{ligo2016tas2mott,qiao2017mottness}, consistent with our density functional theory (DFT) simulations {[Fig.S1]}. However, several monolayer samples showed markedly different patterns in recent STM studies~\cite{lin2020tase2,luic2019tas2,liu2021nbse2}. These experimental observations suggest the emergence of a new, alternative CDW pattern in monolayer 1T-TMD materials. In many CDW materials, it is well established that two or more CDW patterns can closely compete and even appear simultaneously~\cite{gye2019nbse2,liu2020vte2,tan2021kagome,bai2023nbte2}. Monolayer 1T-TMD may be no exception to this. Our recent work puts forth an alternative CDW structure for monolayer 1T-TMD~\cite{park2023accdw}, namely the anion-centered David star (ACDS)~[Fig.\ref{fig:MFig1}(b)], whose simulated STM images closely align with the previous experimental data [Fig.S1]. 

Building upon this recent progress, here we investigate the electronic states and symmetry of the ACDS structure for a few representative materials, namely TaSe$_2$, TaS$_2$, NbSe$_2$, and NbS$_2$. In this pursuit, we show that the ACDS structure breaks the inversion symmetry, leading to flat bands with pronounced spin-orbit couplings (SOC). This distinctive electronic state serves as the foundation for a myriad of emerging quantum phenomena, including giant spin Hall conductivity (SHC), interaction-enabled Chern bands, and strain-engineered spin liquids with spontaneously broken crystalline rotational symmetry. Our investigation provides valuable insights into this novel noncentrosymmetric structure in monolayer 1T-TMD materials.
 
{\textbf{1. Structure.}} The conventional DS structure naturally arises when the transition metal atoms contract toward the cation (Nb or Ta) reference atoms~[Fig.\ref{fig:MFig1}(a)]. However, the ACDS structure arises from the contraction towards an anion (S or Se) atom instead of cation atoms~[Fig.\ref{fig:MFig1}(b)]. Similar to the DS structure, the ACDS structure's unitcell consists of thirteen transition metal atoms and retains the compact CDW cluster, thereby facilitating the generation of flat bands. However, we find that, distinct from the DS, the ACDS structure is noncentrosymmetric. More specifically, unlike the DS structure~[Fig.\ref{fig:MFig1}(c)], where the arrangement of chalcogen atoms maintains inversion symmetry, the ACDS exhibits different bonding network patterns in the top and bottom chalcogenide layers~[Fig.\ref{fig:MFig1}(d)]. This leads to the explicitly broken inversion symmetry. We will show that this broken inversion symmetry, which was unnoticed in our previous work~\cite{park2023accdw}, is essential in understanding the electronic states of the ACDS materials.
 
Notably, our DFT calculations show that the energy difference between DS and ACDS structures is only several meV per atom [Table S1]. Consequently, it is natural to expect that ACDS is readily synthesized experimentally under appropriate conditions. Indeed, we {believe that ACDS is already realized in several monolayer 1T-TMD materials~\cite{lin2020tase2, luic2019tas2,liu2021nbse2}. For example, in \cite{lin2020tase2}, the STM images exhibit a single bright protrusion surrounded by six smaller protrusions. This feature is inconsistent with the DS structure and better explained by the ACDS. Similarly, the ACDS structure can account for several other STM images~\cite{luic2019tas2,liu2021nbse2}, which we summarize in [Fig.S1].}

\begin{figure}[!b]
\includegraphics{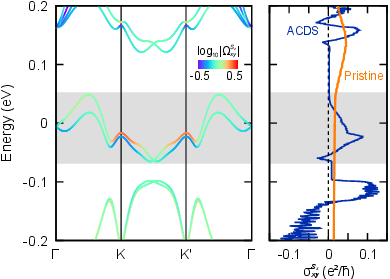}
\caption{\textbf{SHC of ACDS 1T-TaS${}_2$.} We present the band structure and SHC of the ACDS structure in energy. In the band structure, the color scheme represents the strength of the spin Hall Berry curvature $\Omega_{xy}^{S_z}$ \cite{Supp}. The shaded area represents the energy window which is attributed to the flat bands.}
\label{fig:MFig2}
\end{figure}

\begin{figure*}
\includegraphics{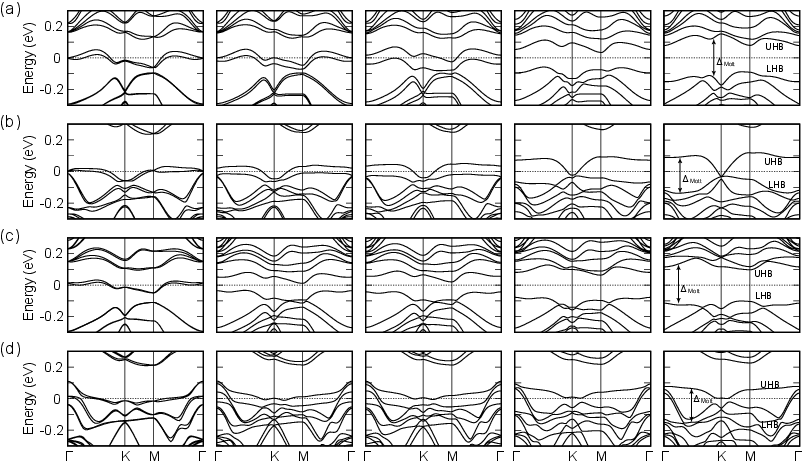}
\caption{\textbf{DFT+U calculation of ACDS 1T-TMD.} We depict the band structures for U= 0, 0.8, 1, 2, and 3 eV (from left to right) for \textbf{(a)} TaS$_2$, \textbf{(b)} TaSe$_2$, \textbf{(c)} NbS$_2$ and \textbf{(d)} NbSe$_2$, respectively.  }
\label{fig:MFig3}
\end{figure*}

 
{\textbf{2. Band Structure.}} The introduction of the ACDS in 1T-MX$_2$ leads to an intriguing electronic band structure with several notable features, as depicted in [Fig.\ref{fig:MFig2}, \ref{fig:MFig3}] and [Fig.S2]. First, one can observe the emergence of flat bands within the CDW gap. The bandwidth of the flat bands is around 80-120 meV, larger than those in the DS cases (around 20 meV). Nonetheless, the flat band states in the ACDS remain well-localized in space [Fig.S2(c,d)] and are half-filled. Furthermore, as apparent from [Fig.\ref{fig:MFig2}], due to the broken inversion symmetry, the flat bands exhibit Rashba-type spin splitting. 



{\textbf{3. Spin Hall Conductivity.}}
An exceptional interplay of the strong SOC and the large density of states of the flat bands in the ACDS structure can give rise to the giant SHC. Due to the dimensionality of the system and the nature of the Rashba SOC, the sole nontrivial component of the SHC tensor is $\sigma_{xy}^{S_z}$. For example, in TaS${}_2$~[Fig.\ref{fig:MFig2}], when the system is gated to tune the Fermi level between the flat bands, the estimated SHC of the monolayer can reach as high as $\sim 0.1~(e^2 /\hbar)$. When converted to values in 3D bulk~\cite{Supp}, it corresponds to $\sim 3 \times 10^4 (\hbar/e)(\Omega  m)^{-1}$, surpassing that of the pristine TaS${}_2$, other TMD and intrinsic spin Hall materials~\cite{xu2020she,feng2012she,safe2019she}. Notably, this giant SHC comes along with the energy splitting between the flat bands, {which is larger than the room temperature. The energy splitting reaches as high as 40 meV near the band maximum points and roughly 30 meV around the saddle-like regions~[Fig.\ref{fig:MFig2}], enabling the effective operation of potential spin devices at room temperature.} Other ACDS TMDs exhibit similar strengths of SHC as TaS${}_2$~[Fig.S4].


{\textbf{4. DFT+$U$ Calculation.}}
In the DS, the $5d_{z^2}$-electron in flat bands experiences substantial onsite Coulomb interaction of 1-2 eV, resulting in a Mott insulator~\cite{dara2014tas2dft,leej2021tas2}. Notably, the effective strength of the correlation $U$ and Mott characteristics can vary significantly, depending on stacking~\cite{lees2019tas2,wang2020tas2stacking,wuzo2022stack}, termination~\cite{butl2020tas2,leej2021tas2,peto2022tas2}, substrates~\cite{chen2020tase2,lin2020tase2}, and strain~\cite{zhang2020mottness}. We expect that the same applies to the ACDS structure while the effect of the electronic correlation is generally expected to be significant due to the presence of the flat bands. Hence, to comprehensively investigate diverse experimental possibilities, we performed DFT+$U$ calculations for a wide range of $U$~[Fig.\ref{fig:MFig3}].



In NbS$_2$ and TaS$_2$, where there are no states adjacent to the flat band, we observe that the flat bands gradually split in energy, leading to the formation of the lower and upper Hubbard bands at the moderate strength of the correlation $U \lesssim 1$ eV~[Fig.\ref{fig:MFig3}(a,c)]. This clearly manifests the transition from a half-filled metal to a trivial Mott insulator. On the other hand, TaSe$_2$ and NbSe$_2$ present distinct possibilities due to the neighboring bands near the flat bands. In TaSe$_2$~[Fig.\ref{fig:MFig3}(b)], the hybridization of the flat bands with neighboring states leads to a topological Mott insulator with the first Chern number of $1$~\cite{Supp}. In NbSe$_2$~[Fig.\ref{fig:MFig3}(d)], the flat bands remain relatively buried beneath other bands, which allows the system to maintain its metallic character. However, {when $U$ reaches the larger values, $U \gtrsim 2$ eV, the flat bands in both TaSe$_2$ and NbSe$_2$ slowly start to develop a trivial energy gap between them. This energy splitting between the flat bands subsequently leads to the suppression of the density of states near the Fermi level~\cite{Supp}.}
 

Remarkably, previous STM experiments~\cite{luic2019tas2,lin2020tase2,liu2021nbse2}, which reported images consistent with the ACDS, have unveiled the presence of substantial spectral gaps at the Fermi level. While a more systematic investigation is desirable to fully comprehend the nature of the observed gaps, these observations indicate the importance of electronic correlation and possibly the emergence of Mott insulators in these half-filled systems. 

\begin{figure*}[!t]
\includegraphics[width=\textwidth]{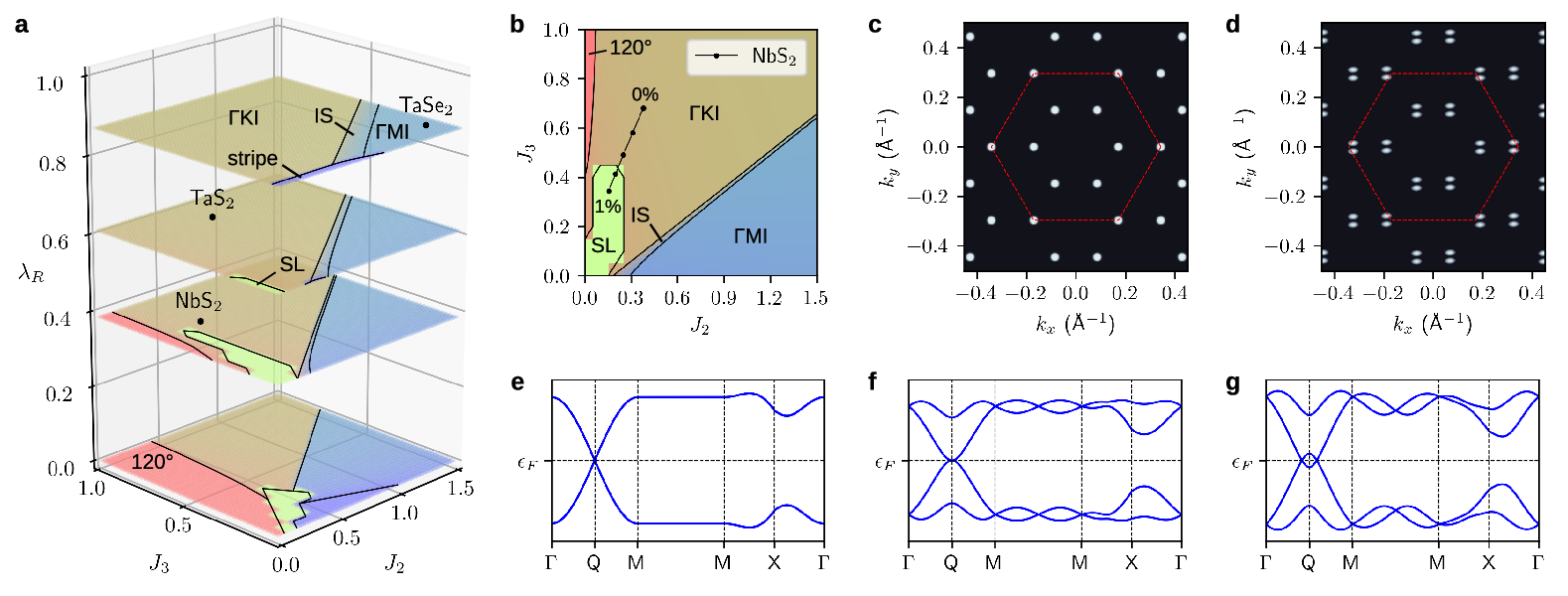}
\caption{\textbf{Magnetic phase diagrams \& Spin liquids.} \textbf{(a)} $\{J_2, J_3, \lambda_R\}$ and \textbf{(b)} $\{J_2, J_3, \lambda_R=0.45 \}$ VMC phase diagrams in units of $J_1^{(0)}=t_1=1$. In (a), the locations of TaS${}_2$, TaSe${}_2$ and NbS${}_2$ are identified as black circles. In (b), NbS$_2$ with the strain 0\%, 0.25\%, 0.5\%, 0.75\%, and 1\% are marked as the black circles. \textbf{(c)} and \textbf{(d)} represent the calculated STM signals (in arbitrary units) of the fully-symmetric U1B11 state and nematic Dirac spin liquid, respectively. We followed \cite{tang2013low} in calculating the STM signals of spin liquids \cite{Supp}. STM signal in (d) clearly breaks the crystalline rotational symmetry unlike (c). When calculating (c) and (d), we used the parameters of NbS$_2$ at 1\% strain and {tuned the voltage bias of STM to be right above the charge gap.} We depict spinon band structures of \textbf{(e)} U1B11 spin liquid without SOC, \textbf{(f)} U1B11 state with SOC Eq.\eqref{eq:Spin_Model}, and \textbf{(g)} nematic Dirac spin liquid.}
\label{fig:MFig4}
\end{figure*}

{\textbf{5. Spin Model.}}
Motivated by our DFT+$U$ calculations and spectral gaps observed in experiments~\cite{luic2019tas2,lin2020tase2,liu2021nbse2}, we derive an effective spin model by expanding around the deep Mott limit $U\gg\mathrm{max}\{t_1,t_2,t_3,\lambda_R\}$ ($t_k$ is the $k$-th neighbor hopping and $\lambda_R$ is the Rashba SOC) \cite{Supp}
\begin{align}
H &= J_1\sum_{\ev{i,j}} \mathbf{S}_i\cdot\mathbf{S}_j+J_2\sum_{\ev{\ev{i,j}}}\mathbf{S}_i\cdot\mathbf{S}_j+J_3\sum_{\ev{\ev{\ev{i,j}}}}\mathbf{S}_i\cdot\mathbf{S}_j \nonumber\\ 
&+\sum_{\ev{i,j}}\left[J_\parallel\left(\mathbf{r}_{ij}\cdot\mathbf{S}_i\right)\left(\mathbf{r}_{ij}\cdot\mathbf{S}_j\right)-D\mathbf{r}_{ij}\cdot\left(\mathbf{S}_i\times\mathbf{S}_j\right)\right].
\label{eq:Spin_Model}
\end{align}
The parameters of Eq.\eqref{eq:Spin_Model} are determined by the hopping and SOC strengths, for example, $J_1 = 4(t_1^2-\lambda_R^2)/U$. Their explicit values can be found in \cite{Supp}. The further neighbor terms and higher-order corrections are typically orders-of-magnitude smaller and thus ignored~\cite{Supp}. $\mathbf{S}_i$ is the spin-1/2 moment at site $i$, and $\ev{i,j}$ refers to nearest neighbor pairs. $\mathbf{r}_{ij}$ is the unit vector directed from the site $i$ to the site $j$. Similarly, $\ev{\ev{i,j}}$ and $\ev{\ev{\ev{i,j}}}$ refer to second- and third-neighbor pairs.

{\textbf{6. Magnetic Phase Diagram.}} Using variational Monte Carlo (VMC)~\cite{gros1988superconductivity, gros1989physics, edegger2007gutzwiller} calculations and Luttinger-Tisza method~\cite{luttinger1946theory, litvin1974luttinger}, we draw the phase diagram [Fig.\ref{fig:MFig4}(a)] of Eq.\eqref{eq:Spin_Model} in terms of $\{ J_2/J_1^{(0)}, J_3/J_1^{(0)}, \lambda_{R}/t_1 \}$ with $J_1^{(0)} = 4t_1^2/U$. This is a standard approach in computing phase diagrams in spin Hamiltonians, see for example \cite{iaconis2018spin}. We append some details on calculation of the phase diagram in \cite{Supp}. The Rashba SOC $\lambda_R$ controls the strength of both the $J_\parallel$- and $D$-terms. We normalize the parameters by setting $J_1^{(0)}=t_1=1$ from here and on.


For sufficiently large values of $\{\lambda_R, J_2, J_3\}$, classical magnetic orders prevail. These orders are denoted as the 120\textdegree, {stripe}, $\Gamma$KI, $\Gamma$MI, and incommensurate spiral (IS) orders in [Fig.\ref{fig:MFig4}(a)]. Their real-space spin configurations can be found in \cite{Supp}. $\Gamma$KI, $\Gamma$MI, and IS are stripy, coplanar magnetic orders whose ordering momentum vary continuously with the parameters of Eq.\eqref{eq:Spin_Model}. For example, in the $\Gamma$KI order, the spins spiral along the momentum, which lies between the $\Gamma$ and $K$ points in the Brillouin zone. The appearance of these magnetic orders are consistent with prior investigations of the $J_1 J_2 J_3$ model~\cite{gong2019chiral,iqbal2016spin,ferrari2019dynamical,iaconis2018spin,zhu2018topography,sherman2023spectral,drescher2022dynamical}, which is closely related to Eq.\eqref{eq:Spin_Model}. 


Remarkably, our VMC simulation has identified the emergence of a spin liquid for lower values of $\{J_2,J_3\}$, labelled as SL in [Fig.\ref{fig:MFig4}(a,b)]. {We employed projected fermionic spinon wavefunctions \cite{wen2002quantum} in the calculation to construct the candidate spin liquid states \cite{Supp}.} Again, our findings regarding the existence and location of the spin liquid phase are in agreement with prior investigations of the $J_1 J_2 J_3$ model~\cite{gong2019chiral,iqbal2016spin,ferrari2019dynamical,iaconis2018spin,zhu2018topography,sherman2023spectral,drescher2022dynamical}. Intriguingly, we observed that introducing a small Rashba SOC {$\lambda_R/t_1\lesssim 0.4$} leads to an expansion of the spin liquid region in the phase diagram. 

It is noteworthy that NbS$_2$ resides in close proximity to the spin liquid in the phase diagram~[Fig.\ref{fig:MFig4}(a)]. The proximity of NbS$_2$ to the spin liquid phase suggests a tantalizing possibility of engineering a spin liquid state in NbS$_2$. Remarkably, under the influence of a small, isotropic tensile strain $\lesssim 1\%$, {we predict} NbS$_2$ undergoes a phase transition and become the spin liquid. As the tensile strain increases from 0\% to 1\%, the second and third neighbor interactions $\{J_2,J_3\}$ decrease and the Rashba SOC increases from $\lambda_R=0.38$ to $\lambda_R=0.45$~\cite{Supp}. While the increment of $\lambda_R$ does cause a slight suppression of the spin liquid region, the reduction of $\{J_2,J_3\}$ is fast enough that NbS$_2$ transits to a spin liquid near 0.75\% strain. See [Fig.\ref{fig:MFig4}(b)] for the trajectory of $\{J_2,J_3\}$ parameters of NbS$_2$ under the strain, projected onto the $\lambda_R=0.45$ plane. {We also confirmed that the ACDS structure in NbS$_2$ is stable against the strain $\lesssim 1\%$ ~\cite{Supp}.} The other TMD materials are deep inside magnetically ordered phases~[Fig.\ref{fig:MFig4}(a)]. 


{\textbf{7. Nature of Spin Liquid.}}
The strong SOC of the ACDS significantly changes the characteristics of the spin liquid state. Specifically, it destabilizes a fully symmetric spin liquid state, leading to an alternative state with the broken crystalline rotational symmetry. 

In the absence of SOC, previous studies established the emergence of a symmetric U(1) Dirac spin liquid, which we refer as the U1B11 state~\cite{Supp}, in the $J_1 J_2 J_3$ model~\cite{iqbal2016spin,ferrari2019dynamical,iaconis2018spin,zhu2018topography,sherman2023spectral,drescher2022dynamical,hu2019dirac}. This state is characterized by two Dirac spinons and artificial gauge photons, with each Dirac cone exhibiting double degeneracy due to spin-rotational symmetry~[Fig.\ref{fig:MFig4}(e)]. We extended this U1B11 state to incorporate SOC and observed that it remains energetically favored over other symmetric U(1) spin liquids. The SOC, however, continuously deforms the spinon band structure from doubly-degenerate Dirac cones~[Fig.\ref{fig:MFig4}(e)] to non-degenerate quadratic band touchings~[Fig.\ref{fig:MFig4}(f)]. It is well known that quadratic band touching is unstable against repulsive interactions, leading to symmetry broken states~\cite{sun2009topological}. Guided by previous literature on the instabilities of the quadratic band touching~\cite{sun2009topological,uebelacker2011instabilities,zhu2016interaction}, we additionally considered five symmetry-broken descendants of the U1B11 state {
and compared their energies within the VMC simulation~\cite{Supp}. Among the candidates, the U(1) nematic Dirac spin liquid, a spin liquid with {spontaneously} broken {crystalline rotational} symmetry, has the lowest energy than the fully symmetry U1B11 state and other candidates.} In particular, strained NbS$_2$ is in the nematic Dirac spin liquid phase~[Fig.\ref{fig:MFig4}(b)], whose broken rotational symmetry can be manifested in STM signals~[Fig.\ref{fig:MFig4}(c,d)]. {The corresponding real-space STM images can be found in \cite{Supp}.}

{\textbf{8. Conclusions.}}
We have demonstrated that ACDS TMDs showcase a fascinating array of intriguing emergent quantum phenomena, notably including giant spin Hall conductivity, interaction-enabled topological Chern bands, and strain-engineered nematic spin liquids. The pivotal factor in understandings of these phenomena lies in the broken inversion symmetry inherent to the ACDS structure, resulting in flat bands with strong SOC. 

{Remarkably, these emergent quantum phenomena are anticipated to manifest within the readily-accessible ACDS samples\cite{lin2020tase2, luic2019tas2,liu2021nbse2}. By juxtaposing the experimentally-probed density of states~\cite{lin2020tase2, luic2019tas2,liu2021nbse2} with our DFT+U calculations, we could deduce the appropriate U values for these ACDS samples, thereby unveiling their emergent properties\cite{Supp}. For instance, our analysis reveals that ACDS TaS$_2$ exhibits a U value of 2.8 eV, resulting in a trivial Mott insulator characterized by $\Gamma KI$ magnetic order. Similarly, TaSe$_2$ realizes a topological Mott insulator, while NbSe$_2$ demonstrates the giant SHC, and NbS$_2$ exhibits a nematic spin liquid~\cite{Supp}. These findings present a compelling opportunity for further experimental exploration to validate our theoretical predictions and shed light on the intriguing physics of ACDS materials.} 


\acknowledgments
We are grateful to Hyun-Woo Lee and Young-Woo Son for valuable discussions. C.-E A. and G.Y.C. are supported by Samsung Science and Technology Foundation under Project Number SSTF-BA2002-05. C.-E A. and G.Y.C. are supported by the NRF of Korea (Grant No. RS-2023-00208291, No. 2023M3K5A1094810, No. 2023M3K5A1094813) funded by the Korean Government (MSIT). K.-H. J. acknowledges the support from the Institute for Basic Science (Grant No. IBS-R014-Y1). C.-E A. and G.Y.C. acknowledge the support by the Air Force Office of Scientific Research under Award No. FA2386-22-1-4061. A.G. is supported by the NRF of Korea (Grant No. 2021R1C1C1010429 and No. 2023M3K5A1094813). Y.B.K. is supported by the Natural Sciences and Engineering Council of Canada, the Center for Quantum Materials at the University of Toronto, Simons Fellowship, and Guggenheim Fellowship. G.Y.C., J.W.P. and H.W.Y. are supported by Institute of Basic Science under project code IBS-R014-D1. 

\bibliographystyle{apsrev4-1}
\bibliography{refs.bib}

\begin{thebibliography}{77}%
\makeatletter
\providecommand \@ifxundefined [1]{%
 \@ifx{#1\undefined}
}%
\providecommand \@ifnum [1]{%
 \ifnum #1\expandafter \@firstoftwo
 \else \expandafter \@secondoftwo
 \fi
}%
\providecommand \@ifx [1]{%
 \ifx #1\expandafter \@firstoftwo
 \else \expandafter \@secondoftwo
 \fi
}%
\providecommand \natexlab [1]{#1}%
\providecommand \enquote  [1]{``#1''}%
\providecommand \bibnamefont  [1]{#1}%
\providecommand \bibfnamefont [1]{#1}%
\providecommand \citenamefont [1]{#1}%
\providecommand \href@noop [0]{\@secondoftwo}%
\providecommand \href [0]{\begingroup \@sanitize@url \@href}%
\providecommand \@href[1]{\@@startlink{#1}\@@href}%
\providecommand \@@href[1]{\endgroup#1\@@endlink}%
\providecommand \@sanitize@url [0]{\catcode `\\12\catcode `\$12\catcode
  `\&12\catcode `\#12\catcode `\^12\catcode `\_12\catcode `\%12\relax}%
\providecommand \@@startlink[1]{}%
\providecommand \@@endlink[0]{}%
\providecommand \url  [0]{\begingroup\@sanitize@url \@url }%
\providecommand \@url [1]{\endgroup\@href {#1}{\urlprefix }}%
\providecommand \urlprefix  [0]{URL }%
\providecommand \Eprint [0]{\href }%
\providecommand \doibase [0]{http://dx.doi.org/}%
\providecommand \selectlanguage [0]{\@gobble}%
\providecommand \bibinfo  [0]{\@secondoftwo}%
\providecommand \bibfield  [0]{\@secondoftwo}%
\providecommand \translation [1]{[#1]}%
\providecommand \BibitemOpen [0]{}%
\providecommand \bibitemStop [0]{}%
\providecommand \bibitemNoStop [0]{.\EOS\space}%
\providecommand \EOS [0]{\spacefactor3000\relax}%
\providecommand \BibitemShut  [1]{\csname bibitem#1\endcsname}%
\let\auto@bib@innerbib\@empty
\bibitem [{\citenamefont {Mak}\ \emph {et~al.}(2012)\citenamefont {Mak},
  \citenamefont {He}, \citenamefont {Shan},\ and\ \citenamefont
  {Heinz}}]{mak2012valley}%
  \BibitemOpen
  \bibfield  {author} {\bibinfo {author} {\bibfnamefont {K.~F.}\ \bibnamefont
  {Mak}}, \bibinfo {author} {\bibfnamefont {K.}~\bibnamefont {He}}, \bibinfo
  {author} {\bibfnamefont {J.}~\bibnamefont {Shan}}, \ and\ \bibinfo {author}
  {\bibfnamefont {T.~F.}\ \bibnamefont {Heinz}},\ }\href {\doibase
  10.1038/NNANO.2012.96} {\bibfield  {journal} {\bibinfo  {journal} {Nat.
  Nanotech.}\ }\textbf {\bibinfo {volume} {7}},\ \bibinfo {pages} {494}
  (\bibinfo {year} {2012})}\BibitemShut {NoStop}%
\bibitem [{\citenamefont {Xu}\ \emph {et~al.}(2014)\citenamefont {Xu},
  \citenamefont {Yao}, \citenamefont {Xiao},\ and\ \citenamefont
  {Heinz}}]{xu2014tmd}%
  \BibitemOpen
  \bibfield  {author} {\bibinfo {author} {\bibfnamefont {X.}~\bibnamefont
  {Xu}}, \bibinfo {author} {\bibfnamefont {W.}~\bibnamefont {Yao}}, \bibinfo
  {author} {\bibfnamefont {D.}~\bibnamefont {Xiao}}, \ and\ \bibinfo {author}
  {\bibfnamefont {T.~F.}\ \bibnamefont {Heinz}},\ }\href {\doibase
  10.1038/NPHYS2942} {\bibfield  {journal} {\bibinfo  {journal} {Nat. Phys.}\
  }\textbf {\bibinfo {volume} {10}},\ \bibinfo {pages} {343} (\bibinfo {year}
  {2014})}\BibitemShut {NoStop}%
\bibitem [{\citenamefont {Lu}\ \emph {et~al.}(2015)\citenamefont {Lu},
  \citenamefont {Zheliuk}, \citenamefont {Leermakers}, \citenamefont {Yuan},
  \citenamefont {Zeitler}, \citenamefont {Law},\ and\ \citenamefont
  {Ye}}]{lu2015isingsc}%
  \BibitemOpen
  \bibfield  {author} {\bibinfo {author} {\bibfnamefont {J.~M.}\ \bibnamefont
  {Lu}}, \bibinfo {author} {\bibfnamefont {O.}~\bibnamefont {Zheliuk}},
  \bibinfo {author} {\bibfnamefont {I.}~\bibnamefont {Leermakers}}, \bibinfo
  {author} {\bibfnamefont {N.~F.~Q.}\ \bibnamefont {Yuan}}, \bibinfo {author}
  {\bibfnamefont {U.}~\bibnamefont {Zeitler}}, \bibinfo {author} {\bibfnamefont
  {K.~T.}\ \bibnamefont {Law}}, \ and\ \bibinfo {author} {\bibfnamefont
  {J.~T.}\ \bibnamefont {Ye}},\ }\href {\doibase 10.1126/science.aab2277}
  {\bibfield  {journal} {\bibinfo  {journal} {Science}\ }\textbf {\bibinfo
  {volume} {350}},\ \bibinfo {pages} {1353} (\bibinfo {year}
  {2015})}\BibitemShut {NoStop}%
\bibitem [{\citenamefont {Yu}\ \emph {et~al.}(2015)\citenamefont {Yu},
  \citenamefont {Yang}, \citenamefont {Lu}, \citenamefont {Yan}, \citenamefont
  {Cho}, \citenamefont {Ma}, \citenamefont {Niu}, \citenamefont {Kim},
  \citenamefont {Son}, \citenamefont {Feng} \emph {et~al.}}]{yu2015gate}%
  \BibitemOpen
  \bibfield  {author} {\bibinfo {author} {\bibfnamefont {Y.}~\bibnamefont
  {Yu}}, \bibinfo {author} {\bibfnamefont {F.}~\bibnamefont {Yang}}, \bibinfo
  {author} {\bibfnamefont {X.~F.}\ \bibnamefont {Lu}}, \bibinfo {author}
  {\bibfnamefont {Y.~J.}\ \bibnamefont {Yan}}, \bibinfo {author} {\bibfnamefont
  {Y.-H.}\ \bibnamefont {Cho}}, \bibinfo {author} {\bibfnamefont
  {L.}~\bibnamefont {Ma}}, \bibinfo {author} {\bibfnamefont {X.}~\bibnamefont
  {Niu}}, \bibinfo {author} {\bibfnamefont {S.}~\bibnamefont {Kim}}, \bibinfo
  {author} {\bibfnamefont {Y.-W.}\ \bibnamefont {Son}}, \bibinfo {author}
  {\bibfnamefont {D.}~\bibnamefont {Feng}},  \emph {et~al.},\ }\href {\doibase
  10.1038/NNANO.2014.323} {\bibfield  {journal} {\bibinfo  {journal} {Nat.
  Nanotech.}\ }\textbf {\bibinfo {volume} {10}},\ \bibinfo {pages} {270}
  (\bibinfo {year} {2015})}\BibitemShut {NoStop}%
\bibitem [{\citenamefont {Park}\ \emph {et~al.}(2019)\citenamefont {Park},
  \citenamefont {Cho}, \citenamefont {Lee},\ and\ \citenamefont
  {Yeom}}]{park2019emergent}%
  \BibitemOpen
  \bibfield  {author} {\bibinfo {author} {\bibfnamefont {J.~W.}\ \bibnamefont
  {Park}}, \bibinfo {author} {\bibfnamefont {G.~Y.}\ \bibnamefont {Cho}},
  \bibinfo {author} {\bibfnamefont {J.}~\bibnamefont {Lee}}, \ and\ \bibinfo
  {author} {\bibfnamefont {H.~W.}\ \bibnamefont {Yeom}},\ }\href {\doibase
  10.1038/s41467-019-11981-5} {\bibfield  {journal} {\bibinfo  {journal} {Nat.
  Commun.}\ }\textbf {\bibinfo {volume} {10}},\ \bibinfo {pages} {4038}
  (\bibinfo {year} {2019})}\BibitemShut {NoStop}%
\bibitem [{\citenamefont {Lee}\ \emph {et~al.}(2020{\natexlab{a}})\citenamefont
  {Lee}, \citenamefont {Geng}, \citenamefont {Park}, \citenamefont {Oshikawa},
  \citenamefont {Lee}, \citenamefont {Yeom},\ and\ \citenamefont
  {Cho}}]{lee2020stable}%
  \BibitemOpen
  \bibfield  {author} {\bibinfo {author} {\bibfnamefont {J.~M.}\ \bibnamefont
  {Lee}}, \bibinfo {author} {\bibfnamefont {C.}~\bibnamefont {Geng}}, \bibinfo
  {author} {\bibfnamefont {J.~W.}\ \bibnamefont {Park}}, \bibinfo {author}
  {\bibfnamefont {M.}~\bibnamefont {Oshikawa}}, \bibinfo {author}
  {\bibfnamefont {S.-S.}\ \bibnamefont {Lee}}, \bibinfo {author} {\bibfnamefont
  {H.~W.}\ \bibnamefont {Yeom}}, \ and\ \bibinfo {author} {\bibfnamefont
  {G.~Y.}\ \bibnamefont {Cho}},\ }\href {\doibase
  10.1103/PhysRevLett.124.137002} {\bibfield  {journal} {\bibinfo  {journal}
  {Phys. Rev. Lett.}\ }\textbf {\bibinfo {volume} {124}},\ \bibinfo {pages}
  {137002} (\bibinfo {year} {2020}{\natexlab{a}})}\BibitemShut {NoStop}%
\bibitem [{\citenamefont {Liu}\ \emph {et~al.}(2016)\citenamefont {Liu},
  \citenamefont {Shao}, \citenamefont {Li}, \citenamefont {Lu}, \citenamefont
  {Zhu}, \citenamefont {Tong}, \citenamefont {Xiao}, \citenamefont {Ling},
  \citenamefont {Xi}, \citenamefont {Pi} \emph {et~al.}}]{liu2016nature}%
  \BibitemOpen
  \bibfield  {author} {\bibinfo {author} {\bibfnamefont {Y.}~\bibnamefont
  {Liu}}, \bibinfo {author} {\bibfnamefont {D.-F.}\ \bibnamefont {Shao}},
  \bibinfo {author} {\bibfnamefont {L.}~\bibnamefont {Li}}, \bibinfo {author}
  {\bibfnamefont {W.}~\bibnamefont {Lu}}, \bibinfo {author} {\bibfnamefont
  {X.}~\bibnamefont {Zhu}}, \bibinfo {author} {\bibfnamefont {P.}~\bibnamefont
  {Tong}}, \bibinfo {author} {\bibfnamefont {R.}~\bibnamefont {Xiao}}, \bibinfo
  {author} {\bibfnamefont {L.}~\bibnamefont {Ling}}, \bibinfo {author}
  {\bibfnamefont {C.}~\bibnamefont {Xi}}, \bibinfo {author} {\bibfnamefont
  {L.}~\bibnamefont {Pi}},  \emph {et~al.},\ }\href {\doibase
  10.1103/PhysRevB.94.045131} {\bibfield  {journal} {\bibinfo  {journal} {Phys.
  Rev. B}\ }\textbf {\bibinfo {volume} {94}},\ \bibinfo {pages} {045131}
  (\bibinfo {year} {2016})}\BibitemShut {NoStop}%
\bibitem [{\citenamefont {Sipos}\ \emph {et~al.}(2008)\citenamefont {Sipos},
  \citenamefont {Kusmartseva}, \citenamefont {Akrap}, \citenamefont {Berger},
  \citenamefont {Forro},\ and\ \citenamefont {Tutis}}]{sipos2008mott}%
  \BibitemOpen
  \bibfield  {author} {\bibinfo {author} {\bibfnamefont {B.}~\bibnamefont
  {Sipos}}, \bibinfo {author} {\bibfnamefont {A.~F.}\ \bibnamefont
  {Kusmartseva}}, \bibinfo {author} {\bibfnamefont {A.}~\bibnamefont {Akrap}},
  \bibinfo {author} {\bibfnamefont {H.}~\bibnamefont {Berger}}, \bibinfo
  {author} {\bibfnamefont {L.}~\bibnamefont {Forro}}, \ and\ \bibinfo {author}
  {\bibfnamefont {E.}~\bibnamefont {Tutis}},\ }\href {\doibase
  10.1038/nmat2318} {\bibfield  {journal} {\bibinfo  {journal} {Nat. Mater.}\
  }\textbf {\bibinfo {volume} {7}},\ \bibinfo {pages} {960} (\bibinfo {year}
  {2008})}\BibitemShut {NoStop}%
\bibitem [{\citenamefont {Rossnagel}(2011)}]{Ross2011tmdmott}%
  \BibitemOpen
  \bibfield  {author} {\bibinfo {author} {\bibfnamefont {K.}~\bibnamefont
  {Rossnagel}},\ }\href {\doibase 10.1088/0953-8984/23/21/213001} {\bibfield
  {journal} {\bibinfo  {journal} {J. Phys. Condens. Matter.}\ }\textbf
  {\bibinfo {volume} {23}},\ \bibinfo {pages} {213001} (\bibinfo {year}
  {2011})}\BibitemShut {NoStop}%
\bibitem [{\citenamefont {Ma}\ \emph {et~al.}(2016)\citenamefont {Ma},
  \citenamefont {Ye}, \citenamefont {Yu}, \citenamefont {Lu}, \citenamefont
  {Niu}, \citenamefont {Kim}, \citenamefont {Feng}, \citenamefont {Tomanek},
  \citenamefont {Son}, \citenamefont {Chen},\ and\ \citenamefont
  {Zhang}}]{ligo2016tas2mott}%
  \BibitemOpen
  \bibfield  {author} {\bibinfo {author} {\bibfnamefont {L.}~\bibnamefont
  {Ma}}, \bibinfo {author} {\bibfnamefont {C.}~\bibnamefont {Ye}}, \bibinfo
  {author} {\bibfnamefont {Y.}~\bibnamefont {Yu}}, \bibinfo {author}
  {\bibfnamefont {X.~F.}\ \bibnamefont {Lu}}, \bibinfo {author} {\bibfnamefont
  {X.}~\bibnamefont {Niu}}, \bibinfo {author} {\bibfnamefont {S.}~\bibnamefont
  {Kim}}, \bibinfo {author} {\bibfnamefont {D.}~\bibnamefont {Feng}}, \bibinfo
  {author} {\bibfnamefont {D.}~\bibnamefont {Tomanek}}, \bibinfo {author}
  {\bibfnamefont {Y.-W.}\ \bibnamefont {Son}}, \bibinfo {author} {\bibfnamefont
  {X.~H.}\ \bibnamefont {Chen}}, \ and\ \bibinfo {author} {\bibfnamefont
  {Y.}~\bibnamefont {Zhang}},\ }\href {\doibase 10.1038/ncomms10956} {\bibfield
   {journal} {\bibinfo  {journal} {Nat. Commun.}\ }\textbf {\bibinfo {volume}
  {7}},\ \bibinfo {pages} {10956} (\bibinfo {year} {2016})}\BibitemShut
  {NoStop}%
\bibitem [{\citenamefont {Nakata}\ \emph {et~al.}(2016)\citenamefont {Nakata},
  \citenamefont {Sugawara}, \citenamefont {Shimizu}, \citenamefont {Okada},
  \citenamefont {Han}, \citenamefont {Hitosugi}, \citenamefont {Ueno},
  \citenamefont {Sato},\ and\ \citenamefont {Takahashi}}]{naka2016nbse2mott}%
  \BibitemOpen
  \bibfield  {author} {\bibinfo {author} {\bibfnamefont {Y.}~\bibnamefont
  {Nakata}}, \bibinfo {author} {\bibfnamefont {K.}~\bibnamefont {Sugawara}},
  \bibinfo {author} {\bibfnamefont {R.}~\bibnamefont {Shimizu}}, \bibinfo
  {author} {\bibfnamefont {Y.}~\bibnamefont {Okada}}, \bibinfo {author}
  {\bibfnamefont {P.}~\bibnamefont {Han}}, \bibinfo {author} {\bibfnamefont
  {T.}~\bibnamefont {Hitosugi}}, \bibinfo {author} {\bibfnamefont
  {K.}~\bibnamefont {Ueno}}, \bibinfo {author} {\bibfnamefont {T.}~\bibnamefont
  {Sato}}, \ and\ \bibinfo {author} {\bibfnamefont {T.}~\bibnamefont
  {Takahashi}},\ }\href {\doibase 10.1038/am.2016.157} {\bibfield  {journal}
  {\bibinfo  {journal} {NPG Asia Mater.}\ }\textbf {\bibinfo {volume} {8}},\
  \bibinfo {pages} {e321} (\bibinfo {year} {2016})}\BibitemShut {NoStop}%
\bibitem [{\citenamefont {Qiao}\ \emph {et~al.}(2017)\citenamefont {Qiao},
  \citenamefont {Li}, \citenamefont {Wang}, \citenamefont {Ruan}, \citenamefont
  {Ye}, \citenamefont {Cai}, \citenamefont {Hao}, \citenamefont {Yao},
  \citenamefont {Chen}, \citenamefont {Wu}, \citenamefont {Wang},\ and\
  \citenamefont {Liu}}]{qiao2017mottness}%
  \BibitemOpen
  \bibfield  {author} {\bibinfo {author} {\bibfnamefont {S.}~\bibnamefont
  {Qiao}}, \bibinfo {author} {\bibfnamefont {X.}~\bibnamefont {Li}}, \bibinfo
  {author} {\bibfnamefont {N.}~\bibnamefont {Wang}}, \bibinfo {author}
  {\bibfnamefont {W.}~\bibnamefont {Ruan}}, \bibinfo {author} {\bibfnamefont
  {C.}~\bibnamefont {Ye}}, \bibinfo {author} {\bibfnamefont {P.}~\bibnamefont
  {Cai}}, \bibinfo {author} {\bibfnamefont {Z.}~\bibnamefont {Hao}}, \bibinfo
  {author} {\bibfnamefont {H.}~\bibnamefont {Yao}}, \bibinfo {author}
  {\bibfnamefont {X.}~\bibnamefont {Chen}}, \bibinfo {author} {\bibfnamefont
  {J.}~\bibnamefont {Wu}}, \bibinfo {author} {\bibfnamefont {Y.}~\bibnamefont
  {Wang}}, \ and\ \bibinfo {author} {\bibfnamefont {Z.}~\bibnamefont {Liu}},\
  }\href {\doibase 10.1103/PhysRevX.7.041054} {\bibfield  {journal} {\bibinfo
  {journal} {Phys. Rev. X}\ }\textbf {\bibinfo {volume} {7}},\ \bibinfo {pages}
  {041054} (\bibinfo {year} {2017})}\BibitemShut {NoStop}%
\bibitem [{\citenamefont {Liu}\ \emph {et~al.}(2021{\natexlab{a}})\citenamefont
  {Liu}, \citenamefont {Yang}, \citenamefont {Huang}, \citenamefont {Song},
  \citenamefont {Zhang}, \citenamefont {Huang}, \citenamefont {Hou},
  \citenamefont {Chen}, \citenamefont {Xu}, \citenamefont {Zhang} \emph
  {et~al.}}]{liu2021direct}%
  \BibitemOpen
  \bibfield  {author} {\bibinfo {author} {\bibfnamefont {L.}~\bibnamefont
  {Liu}}, \bibinfo {author} {\bibfnamefont {H.}~\bibnamefont {Yang}}, \bibinfo
  {author} {\bibfnamefont {Y.}~\bibnamefont {Huang}}, \bibinfo {author}
  {\bibfnamefont {X.}~\bibnamefont {Song}}, \bibinfo {author} {\bibfnamefont
  {Q.}~\bibnamefont {Zhang}}, \bibinfo {author} {\bibfnamefont
  {Z.}~\bibnamefont {Huang}}, \bibinfo {author} {\bibfnamefont
  {Y.}~\bibnamefont {Hou}}, \bibinfo {author} {\bibfnamefont {Y.}~\bibnamefont
  {Chen}}, \bibinfo {author} {\bibfnamefont {Z.}~\bibnamefont {Xu}}, \bibinfo
  {author} {\bibfnamefont {T.}~\bibnamefont {Zhang}},  \emph {et~al.},\ }\href
  {\doibase 10.1038/s41467-021-22233-w} {\bibfield  {journal} {\bibinfo
  {journal} {Nat. Commun.}\ }\textbf {\bibinfo {volume} {12}},\ \bibinfo
  {pages} {1978} (\bibinfo {year} {2021}{\natexlab{a}})}\BibitemShut {NoStop}%
\bibitem [{\citenamefont {Jung}\ \emph {et~al.}(2023)\citenamefont {Jung},
  \citenamefont {Jin}, \citenamefont {Kim},\ and\ \citenamefont
  {Yeom}}]{jung2023tas2mott}%
  \BibitemOpen
  \bibfield  {author} {\bibinfo {author} {\bibfnamefont {J.}~\bibnamefont
  {Jung}}, \bibinfo {author} {\bibfnamefont {K.-H.}\ \bibnamefont {Jin}},
  \bibinfo {author} {\bibfnamefont {J.}~\bibnamefont {Kim}}, \ and\ \bibinfo
  {author} {\bibfnamefont {H.~W.}\ \bibnamefont {Yeom}},\ }\href {\doibase
  10.1021/acs.nanolett.3c02003} {\bibfield  {journal} {\bibinfo  {journal}
  {Nano Letters}\ }\textbf {\bibinfo {volume} {23}},\ \bibinfo {pages} {8029}
  (\bibinfo {year} {2023})}\BibitemShut {NoStop}%
\bibitem [{\citenamefont {Qian}\ \emph {et~al.}(2014)\citenamefont {Qian},
  \citenamefont {Liu}, \citenamefont {Fu},\ and\ \citenamefont
  {Li}}]{li2014toptmd}%
  \BibitemOpen
  \bibfield  {author} {\bibinfo {author} {\bibfnamefont {X.}~\bibnamefont
  {Qian}}, \bibinfo {author} {\bibfnamefont {J.}~\bibnamefont {Liu}}, \bibinfo
  {author} {\bibfnamefont {L.}~\bibnamefont {Fu}}, \ and\ \bibinfo {author}
  {\bibfnamefont {J.}~\bibnamefont {Li}},\ }\href {\doibase
  10.1126/science.1256815} {\bibfield  {journal} {\bibinfo  {journal}
  {Science}\ }\textbf {\bibinfo {volume} {346}},\ \bibinfo {pages} {1344}
  (\bibinfo {year} {2014})}\BibitemShut {NoStop}%
\bibitem [{\citenamefont {Wu}\ \emph {et~al.}(2018)\citenamefont {Wu},
  \citenamefont {Fatemi}, \citenamefont {Gibson}, \citenamefont {Watanabe},
  \citenamefont {Taniguchi}, \citenamefont {Cava},\ and\ \citenamefont
  {Jarillo-Herrero}}]{wata2018topexp}%
  \BibitemOpen
  \bibfield  {author} {\bibinfo {author} {\bibfnamefont {S.}~\bibnamefont
  {Wu}}, \bibinfo {author} {\bibfnamefont {V.}~\bibnamefont {Fatemi}}, \bibinfo
  {author} {\bibfnamefont {Q.~D.}\ \bibnamefont {Gibson}}, \bibinfo {author}
  {\bibfnamefont {K.}~\bibnamefont {Watanabe}}, \bibinfo {author}
  {\bibfnamefont {T.}~\bibnamefont {Taniguchi}}, \bibinfo {author}
  {\bibfnamefont {R.~J.}\ \bibnamefont {Cava}}, \ and\ \bibinfo {author}
  {\bibfnamefont {P.}~\bibnamefont {Jarillo-Herrero}},\ }\href {\doibase
  10.1126/science.aan6003} {\bibfield  {journal} {\bibinfo  {journal}
  {Science}\ }\textbf {\bibinfo {volume} {359}},\ \bibinfo {pages} {76}
  (\bibinfo {year} {2018})}\BibitemShut {NoStop}%
\bibitem [{\citenamefont {Law}\ and\ \citenamefont
  {Lee}(2017)}]{law2017tmdspliq}%
  \BibitemOpen
  \bibfield  {author} {\bibinfo {author} {\bibfnamefont {K.~T.}\ \bibnamefont
  {Law}}\ and\ \bibinfo {author} {\bibfnamefont {P.~A.}\ \bibnamefont {Lee}},\
  }\href {\doibase 10.1073/pnas.1706769114} {\bibfield  {journal} {\bibinfo
  {journal} {Proc. Natl. Acad. Sci. U.S.A.}\ }\textbf {\bibinfo {volume}
  {114}},\ \bibinfo {pages} {6996} (\bibinfo {year} {2017})}\BibitemShut
  {NoStop}%
\bibitem [{\citenamefont {He}\ \emph {et~al.}(2018)\citenamefont {He},
  \citenamefont {Xu}, \citenamefont {Chen}, \citenamefont {Law},\ and\
  \citenamefont {Lee}}]{he2018spinon}%
  \BibitemOpen
  \bibfield  {author} {\bibinfo {author} {\bibfnamefont {W.-Y.}\ \bibnamefont
  {He}}, \bibinfo {author} {\bibfnamefont {X.~Y.}\ \bibnamefont {Xu}}, \bibinfo
  {author} {\bibfnamefont {G.}~\bibnamefont {Chen}}, \bibinfo {author}
  {\bibfnamefont {K.~T.}\ \bibnamefont {Law}}, \ and\ \bibinfo {author}
  {\bibfnamefont {P.~A.}\ \bibnamefont {Lee}},\ }\href {\doibase
  10.1103/PhysRevLett.121.046401} {\bibfield  {journal} {\bibinfo  {journal}
  {Phys. Rev. Lett.}\ }\textbf {\bibinfo {volume} {121}},\ \bibinfo {pages}
  {046401} (\bibinfo {year} {2018})}\BibitemShut {NoStop}%
\bibitem [{\citenamefont {Chen}\ \emph {et~al.}(2020)\citenamefont {Chen},
  \citenamefont {Ruan}, \citenamefont {Wu}, \citenamefont {Tang}, \citenamefont
  {Ryu}, \citenamefont {Tsai}, \citenamefont {Lee}, \citenamefont {Kahn},
  \citenamefont {Liou}, \citenamefont {Jia}, \citenamefont {Albertini},
  \citenamefont {Xiong}, \citenamefont {Jia}, \citenamefont {Liu},
  \citenamefont {Sobota}, \citenamefont {Liu}, \citenamefont {Moore},
  \citenamefont {Shen}, \citenamefont {Louie}, \citenamefont {Mo},\ and\
  \citenamefont {Crommie}}]{chen2020tase2}%
  \BibitemOpen
  \bibfield  {author} {\bibinfo {author} {\bibfnamefont {Y.}~\bibnamefont
  {Chen}}, \bibinfo {author} {\bibfnamefont {W.}~\bibnamefont {Ruan}}, \bibinfo
  {author} {\bibfnamefont {M.}~\bibnamefont {Wu}}, \bibinfo {author}
  {\bibfnamefont {S.}~\bibnamefont {Tang}}, \bibinfo {author} {\bibfnamefont
  {H.}~\bibnamefont {Ryu}}, \bibinfo {author} {\bibfnamefont {H.-Z.}\
  \bibnamefont {Tsai}}, \bibinfo {author} {\bibfnamefont {R.}~\bibnamefont
  {Lee}}, \bibinfo {author} {\bibfnamefont {S.}~\bibnamefont {Kahn}}, \bibinfo
  {author} {\bibfnamefont {F.}~\bibnamefont {Liou}}, \bibinfo {author}
  {\bibfnamefont {C.}~\bibnamefont {Jia}}, \bibinfo {author} {\bibfnamefont
  {O.~R.}\ \bibnamefont {Albertini}}, \bibinfo {author} {\bibfnamefont
  {H.}~\bibnamefont {Xiong}}, \bibinfo {author} {\bibfnamefont
  {T.}~\bibnamefont {Jia}}, \bibinfo {author} {\bibfnamefont {Z.}~\bibnamefont
  {Liu}}, \bibinfo {author} {\bibfnamefont {J.~A.}\ \bibnamefont {Sobota}},
  \bibinfo {author} {\bibfnamefont {A.~Y.}\ \bibnamefont {Liu}}, \bibinfo
  {author} {\bibfnamefont {J.~E.}\ \bibnamefont {Moore}}, \bibinfo {author}
  {\bibfnamefont {Z.-X.}\ \bibnamefont {Shen}}, \bibinfo {author}
  {\bibfnamefont {S.~G.}\ \bibnamefont {Louie}}, \bibinfo {author}
  {\bibfnamefont {S.-K.}\ \bibnamefont {Mo}}, \ and\ \bibinfo {author}
  {\bibfnamefont {M.~F.}\ \bibnamefont {Crommie}},\ }\href {\doibase
  10.1038/s41567-019-0744-9} {\bibfield  {journal} {\bibinfo  {journal} {Nat.
  Phys.}\ }\textbf {\bibinfo {volume} {16}},\ \bibinfo {pages} {218} (\bibinfo
  {year} {2020})}\BibitemShut {NoStop}%
\bibitem [{\citenamefont {Ruan}\ \emph {et~al.}(2021)\citenamefont {Ruan},
  \citenamefont {Chen}, \citenamefont {Tang}, \citenamefont {Hwang},
  \citenamefont {Tsai}, \citenamefont {Lee}, \citenamefont {Wu}, \citenamefont
  {Ryu}, \citenamefont {Kahn}, \citenamefont {Liou}, \citenamefont {Jia},
  \citenamefont {Aikawa}, \citenamefont {Hwang}, \citenamefont {Wang},
  \citenamefont {Choi}, \citenamefont {Louie}, \citenamefont {Lee},
  \citenamefont {Shen}, \citenamefont {Mo},\ and\ \citenamefont
  {Crommie}}]{ruan2021tase2}%
  \BibitemOpen
  \bibfield  {author} {\bibinfo {author} {\bibfnamefont {W.}~\bibnamefont
  {Ruan}}, \bibinfo {author} {\bibfnamefont {Y.}~\bibnamefont {Chen}}, \bibinfo
  {author} {\bibfnamefont {S.}~\bibnamefont {Tang}}, \bibinfo {author}
  {\bibfnamefont {J.}~\bibnamefont {Hwang}}, \bibinfo {author} {\bibfnamefont
  {H.-Z.}\ \bibnamefont {Tsai}}, \bibinfo {author} {\bibfnamefont {R.~L.}\
  \bibnamefont {Lee}}, \bibinfo {author} {\bibfnamefont {M.}~\bibnamefont
  {Wu}}, \bibinfo {author} {\bibfnamefont {H.}~\bibnamefont {Ryu}}, \bibinfo
  {author} {\bibfnamefont {S.}~\bibnamefont {Kahn}}, \bibinfo {author}
  {\bibfnamefont {F.}~\bibnamefont {Liou}}, \bibinfo {author} {\bibfnamefont
  {C.}~\bibnamefont {Jia}}, \bibinfo {author} {\bibfnamefont {A.}~\bibnamefont
  {Aikawa}}, \bibinfo {author} {\bibfnamefont {C.}~\bibnamefont {Hwang}},
  \bibinfo {author} {\bibfnamefont {F.}~\bibnamefont {Wang}}, \bibinfo {author}
  {\bibfnamefont {Y.}~\bibnamefont {Choi}}, \bibinfo {author} {\bibfnamefont
  {S.~G.}\ \bibnamefont {Louie}}, \bibinfo {author} {\bibfnamefont {P.~A.}\
  \bibnamefont {Lee}}, \bibinfo {author} {\bibfnamefont {Z.-X.}\ \bibnamefont
  {Shen}}, \bibinfo {author} {\bibfnamefont {S.-K.}\ \bibnamefont {Mo}}, \ and\
  \bibinfo {author} {\bibfnamefont {M.~F.}\ \bibnamefont {Crommie}},\ }\href
  {\doibase 10.1038/s41567-021-01321-0} {\bibfield  {journal} {\bibinfo
  {journal} {Nat. Phys.}\ }\textbf {\bibinfo {volume} {17}},\ \bibinfo {pages}
  {1154} (\bibinfo {year} {2021})}\BibitemShut {NoStop}%
\bibitem [{\citenamefont {Nakata}\ \emph {et~al.}(2021)\citenamefont {Nakata},
  \citenamefont {Sugawara}, \citenamefont {Chainani}, \citenamefont {Oka},
  \citenamefont {Bao}, \citenamefont {Zhou}, \citenamefont {Chuang},
  \citenamefont {Cheng}, \citenamefont {Kawakami}, \citenamefont {Saruta},
  \citenamefont {Fukumura}, \citenamefont {Zhou}, \citenamefont {Takahashi},\
  and\ \citenamefont {Sato}}]{naka2021tanb}%
  \BibitemOpen
  \bibfield  {author} {\bibinfo {author} {\bibfnamefont {Y.}~\bibnamefont
  {Nakata}}, \bibinfo {author} {\bibfnamefont {K.}~\bibnamefont {Sugawara}},
  \bibinfo {author} {\bibfnamefont {A.}~\bibnamefont {Chainani}}, \bibinfo
  {author} {\bibfnamefont {H.}~\bibnamefont {Oka}}, \bibinfo {author}
  {\bibfnamefont {C.}~\bibnamefont {Bao}}, \bibinfo {author} {\bibfnamefont
  {S.}~\bibnamefont {Zhou}}, \bibinfo {author} {\bibfnamefont {P.-Y.}\
  \bibnamefont {Chuang}}, \bibinfo {author} {\bibfnamefont {C.-M.}\
  \bibnamefont {Cheng}}, \bibinfo {author} {\bibfnamefont {T.}~\bibnamefont
  {Kawakami}}, \bibinfo {author} {\bibfnamefont {Y.}~\bibnamefont {Saruta}},
  \bibinfo {author} {\bibfnamefont {T.}~\bibnamefont {Fukumura}}, \bibinfo
  {author} {\bibfnamefont {S.}~\bibnamefont {Zhou}}, \bibinfo {author}
  {\bibfnamefont {T.}~\bibnamefont {Takahashi}}, \ and\ \bibinfo {author}
  {\bibfnamefont {T.}~\bibnamefont {Sato}},\ }\href {\doibase
  10.1038/s41467-021-26105-1} {\bibfield  {journal} {\bibinfo  {journal} {Nat.
  Commun.}\ }\textbf {\bibinfo {volume} {12}},\ \bibinfo {pages} {5873}
  (\bibinfo {year} {2021})}\BibitemShut {NoStop}%
\bibitem [{\citenamefont {Wilson}\ \emph {et~al.}(1975)\citenamefont {Wilson},
  \citenamefont {Salvo},\ and\ \citenamefont {Mahajan}}]{wil1975cdw}%
  \BibitemOpen
  \bibfield  {author} {\bibinfo {author} {\bibfnamefont {J.}~\bibnamefont
  {Wilson}}, \bibinfo {author} {\bibfnamefont {F.~D.}\ \bibnamefont {Salvo}}, \
  and\ \bibinfo {author} {\bibfnamefont {S.}~\bibnamefont {Mahajan}},\ }\href
  {\doibase 10.1080/00018737500101391} {\bibfield  {journal} {\bibinfo
  {journal} {Adv. Phys.}\ }\textbf {\bibinfo {volume} {24}},\ \bibinfo {pages}
  {117} (\bibinfo {year} {1975})}\BibitemShut {NoStop}%
\bibitem [{\citenamefont {Fazekas}\ and\ \citenamefont
  {Tosatti}(1980)}]{faze1980tas2}%
  \BibitemOpen
  \bibfield  {author} {\bibinfo {author} {\bibfnamefont {P.}~\bibnamefont
  {Fazekas}}\ and\ \bibinfo {author} {\bibfnamefont {E.}~\bibnamefont
  {Tosatti}},\ }\href {\doibase https://doi.org/10.1016/0378-4363(80)90229-6}
  {\bibfield  {journal} {\bibinfo  {journal} {Physica B+C}\ }\textbf {\bibinfo
  {volume} {99}},\ \bibinfo {pages} {183} (\bibinfo {year} {1980})}\BibitemShut
  {NoStop}%
\bibitem [{\citenamefont {Stojchevska}\ \emph {et~al.}(2014)\citenamefont
  {Stojchevska}, \citenamefont {Vaskivskyi}, \citenamefont {Mertelj},
  \citenamefont {Kusar}, \citenamefont {Svetin}, \citenamefont {Brazovskii},\
  and\ \citenamefont {Mihailovic}}]{stoj2014tas2}%
  \BibitemOpen
  \bibfield  {author} {\bibinfo {author} {\bibfnamefont {L.}~\bibnamefont
  {Stojchevska}}, \bibinfo {author} {\bibfnamefont {I.}~\bibnamefont
  {Vaskivskyi}}, \bibinfo {author} {\bibfnamefont {T.}~\bibnamefont {Mertelj}},
  \bibinfo {author} {\bibfnamefont {P.}~\bibnamefont {Kusar}}, \bibinfo
  {author} {\bibfnamefont {D.}~\bibnamefont {Svetin}}, \bibinfo {author}
  {\bibfnamefont {S.}~\bibnamefont {Brazovskii}}, \ and\ \bibinfo {author}
  {\bibfnamefont {D.}~\bibnamefont {Mihailovic}},\ }\href {\doibase
  10.1126/science.1241591} {\bibfield  {journal} {\bibinfo  {journal}
  {Science}\ }\textbf {\bibinfo {volume} {344}},\ \bibinfo {pages} {177}
  (\bibinfo {year} {2014})}\BibitemShut {NoStop}%
\bibitem [{\citenamefont {Ritschel}\ \emph {et~al.}(2015)\citenamefont
  {Ritschel}, \citenamefont {Trinckauf}, \citenamefont {Koepernik},
  \citenamefont {Buechner}, \citenamefont {Zimmermann}, \citenamefont {Berger},
  \citenamefont {Joe}, \citenamefont {Abbamonte},\ and\ \citenamefont
  {Geck}}]{rits2015tas2arpes}%
  \BibitemOpen
  \bibfield  {author} {\bibinfo {author} {\bibfnamefont {T.}~\bibnamefont
  {Ritschel}}, \bibinfo {author} {\bibfnamefont {J.}~\bibnamefont {Trinckauf}},
  \bibinfo {author} {\bibfnamefont {K.}~\bibnamefont {Koepernik}}, \bibinfo
  {author} {\bibfnamefont {B.}~\bibnamefont {Buechner}}, \bibinfo {author}
  {\bibfnamefont {M.~V.}\ \bibnamefont {Zimmermann}}, \bibinfo {author}
  {\bibfnamefont {H.}~\bibnamefont {Berger}}, \bibinfo {author} {\bibfnamefont
  {Y.~I.}\ \bibnamefont {Joe}}, \bibinfo {author} {\bibfnamefont
  {P.}~\bibnamefont {Abbamonte}}, \ and\ \bibinfo {author} {\bibfnamefont
  {J.}~\bibnamefont {Geck}},\ }\href {\doibase 10.1038/NPHYS3267} {\bibfield
  {journal} {\bibinfo  {journal} {Nat. Phys.}\ }\textbf {\bibinfo {volume}
  {11}},\ \bibinfo {pages} {328} (\bibinfo {year} {2015})}\BibitemShut
  {NoStop}%
\bibitem [{\citenamefont {Cho}\ \emph {et~al.}(2016)\citenamefont {Cho},
  \citenamefont {Cheon}, \citenamefont {Kim}, \citenamefont {Lee},
  \citenamefont {Cho}, \citenamefont {Cheong},\ and\ \citenamefont
  {Yeom}}]{cho2016tas2}%
  \BibitemOpen
  \bibfield  {author} {\bibinfo {author} {\bibfnamefont {D.}~\bibnamefont
  {Cho}}, \bibinfo {author} {\bibfnamefont {S.}~\bibnamefont {Cheon}}, \bibinfo
  {author} {\bibfnamefont {K.-S.}\ \bibnamefont {Kim}}, \bibinfo {author}
  {\bibfnamefont {S.-H.}\ \bibnamefont {Lee}}, \bibinfo {author} {\bibfnamefont
  {Y.-H.}\ \bibnamefont {Cho}}, \bibinfo {author} {\bibfnamefont {S.-W.}\
  \bibnamefont {Cheong}}, \ and\ \bibinfo {author} {\bibfnamefont {H.~W.}\
  \bibnamefont {Yeom}},\ }\href {\doibase 10.1038/ncomms10453} {\bibfield
  {journal} {\bibinfo  {journal} {Nat. Commun.}\ }\textbf {\bibinfo {volume}
  {7}},\ \bibinfo {pages} {10453} (\bibinfo {year} {2016})}\BibitemShut
  {NoStop}%
\bibitem [{\citenamefont {Lin}\ \emph {et~al.}(2020)\citenamefont {Lin},
  \citenamefont {Huang}, \citenamefont {Zhao}, \citenamefont {Qiao},
  \citenamefont {Liu}, \citenamefont {Wu}, \citenamefont {Chen},\ and\
  \citenamefont {Ji}}]{lin2020tase2}%
  \BibitemOpen
  \bibfield  {author} {\bibinfo {author} {\bibfnamefont {H.}~\bibnamefont
  {Lin}}, \bibinfo {author} {\bibfnamefont {W.}~\bibnamefont {Huang}}, \bibinfo
  {author} {\bibfnamefont {K.}~\bibnamefont {Zhao}}, \bibinfo {author}
  {\bibfnamefont {S.}~\bibnamefont {Qiao}}, \bibinfo {author} {\bibfnamefont
  {Z.}~\bibnamefont {Liu}}, \bibinfo {author} {\bibfnamefont {J.}~\bibnamefont
  {Wu}}, \bibinfo {author} {\bibfnamefont {X.}~\bibnamefont {Chen}}, \ and\
  \bibinfo {author} {\bibfnamefont {S.-H.}\ \bibnamefont {Ji}},\ }\href
  {\doibase 10.1007/s12274-019-2584-4} {\bibfield  {journal} {\bibinfo
  {journal} {Nano Res.}\ }\textbf {\bibinfo {volume} {13}},\ \bibinfo {pages}
  {133} (\bibinfo {year} {2020})}\BibitemShut {NoStop}%
\bibitem [{\citenamefont {Luican-Mayer}\ \emph {et~al.}(2019)\citenamefont
  {Luican-Mayer}, \citenamefont {Zhang}, \citenamefont {DiLullo}, \citenamefont
  {Li}, \citenamefont {Fisher}, \citenamefont {Ulloa},\ and\ \citenamefont
  {Hla}}]{luic2019tas2}%
  \BibitemOpen
  \bibfield  {author} {\bibinfo {author} {\bibfnamefont {A.}~\bibnamefont
  {Luican-Mayer}}, \bibinfo {author} {\bibfnamefont {Y.}~\bibnamefont {Zhang}},
  \bibinfo {author} {\bibfnamefont {A.}~\bibnamefont {DiLullo}}, \bibinfo
  {author} {\bibfnamefont {Y.}~\bibnamefont {Li}}, \bibinfo {author}
  {\bibfnamefont {B.}~\bibnamefont {Fisher}}, \bibinfo {author} {\bibfnamefont
  {S.~E.}\ \bibnamefont {Ulloa}}, \ and\ \bibinfo {author} {\bibfnamefont
  {S.-W.}\ \bibnamefont {Hla}},\ }\href {\doibase 10.1039/c9nr07857f}
  {\bibfield  {journal} {\bibinfo  {journal} {Nanoscale}\ }\textbf {\bibinfo
  {volume} {11}},\ \bibinfo {pages} {22351} (\bibinfo {year}
  {2019})}\BibitemShut {NoStop}%
\bibitem [{\citenamefont {Liu}\ \emph {et~al.}(2021{\natexlab{b}})\citenamefont
  {Liu}, \citenamefont {Qiao}, \citenamefont {Huang}, \citenamefont {Tang},
  \citenamefont {Ling}, \citenamefont {Zhang}, \citenamefont {Xia},
  \citenamefont {Liao}, \citenamefont {Shi}, \citenamefont {Mao}, \citenamefont
  {Zhu}, \citenamefont {Lu},\ and\ \citenamefont {Fu}}]{liu2021nbse2}%
  \BibitemOpen
  \bibfield  {author} {\bibinfo {author} {\bibfnamefont {Z.-Y.}\ \bibnamefont
  {Liu}}, \bibinfo {author} {\bibfnamefont {S.}~\bibnamefont {Qiao}}, \bibinfo
  {author} {\bibfnamefont {B.}~\bibnamefont {Huang}}, \bibinfo {author}
  {\bibfnamefont {Q.-Y.}\ \bibnamefont {Tang}}, \bibinfo {author}
  {\bibfnamefont {Z.-H.}\ \bibnamefont {Ling}}, \bibinfo {author}
  {\bibfnamefont {W.-H.}\ \bibnamefont {Zhang}}, \bibinfo {author}
  {\bibfnamefont {H.-N.}\ \bibnamefont {Xia}}, \bibinfo {author} {\bibfnamefont
  {X.}~\bibnamefont {Liao}}, \bibinfo {author} {\bibfnamefont {H.}~\bibnamefont
  {Shi}}, \bibinfo {author} {\bibfnamefont {W.-H.}\ \bibnamefont {Mao}},
  \bibinfo {author} {\bibfnamefont {G.-L.}\ \bibnamefont {Zhu}}, \bibinfo
  {author} {\bibfnamefont {J.-T.}\ \bibnamefont {Lu}}, \ and\ \bibinfo {author}
  {\bibfnamefont {Y.-S.}\ \bibnamefont {Fu}},\ }\href {\doibase
  10.1021/acs.nanolett.1c02348} {\bibfield  {journal} {\bibinfo  {journal}
  {Nano Lett.}\ }\textbf {\bibinfo {volume} {21}},\ \bibinfo {pages} {7005}
  (\bibinfo {year} {2021}{\natexlab{b}})}\BibitemShut {NoStop}%
\bibitem [{\citenamefont {Gye}\ \emph {et~al.}(2019)\citenamefont {Gye},
  \citenamefont {Oh},\ and\ \citenamefont {Yeom}}]{gye2019nbse2}%
  \BibitemOpen
  \bibfield  {author} {\bibinfo {author} {\bibfnamefont {G.}~\bibnamefont
  {Gye}}, \bibinfo {author} {\bibfnamefont {E.}~\bibnamefont {Oh}}, \ and\
  \bibinfo {author} {\bibfnamefont {H.~W.}\ \bibnamefont {Yeom}},\ }\href
  {\doibase 10.1103/PhysRevLett.122.016403} {\bibfield  {journal} {\bibinfo
  {journal} {Phys. Rev Lett.}\ }\textbf {\bibinfo {volume} {122}},\ \bibinfo
  {pages} {016403} (\bibinfo {year} {2019})}\BibitemShut {NoStop}%
\bibitem [{\citenamefont {Liu}\ \emph {et~al.}(2020)\citenamefont {Liu},
  \citenamefont {Wu}, \citenamefont {Liu}, \citenamefont {Wang}, \citenamefont
  {Yao},\ and\ \citenamefont {Zhong}}]{liu2020vte2}%
  \BibitemOpen
  \bibfield  {author} {\bibinfo {author} {\bibfnamefont {M.}~\bibnamefont
  {Liu}}, \bibinfo {author} {\bibfnamefont {C.}~\bibnamefont {Wu}}, \bibinfo
  {author} {\bibfnamefont {Z.}~\bibnamefont {Liu}}, \bibinfo {author}
  {\bibfnamefont {Z.}~\bibnamefont {Wang}}, \bibinfo {author} {\bibfnamefont
  {D.-X.}\ \bibnamefont {Yao}}, \ and\ \bibinfo {author} {\bibfnamefont
  {D.}~\bibnamefont {Zhong}},\ }\href {\doibase 10.1007/s12274-020-2799-4}
  {\bibfield  {journal} {\bibinfo  {journal} {Nano Res.}\ }\textbf {\bibinfo
  {volume} {13}},\ \bibinfo {pages} {1733} (\bibinfo {year}
  {2020})}\BibitemShut {NoStop}%
\bibitem [{\citenamefont {Tan}\ \emph {et~al.}(2021)\citenamefont {Tan},
  \citenamefont {Liu}, \citenamefont {Wang},\ and\ \citenamefont
  {Yan}}]{tan2021kagome}%
  \BibitemOpen
  \bibfield  {author} {\bibinfo {author} {\bibfnamefont {H.}~\bibnamefont
  {Tan}}, \bibinfo {author} {\bibfnamefont {Y.}~\bibnamefont {Liu}}, \bibinfo
  {author} {\bibfnamefont {Z.}~\bibnamefont {Wang}}, \ and\ \bibinfo {author}
  {\bibfnamefont {B.}~\bibnamefont {Yan}},\ }\href {\doibase
  10.1103/PhysRevLett.127.046401} {\bibfield  {journal} {\bibinfo  {journal}
  {Phys. Rev Lett.}\ }\textbf {\bibinfo {volume} {127}},\ \bibinfo {pages}
  {046401} (\bibinfo {year} {2021})}\BibitemShut {NoStop}%
\bibitem [{\citenamefont {Bai}\ \emph {et~al.}(2023)\citenamefont {Bai},
  \citenamefont {Jian}, \citenamefont {Pan}, \citenamefont {Deng},
  \citenamefont {Lin}, \citenamefont {Zhu}, \citenamefont {Huo}, \citenamefont
  {Cheng}, \citenamefont {Liu}, \citenamefont {Cui}, \citenamefont {Zhang},
  \citenamefont {Zou},\ and\ \citenamefont {Zhang}}]{bai2023nbte2}%
  \BibitemOpen
  \bibfield  {author} {\bibinfo {author} {\bibfnamefont {Y.}~\bibnamefont
  {Bai}}, \bibinfo {author} {\bibfnamefont {T.}~\bibnamefont {Jian}}, \bibinfo
  {author} {\bibfnamefont {Z.}~\bibnamefont {Pan}}, \bibinfo {author}
  {\bibfnamefont {J.}~\bibnamefont {Deng}}, \bibinfo {author} {\bibfnamefont
  {X.}~\bibnamefont {Lin}}, \bibinfo {author} {\bibfnamefont {C.}~\bibnamefont
  {Zhu}}, \bibinfo {author} {\bibfnamefont {D.}~\bibnamefont {Huo}}, \bibinfo
  {author} {\bibfnamefont {Z.}~\bibnamefont {Cheng}}, \bibinfo {author}
  {\bibfnamefont {Y.}~\bibnamefont {Liu}}, \bibinfo {author} {\bibfnamefont
  {P.}~\bibnamefont {Cui}}, \bibinfo {author} {\bibfnamefont {Z.}~\bibnamefont
  {Zhang}}, \bibinfo {author} {\bibfnamefont {Q.}~\bibnamefont {Zou}}, \ and\
  \bibinfo {author} {\bibfnamefont {C.}~\bibnamefont {Zhang}},\ }\href
  {\doibase 10.1021/acs.nanolett.2c04306} {\bibfield  {journal} {\bibinfo
  {journal} {Nano Lett.}\ }\textbf {\bibinfo {volume} {23}},\ \bibinfo {pages}
  {2107} (\bibinfo {year} {2023})}\BibitemShut {NoStop}%
\bibitem [{\citenamefont {Park}\ and\ \citenamefont
  {Yeom}(2023)}]{park2023accdw}%
  \BibitemOpen
  \bibfield  {author} {\bibinfo {author} {\bibfnamefont {J.~W.}\ \bibnamefont
  {Park}}\ and\ \bibinfo {author} {\bibfnamefont {H.~W.}\ \bibnamefont
  {Yeom}},\ }\href {\doibase 10.1021/acsnano.3c04398} {\bibfield  {journal}
  {\bibinfo  {journal} {ACS Nano}\ }\textbf {\bibinfo {volume} {17}},\ \bibinfo
  {pages} {17041} (\bibinfo {year} {2023})}\BibitemShut {NoStop}%
\bibitem [{Sup()}]{Supp}%
  \BibitemOpen
  \href@noop {} {}\bibinfo {note} {See Supplemental Material at
  \url{https://journals.aps.org/prl/}, which includes Refs.\cite{kres1996dft,
  perd1996pbe, duda1998plu, most2008wann, grad2012she, guo2005she,
  macdonald1988t, weinberg2017quspin, dodds2013quantum, bieri2016projective,
  fukui2005chern, mengke2021nbs2, lee2020tas2}, for detailed discussion of
  previous experiments, numerical simulation methodology, and our
  results.}\BibitemShut {Stop}%
\bibitem [{\citenamefont {Xu}\ \emph {et~al.}(2020)\citenamefont {Xu},
  \citenamefont {Wei}, \citenamefont {Zhou}, \citenamefont {Feng},
  \citenamefont {Xu}, \citenamefont {Du}, \citenamefont {He}, \citenamefont
  {Huang}, \citenamefont {Zhang}, \citenamefont {Liu}, \citenamefont {Wu},
  \citenamefont {Guo}, \citenamefont {Wang}, \citenamefont {Guang},
  \citenamefont {Wei}, \citenamefont {Peng}, \citenamefont {Jiang},
  \citenamefont {Yu},\ and\ \citenamefont {Han}}]{xu2020she}%
  \BibitemOpen
  \bibfield  {author} {\bibinfo {author} {\bibfnamefont {H.}~\bibnamefont
  {Xu}}, \bibinfo {author} {\bibfnamefont {J.}~\bibnamefont {Wei}}, \bibinfo
  {author} {\bibfnamefont {H.}~\bibnamefont {Zhou}}, \bibinfo {author}
  {\bibfnamefont {J.}~\bibnamefont {Feng}}, \bibinfo {author} {\bibfnamefont
  {T.}~\bibnamefont {Xu}}, \bibinfo {author} {\bibfnamefont {H.}~\bibnamefont
  {Du}}, \bibinfo {author} {\bibfnamefont {C.}~\bibnamefont {He}}, \bibinfo
  {author} {\bibfnamefont {Y.}~\bibnamefont {Huang}}, \bibinfo {author}
  {\bibfnamefont {J.}~\bibnamefont {Zhang}}, \bibinfo {author} {\bibfnamefont
  {Y.}~\bibnamefont {Liu}}, \bibinfo {author} {\bibfnamefont {H.-C.}\
  \bibnamefont {Wu}}, \bibinfo {author} {\bibfnamefont {C.}~\bibnamefont
  {Guo}}, \bibinfo {author} {\bibfnamefont {X.}~\bibnamefont {Wang}}, \bibinfo
  {author} {\bibfnamefont {Y.}~\bibnamefont {Guang}}, \bibinfo {author}
  {\bibfnamefont {H.}~\bibnamefont {Wei}}, \bibinfo {author} {\bibfnamefont
  {Y.}~\bibnamefont {Peng}}, \bibinfo {author} {\bibfnamefont {W.}~\bibnamefont
  {Jiang}}, \bibinfo {author} {\bibfnamefont {G.}~\bibnamefont {Yu}}, \ and\
  \bibinfo {author} {\bibfnamefont {X.}~\bibnamefont {Han}},\ }\href {\doibase
  10.1002/adma.202000513} {\bibfield  {journal} {\bibinfo  {journal} {Adv.
  Mater.}\ }\textbf {\bibinfo {volume} {32}},\ \bibinfo {pages} {2000513}
  (\bibinfo {year} {2020})}\BibitemShut {NoStop}%
\bibitem [{\citenamefont {Feng}\ \emph {et~al.}(2012)\citenamefont {Feng},
  \citenamefont {Yao}, \citenamefont {Zhu}, \citenamefont {Zhou}, \citenamefont
  {Yao},\ and\ \citenamefont {Xiao}}]{feng2012she}%
  \BibitemOpen
  \bibfield  {author} {\bibinfo {author} {\bibfnamefont {W.}~\bibnamefont
  {Feng}}, \bibinfo {author} {\bibfnamefont {Y.}~\bibnamefont {Yao}}, \bibinfo
  {author} {\bibfnamefont {W.}~\bibnamefont {Zhu}}, \bibinfo {author}
  {\bibfnamefont {J.}~\bibnamefont {Zhou}}, \bibinfo {author} {\bibfnamefont
  {W.}~\bibnamefont {Yao}}, \ and\ \bibinfo {author} {\bibfnamefont
  {D.}~\bibnamefont {Xiao}},\ }\href {\doibase 10.1103/PhysRevB.86.165108}
  {\bibfield  {journal} {\bibinfo  {journal} {Phys. Rev. B}\ }\textbf {\bibinfo
  {volume} {86}},\ \bibinfo {pages} {165108} (\bibinfo {year}
  {2012})}\BibitemShut {NoStop}%
\bibitem [{\citenamefont {Safeer}\ \emph {et~al.}(2019)\citenamefont {Safeer},
  \citenamefont {Ingla-Aynes}, \citenamefont {Herling}, \citenamefont {Garcia},
  \citenamefont {Vila}, \citenamefont {Ontoso}, \citenamefont {Reyes~Calvo},
  \citenamefont {Roche}, \citenamefont {Hueso},\ and\ \citenamefont
  {Casanova}}]{safe2019she}%
  \BibitemOpen
  \bibfield  {author} {\bibinfo {author} {\bibfnamefont {C.~K.}\ \bibnamefont
  {Safeer}}, \bibinfo {author} {\bibfnamefont {J.}~\bibnamefont {Ingla-Aynes}},
  \bibinfo {author} {\bibfnamefont {F.}~\bibnamefont {Herling}}, \bibinfo
  {author} {\bibfnamefont {J.~H.}\ \bibnamefont {Garcia}}, \bibinfo {author}
  {\bibfnamefont {M.}~\bibnamefont {Vila}}, \bibinfo {author} {\bibfnamefont
  {N.}~\bibnamefont {Ontoso}}, \bibinfo {author} {\bibfnamefont
  {M.}~\bibnamefont {Reyes~Calvo}}, \bibinfo {author} {\bibfnamefont
  {S.}~\bibnamefont {Roche}}, \bibinfo {author} {\bibfnamefont {L.~E.}\
  \bibnamefont {Hueso}}, \ and\ \bibinfo {author} {\bibfnamefont
  {F.}~\bibnamefont {Casanova}},\ }\href {\doibase
  10.1021/acs.nanolett.8b04368} {\bibfield  {journal} {\bibinfo  {journal}
  {Nano Lett.}\ }\textbf {\bibinfo {volume} {19}},\ \bibinfo {pages} {1074}
  (\bibinfo {year} {2019})}\BibitemShut {NoStop}%
\bibitem [{\citenamefont {Darancet}\ \emph {et~al.}(2014)\citenamefont
  {Darancet}, \citenamefont {Millis},\ and\ \citenamefont
  {Marianetti}}]{dara2014tas2dft}%
  \BibitemOpen
  \bibfield  {author} {\bibinfo {author} {\bibfnamefont {P.}~\bibnamefont
  {Darancet}}, \bibinfo {author} {\bibfnamefont {A.~J.}\ \bibnamefont
  {Millis}}, \ and\ \bibinfo {author} {\bibfnamefont {C.~A.}\ \bibnamefont
  {Marianetti}},\ }\href {\doibase 10.1103/PhysRevB.90.045134} {\bibfield
  {journal} {\bibinfo  {journal} {Phys. Rev. B}\ }\textbf {\bibinfo {volume}
  {90}},\ \bibinfo {pages} {045134} (\bibinfo {year} {2014})}\BibitemShut
  {NoStop}%
\bibitem [{\citenamefont {Lee}\ \emph {et~al.}(2021)\citenamefont {Lee},
  \citenamefont {Jin},\ and\ \citenamefont {Yeom}}]{leej2021tas2}%
  \BibitemOpen
  \bibfield  {author} {\bibinfo {author} {\bibfnamefont {J.}~\bibnamefont
  {Lee}}, \bibinfo {author} {\bibfnamefont {K.-H.}\ \bibnamefont {Jin}}, \ and\
  \bibinfo {author} {\bibfnamefont {H.~W.}\ \bibnamefont {Yeom}},\ }\href
  {\doibase 10.1103/PhysRevLett.126.196405} {\bibfield  {journal} {\bibinfo
  {journal} {Phys. Rev. Lett.}\ }\textbf {\bibinfo {volume} {126}},\ \bibinfo
  {pages} {196405} (\bibinfo {year} {2021})}\BibitemShut {NoStop}%
\bibitem [{\citenamefont {Lee}\ \emph {et~al.}(2019)\citenamefont {Lee},
  \citenamefont {Goh},\ and\ \citenamefont {Cho}}]{lees2019tas2}%
  \BibitemOpen
  \bibfield  {author} {\bibinfo {author} {\bibfnamefont {S.-H.}\ \bibnamefont
  {Lee}}, \bibinfo {author} {\bibfnamefont {J.~S.}\ \bibnamefont {Goh}}, \ and\
  \bibinfo {author} {\bibfnamefont {D.}~\bibnamefont {Cho}},\ }\href {\doibase
  10.1103/PhysRevLett.122.106404} {\bibfield  {journal} {\bibinfo  {journal}
  {Phys. Rev. Lett.}\ }\textbf {\bibinfo {volume} {122}},\ \bibinfo {pages}
  {106404} (\bibinfo {year} {2019})}\BibitemShut {NoStop}%
\bibitem [{\citenamefont {Wang}\ \emph {et~al.}(2020)\citenamefont {Wang},
  \citenamefont {Yao}, \citenamefont {Xin}, \citenamefont {Han}, \citenamefont
  {Wang}, \citenamefont {Chen}, \citenamefont {Cai}, \citenamefont {Li},\ and\
  \citenamefont {Zhang}}]{wang2020tas2stacking}%
  \BibitemOpen
  \bibfield  {author} {\bibinfo {author} {\bibfnamefont {Y.~D.}\ \bibnamefont
  {Wang}}, \bibinfo {author} {\bibfnamefont {W.~L.}\ \bibnamefont {Yao}},
  \bibinfo {author} {\bibfnamefont {Z.~M.}\ \bibnamefont {Xin}}, \bibinfo
  {author} {\bibfnamefont {T.~T.}\ \bibnamefont {Han}}, \bibinfo {author}
  {\bibfnamefont {Z.~G.}\ \bibnamefont {Wang}}, \bibinfo {author}
  {\bibfnamefont {L.}~\bibnamefont {Chen}}, \bibinfo {author} {\bibfnamefont
  {C.}~\bibnamefont {Cai}}, \bibinfo {author} {\bibfnamefont {Y.}~\bibnamefont
  {Li}}, \ and\ \bibinfo {author} {\bibfnamefont {Y.}~\bibnamefont {Zhang}},\
  }\href {\doibase 10.1038/s41467-020-18040-4} {\bibfield  {journal} {\bibinfo
  {journal} {Nat. Commun.}\ }\textbf {\bibinfo {volume} {11}},\ \bibinfo
  {pages} {4215} (\bibinfo {year} {2020})}\BibitemShut {NoStop}%
\bibitem [{\citenamefont {Wu}\ \emph {et~al.}(2022)\citenamefont {Wu},
  \citenamefont {Bu}, \citenamefont {Zhang}, \citenamefont {Fei}, \citenamefont
  {Zheng}, \citenamefont {Gao}, \citenamefont {Luo}, \citenamefont {Liu},
  \citenamefont {Sun},\ and\ \citenamefont {Yin}}]{wuzo2022stack}%
  \BibitemOpen
  \bibfield  {author} {\bibinfo {author} {\bibfnamefont {Z.}~\bibnamefont
  {Wu}}, \bibinfo {author} {\bibfnamefont {K.}~\bibnamefont {Bu}}, \bibinfo
  {author} {\bibfnamefont {W.}~\bibnamefont {Zhang}}, \bibinfo {author}
  {\bibfnamefont {Y.}~\bibnamefont {Fei}}, \bibinfo {author} {\bibfnamefont
  {Y.}~\bibnamefont {Zheng}}, \bibinfo {author} {\bibfnamefont
  {J.}~\bibnamefont {Gao}}, \bibinfo {author} {\bibfnamefont {X.}~\bibnamefont
  {Luo}}, \bibinfo {author} {\bibfnamefont {Z.}~\bibnamefont {Liu}}, \bibinfo
  {author} {\bibfnamefont {Y.-P.}\ \bibnamefont {Sun}}, \ and\ \bibinfo
  {author} {\bibfnamefont {Y.}~\bibnamefont {Yin}},\ }\href {\doibase
  10.1103/PhysRevB.105.035109} {\bibfield  {journal} {\bibinfo  {journal}
  {Phys. Rev. B}\ }\textbf {\bibinfo {volume} {105}},\ \bibinfo {pages}
  {035109} (\bibinfo {year} {2022})}\BibitemShut {NoStop}%
\bibitem [{\citenamefont {Butler}\ \emph {et~al.}(2020)\citenamefont {Butler},
  \citenamefont {Yoshida}, \citenamefont {Hanaguri},\ and\ \citenamefont
  {Iwasa}}]{butl2020tas2}%
  \BibitemOpen
  \bibfield  {author} {\bibinfo {author} {\bibfnamefont {C.~J.}\ \bibnamefont
  {Butler}}, \bibinfo {author} {\bibfnamefont {M.}~\bibnamefont {Yoshida}},
  \bibinfo {author} {\bibfnamefont {T.}~\bibnamefont {Hanaguri}}, \ and\
  \bibinfo {author} {\bibfnamefont {Y.}~\bibnamefont {Iwasa}},\ }\href
  {\doibase 10.1038/s41467-020-16132-9} {\bibfield  {journal} {\bibinfo
  {journal} {Nat. Commun.}\ }\textbf {\bibinfo {volume} {11}},\ \bibinfo
  {pages} {2477} (\bibinfo {year} {2020})}\BibitemShut {NoStop}%
\bibitem [{\citenamefont {Petocchi}\ \emph {et~al.}(2022)\citenamefont
  {Petocchi}, \citenamefont {Nicholson}, \citenamefont {Salzmann},
  \citenamefont {Pasquier}, \citenamefont {Yazyev}, \citenamefont {Monney},\
  and\ \citenamefont {Werner}}]{peto2022tas2}%
  \BibitemOpen
  \bibfield  {author} {\bibinfo {author} {\bibfnamefont {F.}~\bibnamefont
  {Petocchi}}, \bibinfo {author} {\bibfnamefont {C.~W.}\ \bibnamefont
  {Nicholson}}, \bibinfo {author} {\bibfnamefont {B.}~\bibnamefont {Salzmann}},
  \bibinfo {author} {\bibfnamefont {D.}~\bibnamefont {Pasquier}}, \bibinfo
  {author} {\bibfnamefont {O.~V.}\ \bibnamefont {Yazyev}}, \bibinfo {author}
  {\bibfnamefont {C.}~\bibnamefont {Monney}}, \ and\ \bibinfo {author}
  {\bibfnamefont {P.}~\bibnamefont {Werner}},\ }\href {\doibase
  10.1103/PhysRevLett.129.016402} {\bibfield  {journal} {\bibinfo  {journal}
  {Phys. Rev. Lett.}\ }\textbf {\bibinfo {volume} {129}},\ \bibinfo {pages}
  {016402} (\bibinfo {year} {2022})}\BibitemShut {NoStop}%
\bibitem [{\citenamefont {Zhang}\ \emph {et~al.}(2020)\citenamefont {Zhang},
  \citenamefont {Si}, \citenamefont {Lian}, \citenamefont {Zhou},\ and\
  \citenamefont {Sun}}]{zhang2020mottness}%
  \BibitemOpen
  \bibfield  {author} {\bibinfo {author} {\bibfnamefont {K.}~\bibnamefont
  {Zhang}}, \bibinfo {author} {\bibfnamefont {C.}~\bibnamefont {Si}}, \bibinfo
  {author} {\bibfnamefont {C.-S.}\ \bibnamefont {Lian}}, \bibinfo {author}
  {\bibfnamefont {J.}~\bibnamefont {Zhou}}, \ and\ \bibinfo {author}
  {\bibfnamefont {Z.}~\bibnamefont {Sun}},\ }\href {\doibase
  10.1039/D0TC01719A} {\bibfield  {journal} {\bibinfo  {journal} {J. Mater.
  Chem. C}\ }\textbf {\bibinfo {volume} {8}},\ \bibinfo {pages} {9742}
  (\bibinfo {year} {2020})}\BibitemShut {NoStop}%
\bibitem [{\citenamefont {Tang}\ \emph {et~al.}(2013)\citenamefont {Tang},
  \citenamefont {Fisher},\ and\ \citenamefont {Lee}}]{tang2013low}%
  \BibitemOpen
  \bibfield  {author} {\bibinfo {author} {\bibfnamefont {E.}~\bibnamefont
  {Tang}}, \bibinfo {author} {\bibfnamefont {M.~P.}\ \bibnamefont {Fisher}}, \
  and\ \bibinfo {author} {\bibfnamefont {P.~A.}\ \bibnamefont {Lee}},\ }\href
  {\doibase 10.1103/PhysRevB.87.045119} {\bibfield  {journal} {\bibinfo
  {journal} {Phys. Rev. B}\ }\textbf {\bibinfo {volume} {87}},\ \bibinfo
  {pages} {045119} (\bibinfo {year} {2013})}\BibitemShut {NoStop}%
\bibitem [{\citenamefont {Gros}\ \emph {et~al.}(1988)\citenamefont {Gros},
  \citenamefont {Poilblanc}, \citenamefont {Rice},\ and\ \citenamefont
  {Zhang}}]{gros1988superconductivity}%
  \BibitemOpen
  \bibfield  {author} {\bibinfo {author} {\bibfnamefont {C.}~\bibnamefont
  {Gros}}, \bibinfo {author} {\bibfnamefont {D.}~\bibnamefont {Poilblanc}},
  \bibinfo {author} {\bibfnamefont {T.}~\bibnamefont {Rice}}, \ and\ \bibinfo
  {author} {\bibfnamefont {F.}~\bibnamefont {Zhang}},\ }\href {\doibase
  https://doi.org/10.1016/0921-4534(88)90715-0} {\bibfield  {journal} {\bibinfo
   {journal} {Phys. C: Supercond.}\ }\textbf {\bibinfo {volume} {153}},\
  \bibinfo {pages} {543} (\bibinfo {year} {1988})}\BibitemShut {NoStop}%
\bibitem [{\citenamefont {Gros}(1989)}]{gros1989physics}%
  \BibitemOpen
  \bibfield  {author} {\bibinfo {author} {\bibfnamefont {C.}~\bibnamefont
  {Gros}},\ }\href {\doibase 10.1016/0003-4916(89)90077-8} {\bibfield
  {journal} {\bibinfo  {journal} {Ann. Phys.}\ }\textbf {\bibinfo {volume}
  {189}},\ \bibinfo {pages} {53} (\bibinfo {year} {1989})}\BibitemShut
  {NoStop}%
\bibitem [{\citenamefont {Edegger}\ \emph {et~al.}(2007)\citenamefont
  {Edegger}, \citenamefont {Muthukumar},\ and\ \citenamefont
  {Gros}}]{edegger2007gutzwiller}%
  \BibitemOpen
  \bibfield  {author} {\bibinfo {author} {\bibfnamefont {B.}~\bibnamefont
  {Edegger}}, \bibinfo {author} {\bibfnamefont {V.~N.}\ \bibnamefont
  {Muthukumar}}, \ and\ \bibinfo {author} {\bibfnamefont {C.}~\bibnamefont
  {Gros}},\ }\href {\doibase 10.1080/00018730701627707} {\bibfield  {journal}
  {\bibinfo  {journal} {Adv. Phys.}\ }\textbf {\bibinfo {volume} {56}},\
  \bibinfo {pages} {927} (\bibinfo {year} {2007})}\BibitemShut {NoStop}%
\bibitem [{\citenamefont {Luttinger}\ and\ \citenamefont
  {Tisza}(1946)}]{luttinger1946theory}%
  \BibitemOpen
  \bibfield  {author} {\bibinfo {author} {\bibfnamefont {J.}~\bibnamefont
  {Luttinger}}\ and\ \bibinfo {author} {\bibfnamefont {L.}~\bibnamefont
  {Tisza}},\ }\href@noop {} {\bibfield  {journal} {\bibinfo  {journal}
  {Physical Review}\ }\textbf {\bibinfo {volume} {70}},\ \bibinfo {pages} {954}
  (\bibinfo {year} {1946})}\BibitemShut {NoStop}%
\bibitem [{\citenamefont {Litvin}(1974)}]{litvin1974luttinger}%
  \BibitemOpen
  \bibfield  {author} {\bibinfo {author} {\bibfnamefont {D.~B.}\ \bibnamefont
  {Litvin}},\ }\href {\doibase https://doi.org/10.1016/0031-8914(74)90257-2}
  {\bibfield  {journal} {\bibinfo  {journal} {Physica}\ }\textbf {\bibinfo
  {volume} {77}},\ \bibinfo {pages} {205} (\bibinfo {year} {1974})}\BibitemShut
  {NoStop}%
\bibitem [{\citenamefont {Iaconis}\ \emph {et~al.}(2018)\citenamefont
  {Iaconis}, \citenamefont {Liu}, \citenamefont {Hal{\'a}sz},\ and\
  \citenamefont {Balents}}]{iaconis2018spin}%
  \BibitemOpen
  \bibfield  {author} {\bibinfo {author} {\bibfnamefont {J.}~\bibnamefont
  {Iaconis}}, \bibinfo {author} {\bibfnamefont {C.-X.}\ \bibnamefont {Liu}},
  \bibinfo {author} {\bibfnamefont {G.}~\bibnamefont {Hal{\'a}sz}}, \ and\
  \bibinfo {author} {\bibfnamefont {L.}~\bibnamefont {Balents}},\ }\href
  {\doibase 10.21468/SciPostPhys.4.1.003} {\bibfield  {journal} {\bibinfo
  {journal} {SciPost Phys.}\ }\textbf {\bibinfo {volume} {4}},\ \bibinfo
  {pages} {003} (\bibinfo {year} {2018})}\BibitemShut {NoStop}%
\bibitem [{\citenamefont {Gong}\ \emph {et~al.}(2019)\citenamefont {Gong},
  \citenamefont {Zheng}, \citenamefont {Lee}, \citenamefont {Lu},\ and\
  \citenamefont {Sheng}}]{gong2019chiral}%
  \BibitemOpen
  \bibfield  {author} {\bibinfo {author} {\bibfnamefont {S.-S.}\ \bibnamefont
  {Gong}}, \bibinfo {author} {\bibfnamefont {W.}~\bibnamefont {Zheng}},
  \bibinfo {author} {\bibfnamefont {M.}~\bibnamefont {Lee}}, \bibinfo {author}
  {\bibfnamefont {Y.-M.}\ \bibnamefont {Lu}}, \ and\ \bibinfo {author}
  {\bibfnamefont {D.~N.}\ \bibnamefont {Sheng}},\ }\href {\doibase
  10.1103/PhysRevB.100.241111} {\bibfield  {journal} {\bibinfo  {journal}
  {Phys. Rev. B}\ }\textbf {\bibinfo {volume} {100}},\ \bibinfo {pages}
  {241111} (\bibinfo {year} {2019})}\BibitemShut {NoStop}%
\bibitem [{\citenamefont {Iqbal}\ \emph {et~al.}(2016)\citenamefont {Iqbal},
  \citenamefont {Hu}, \citenamefont {Thomale}, \citenamefont {Poilblanc},\ and\
  \citenamefont {Becca}}]{iqbal2016spin}%
  \BibitemOpen
  \bibfield  {author} {\bibinfo {author} {\bibfnamefont {Y.}~\bibnamefont
  {Iqbal}}, \bibinfo {author} {\bibfnamefont {W.-J.}\ \bibnamefont {Hu}},
  \bibinfo {author} {\bibfnamefont {R.}~\bibnamefont {Thomale}}, \bibinfo
  {author} {\bibfnamefont {D.}~\bibnamefont {Poilblanc}}, \ and\ \bibinfo
  {author} {\bibfnamefont {F.}~\bibnamefont {Becca}},\ }\href {\doibase
  10.1103/PhysRevB.93.144411} {\bibfield  {journal} {\bibinfo  {journal} {Phys.
  Rev. B}\ }\textbf {\bibinfo {volume} {93}},\ \bibinfo {pages} {144411}
  (\bibinfo {year} {2016})}\BibitemShut {NoStop}%
\bibitem [{\citenamefont {Ferrari}\ and\ \citenamefont
  {Becca}(2019)}]{ferrari2019dynamical}%
  \BibitemOpen
  \bibfield  {author} {\bibinfo {author} {\bibfnamefont {F.}~\bibnamefont
  {Ferrari}}\ and\ \bibinfo {author} {\bibfnamefont {F.}~\bibnamefont
  {Becca}},\ }\href {\doibase 10.1103/PhysRevX.9.031026} {\bibfield  {journal}
  {\bibinfo  {journal} {Phys. Rev. X}\ }\textbf {\bibinfo {volume} {9}},\
  \bibinfo {pages} {031026} (\bibinfo {year} {2019})}\BibitemShut {NoStop}%
\bibitem [{\citenamefont {Zhu}\ \emph {et~al.}(2018)\citenamefont {Zhu},
  \citenamefont {Maksimov}, \citenamefont {White},\ and\ \citenamefont
  {Chernyshev}}]{zhu2018topography}%
  \BibitemOpen
  \bibfield  {author} {\bibinfo {author} {\bibfnamefont {Z.}~\bibnamefont
  {Zhu}}, \bibinfo {author} {\bibfnamefont {P.}~\bibnamefont {Maksimov}},
  \bibinfo {author} {\bibfnamefont {S.~R.}\ \bibnamefont {White}}, \ and\
  \bibinfo {author} {\bibfnamefont {A.}~\bibnamefont {Chernyshev}},\ }\href
  {\doibase 10.1103/PhysRevLett.120.207203} {\bibfield  {journal} {\bibinfo
  {journal} {Phys. Rev. Lett.}\ }\textbf {\bibinfo {volume} {120}},\ \bibinfo
  {pages} {207203} (\bibinfo {year} {2018})}\BibitemShut {NoStop}%
\bibitem [{\citenamefont {Sherman}\ \emph {et~al.}(2023)\citenamefont
  {Sherman}, \citenamefont {Dupont},\ and\ \citenamefont
  {Moore}}]{sherman2023spectral}%
  \BibitemOpen
  \bibfield  {author} {\bibinfo {author} {\bibfnamefont {N.~E.}\ \bibnamefont
  {Sherman}}, \bibinfo {author} {\bibfnamefont {M.}~\bibnamefont {Dupont}}, \
  and\ \bibinfo {author} {\bibfnamefont {J.~E.}\ \bibnamefont {Moore}},\ }\href
  {\doibase 10.1103/PhysRevB.107.165146} {\bibfield  {journal} {\bibinfo
  {journal} {Phys. Rev. B}\ }\textbf {\bibinfo {volume} {107}},\ \bibinfo
  {pages} {165146} (\bibinfo {year} {2023})}\BibitemShut {NoStop}%
\bibitem [{\citenamefont {Drescher}\ \emph {et~al.}(2022)\citenamefont
  {Drescher}, \citenamefont {Vanderstraeten}, \citenamefont {Moessner},\ and\
  \citenamefont {Pollmann}}]{drescher2022dynamical}%
  \BibitemOpen
  \bibfield  {author} {\bibinfo {author} {\bibfnamefont {M.}~\bibnamefont
  {Drescher}}, \bibinfo {author} {\bibfnamefont {L.}~\bibnamefont
  {Vanderstraeten}}, \bibinfo {author} {\bibfnamefont {R.}~\bibnamefont
  {Moessner}}, \ and\ \bibinfo {author} {\bibfnamefont {F.}~\bibnamefont
  {Pollmann}},\ }\href@noop {} {\bibfield  {journal} {\bibinfo  {journal}
  {arXiv preprint arXiv:2209.03344}\ } (\bibinfo {year} {2022})}\BibitemShut
  {NoStop}%
\bibitem [{\citenamefont {Wen}(2002)}]{wen2002quantum}%
  \BibitemOpen
  \bibfield  {author} {\bibinfo {author} {\bibfnamefont {X.-G.}\ \bibnamefont
  {Wen}},\ }\href {\doibase 10.1103/PhysRevB.65.165113} {\bibfield  {journal}
  {\bibinfo  {journal} {Phys. Rev. B}\ }\textbf {\bibinfo {volume} {65}},\
  \bibinfo {pages} {165113} (\bibinfo {year} {2002})}\BibitemShut {NoStop}%
\bibitem [{\citenamefont {Hu}\ \emph {et~al.}(2019)\citenamefont {Hu},
  \citenamefont {Zhu}, \citenamefont {Eggert},\ and\ \citenamefont
  {He}}]{hu2019dirac}%
  \BibitemOpen
  \bibfield  {author} {\bibinfo {author} {\bibfnamefont {S.}~\bibnamefont
  {Hu}}, \bibinfo {author} {\bibfnamefont {W.}~\bibnamefont {Zhu}}, \bibinfo
  {author} {\bibfnamefont {S.}~\bibnamefont {Eggert}}, \ and\ \bibinfo {author}
  {\bibfnamefont {Y.-C.}\ \bibnamefont {He}},\ }\href {\doibase
  10.1103/PhysRevLett.123.207203} {\bibfield  {journal} {\bibinfo  {journal}
  {Phys. Rev. Lett.}\ }\textbf {\bibinfo {volume} {123}},\ \bibinfo {pages}
  {207203} (\bibinfo {year} {2019})}\BibitemShut {NoStop}%
\bibitem [{\citenamefont {Sun}\ \emph {et~al.}(2009)\citenamefont {Sun},
  \citenamefont {Yao}, \citenamefont {Fradkin},\ and\ \citenamefont
  {Kivelson}}]{sun2009topological}%
  \BibitemOpen
  \bibfield  {author} {\bibinfo {author} {\bibfnamefont {K.}~\bibnamefont
  {Sun}}, \bibinfo {author} {\bibfnamefont {H.}~\bibnamefont {Yao}}, \bibinfo
  {author} {\bibfnamefont {E.}~\bibnamefont {Fradkin}}, \ and\ \bibinfo
  {author} {\bibfnamefont {S.~A.}\ \bibnamefont {Kivelson}},\ }\href {\doibase
  10.1103/PhysRevLett.103.046811} {\bibfield  {journal} {\bibinfo  {journal}
  {Phys. Rev. Lett.}\ }\textbf {\bibinfo {volume} {103}},\ \bibinfo {pages}
  {046811} (\bibinfo {year} {2009})}\BibitemShut {NoStop}%
\bibitem [{\citenamefont {Uebelacker}\ and\ \citenamefont
  {Honerkamp}(2011)}]{uebelacker2011instabilities}%
  \BibitemOpen
  \bibfield  {author} {\bibinfo {author} {\bibfnamefont {S.}~\bibnamefont
  {Uebelacker}}\ and\ \bibinfo {author} {\bibfnamefont {C.}~\bibnamefont
  {Honerkamp}},\ }\href {\doibase 10.1103/PhysRevB.84.205122} {\bibfield
  {journal} {\bibinfo  {journal} {Phys. Rev. B}\ }\textbf {\bibinfo {volume}
  {84}},\ \bibinfo {pages} {205122} (\bibinfo {year} {2011})}\BibitemShut
  {NoStop}%
\bibitem [{\citenamefont {Zhu}\ \emph {et~al.}(2016)\citenamefont {Zhu},
  \citenamefont {Gong}, \citenamefont {Zeng}, \citenamefont {Fu},\ and\
  \citenamefont {Sheng}}]{zhu2016interaction}%
  \BibitemOpen
  \bibfield  {author} {\bibinfo {author} {\bibfnamefont {W.}~\bibnamefont
  {Zhu}}, \bibinfo {author} {\bibfnamefont {S.-S.}\ \bibnamefont {Gong}},
  \bibinfo {author} {\bibfnamefont {T.-S.}\ \bibnamefont {Zeng}}, \bibinfo
  {author} {\bibfnamefont {L.}~\bibnamefont {Fu}}, \ and\ \bibinfo {author}
  {\bibfnamefont {D.}~\bibnamefont {Sheng}},\ }\href {\doibase
  10.1103/PhysRevLett.117.096402} {\bibfield  {journal} {\bibinfo  {journal}
  {Phys. Rev. Lett.}\ }\textbf {\bibinfo {volume} {117}},\ \bibinfo {pages}
  {096402} (\bibinfo {year} {2016})}\BibitemShut {NoStop}%
\bibitem [{\citenamefont {Kresse}\ and\ \citenamefont
  {Furthm\"uller}(1996)}]{kres1996dft}%
  \BibitemOpen
  \bibfield  {author} {\bibinfo {author} {\bibfnamefont {G.}~\bibnamefont
  {Kresse}}\ and\ \bibinfo {author} {\bibfnamefont {J.}~\bibnamefont
  {Furthm\"uller}},\ }\href {\doibase 10.1103/PhysRevB.54.11169} {\bibfield
  {journal} {\bibinfo  {journal} {Phys. Rev. B}\ }\textbf {\bibinfo {volume}
  {54}},\ \bibinfo {pages} {11169} (\bibinfo {year} {1996})}\BibitemShut
  {NoStop}%
\bibitem [{\citenamefont {Perdew}\ \emph {et~al.}(1996)\citenamefont {Perdew},
  \citenamefont {Burke},\ and\ \citenamefont {Ernzerhof}}]{perd1996pbe}%
  \BibitemOpen
  \bibfield  {author} {\bibinfo {author} {\bibfnamefont {J.~P.}\ \bibnamefont
  {Perdew}}, \bibinfo {author} {\bibfnamefont {K.}~\bibnamefont {Burke}}, \
  and\ \bibinfo {author} {\bibfnamefont {M.}~\bibnamefont {Ernzerhof}},\ }\href
  {\doibase 10.1103/PhysRevLett.77.3865} {\bibfield  {journal} {\bibinfo
  {journal} {Phys. Rev. Lett.}\ }\textbf {\bibinfo {volume} {77}},\ \bibinfo
  {pages} {3865} (\bibinfo {year} {1996})}\BibitemShut {NoStop}%
\bibitem [{\citenamefont {Dudarev}\ \emph {et~al.}(1998)\citenamefont
  {Dudarev}, \citenamefont {Botton}, \citenamefont {Savrasov}, \citenamefont
  {Humphreys},\ and\ \citenamefont {Sutton}}]{duda1998plu}%
  \BibitemOpen
  \bibfield  {author} {\bibinfo {author} {\bibfnamefont {S.~L.}\ \bibnamefont
  {Dudarev}}, \bibinfo {author} {\bibfnamefont {G.~A.}\ \bibnamefont {Botton}},
  \bibinfo {author} {\bibfnamefont {S.~Y.}\ \bibnamefont {Savrasov}}, \bibinfo
  {author} {\bibfnamefont {C.~J.}\ \bibnamefont {Humphreys}}, \ and\ \bibinfo
  {author} {\bibfnamefont {A.~P.}\ \bibnamefont {Sutton}},\ }\href {\doibase
  10.1103/PhysRevB.57.1505} {\bibfield  {journal} {\bibinfo  {journal} {Phys.
  Rev. B}\ }\textbf {\bibinfo {volume} {57}},\ \bibinfo {pages} {1505}
  (\bibinfo {year} {1998})}\BibitemShut {NoStop}%
\bibitem [{\citenamefont {Mostofi}\ \emph {et~al.}(2008)\citenamefont
  {Mostofi}, \citenamefont {Yates}, \citenamefont {Lee}, \citenamefont {Souza},
  \citenamefont {Vanderbilt},\ and\ \citenamefont {Marzari}}]{most2008wann}%
  \BibitemOpen
  \bibfield  {author} {\bibinfo {author} {\bibfnamefont {A.~A.}\ \bibnamefont
  {Mostofi}}, \bibinfo {author} {\bibfnamefont {J.~R.}\ \bibnamefont {Yates}},
  \bibinfo {author} {\bibfnamefont {Y.-S.}\ \bibnamefont {Lee}}, \bibinfo
  {author} {\bibfnamefont {I.}~\bibnamefont {Souza}}, \bibinfo {author}
  {\bibfnamefont {D.}~\bibnamefont {Vanderbilt}}, \ and\ \bibinfo {author}
  {\bibfnamefont {N.}~\bibnamefont {Marzari}},\ }\href {\doibase
  10.1016/j.cpc.2007.11.016} {\bibfield  {journal} {\bibinfo  {journal}
  {Comput. Phys. Commun.}\ }\textbf {\bibinfo {volume} {178}},\ \bibinfo
  {pages} {685} (\bibinfo {year} {2008})}\BibitemShut {NoStop}%
\bibitem [{\citenamefont {Gradhand}\ \emph {et~al.}(2012)\citenamefont
  {Gradhand}, \citenamefont {Fedorov}, \citenamefont {Pientka}, \citenamefont
  {Zahn}, \citenamefont {Mertig},\ and\ \citenamefont
  {Gyoerffy}}]{grad2012she}%
  \BibitemOpen
  \bibfield  {author} {\bibinfo {author} {\bibfnamefont {M.}~\bibnamefont
  {Gradhand}}, \bibinfo {author} {\bibfnamefont {D.~V.}\ \bibnamefont
  {Fedorov}}, \bibinfo {author} {\bibfnamefont {F.}~\bibnamefont {Pientka}},
  \bibinfo {author} {\bibfnamefont {P.}~\bibnamefont {Zahn}}, \bibinfo {author}
  {\bibfnamefont {I.}~\bibnamefont {Mertig}}, \ and\ \bibinfo {author}
  {\bibfnamefont {B.~L.}\ \bibnamefont {Gyoerffy}},\ }\href {\doibase
  10.1088/0953-8984/24/21/213202} {\bibfield  {journal} {\bibinfo  {journal}
  {J. Phys.: Condens. Mater.}\ }\textbf {\bibinfo {volume} {24}},\ \bibinfo
  {pages} {213202} (\bibinfo {year} {2012})}\BibitemShut {NoStop}%
\bibitem [{\citenamefont {Guo}\ \emph {et~al.}(2005)\citenamefont {Guo},
  \citenamefont {Yao},\ and\ \citenamefont {Niu}}]{guo2005she}%
  \BibitemOpen
  \bibfield  {author} {\bibinfo {author} {\bibfnamefont {G.~Y.}\ \bibnamefont
  {Guo}}, \bibinfo {author} {\bibfnamefont {Y.}~\bibnamefont {Yao}}, \ and\
  \bibinfo {author} {\bibfnamefont {Q.}~\bibnamefont {Niu}},\ }\href {\doibase
  10.1103/PhysRevLett.94.226601} {\bibfield  {journal} {\bibinfo  {journal}
  {Phys. Rev. Lett.}\ }\textbf {\bibinfo {volume} {94}},\ \bibinfo {pages}
  {226601} (\bibinfo {year} {2005})}\BibitemShut {NoStop}%
\bibitem [{\citenamefont {MacDonald}\ \emph {et~al.}(1988)\citenamefont
  {MacDonald}, \citenamefont {Girvin},\ and\ \citenamefont
  {Yoshioka}}]{macdonald1988t}%
  \BibitemOpen
  \bibfield  {author} {\bibinfo {author} {\bibfnamefont {A.~H.}\ \bibnamefont
  {MacDonald}}, \bibinfo {author} {\bibfnamefont {S.}~\bibnamefont {Girvin}}, \
  and\ \bibinfo {author} {\bibfnamefont {D.}~\bibnamefont {Yoshioka}},\ }\href
  {\doibase 10.1103/PhysRevB.37.9753} {\bibfield  {journal} {\bibinfo
  {journal} {Phys. Rev. B}\ }\textbf {\bibinfo {volume} {37}},\ \bibinfo
  {pages} {9753} (\bibinfo {year} {1988})}\BibitemShut {NoStop}%
\bibitem [{\citenamefont {Weinberg}\ and\ \citenamefont
  {Bukov}(2017)}]{weinberg2017quspin}%
  \BibitemOpen
  \bibfield  {author} {\bibinfo {author} {\bibfnamefont {P.}~\bibnamefont
  {Weinberg}}\ and\ \bibinfo {author} {\bibfnamefont {M.}~\bibnamefont
  {Bukov}},\ }\href {\doibase 10.21468/SciPostPhys.2.1.003} {\bibfield
  {journal} {\bibinfo  {journal} {SciPost Phys.}\ }\textbf {\bibinfo {volume}
  {2}},\ \bibinfo {pages} {003} (\bibinfo {year} {2017})}\BibitemShut {NoStop}%
\bibitem [{\citenamefont {Dodds}\ \emph {et~al.}(2013)\citenamefont {Dodds},
  \citenamefont {Bhattacharjee},\ and\ \citenamefont {Kim}}]{dodds2013quantum}%
  \BibitemOpen
  \bibfield  {author} {\bibinfo {author} {\bibfnamefont {T.}~\bibnamefont
  {Dodds}}, \bibinfo {author} {\bibfnamefont {S.}~\bibnamefont
  {Bhattacharjee}}, \ and\ \bibinfo {author} {\bibfnamefont {Y.~B.}\
  \bibnamefont {Kim}},\ }\href {\doibase 10.1103/PhysRevB.88.224413} {\bibfield
   {journal} {\bibinfo  {journal} {Phys. Rev. B}\ }\textbf {\bibinfo {volume}
  {88}},\ \bibinfo {pages} {224413} (\bibinfo {year} {2013})}\BibitemShut
  {NoStop}%
\bibitem [{\citenamefont {Bieri}\ \emph {et~al.}(2016)\citenamefont {Bieri},
  \citenamefont {Lhuillier},\ and\ \citenamefont
  {Messio}}]{bieri2016projective}%
  \BibitemOpen
  \bibfield  {author} {\bibinfo {author} {\bibfnamefont {S.}~\bibnamefont
  {Bieri}}, \bibinfo {author} {\bibfnamefont {C.}~\bibnamefont {Lhuillier}}, \
  and\ \bibinfo {author} {\bibfnamefont {L.}~\bibnamefont {Messio}},\ }\href
  {\doibase 10.1103/PhysRevB.93.094437} {\bibfield  {journal} {\bibinfo
  {journal} {Phys. Rev. B}\ }\textbf {\bibinfo {volume} {93}},\ \bibinfo
  {pages} {094437} (\bibinfo {year} {2016})}\BibitemShut {NoStop}%
\bibitem [{\citenamefont {Fukui}\ \emph {et~al.}(2005)\citenamefont {Fukui},
  \citenamefont {Hatsugai},\ and\ \citenamefont {Suzuki}}]{fukui2005chern}%
  \BibitemOpen
  \bibfield  {author} {\bibinfo {author} {\bibfnamefont {T.}~\bibnamefont
  {Fukui}}, \bibinfo {author} {\bibfnamefont {Y.}~\bibnamefont {Hatsugai}}, \
  and\ \bibinfo {author} {\bibfnamefont {H.}~\bibnamefont {Suzuki}},\ }\href
  {\doibase 10.1143/JPSJ.74.1674} {\bibfield  {journal} {\bibinfo  {journal}
  {Journal of the Physical Society of Japan}\ }\textbf {\bibinfo {volume}
  {74}},\ \bibinfo {pages} {1674} (\bibinfo {year} {2005})}\BibitemShut
  {NoStop}%
\bibitem [{\citenamefont {Liu}\ \emph {et~al.}(2021{\natexlab{c}})\citenamefont
  {Liu}, \citenamefont {Leveillee}, \citenamefont {Lu}, \citenamefont {Yu},
  \citenamefont {Kim}, \citenamefont {Tian}, \citenamefont {Shi}, \citenamefont
  {Lai}, \citenamefont {Zhang}, \citenamefont {Giustino},\ and\ \citenamefont
  {Shih}}]{mengke2021nbs2}%
  \BibitemOpen
  \bibfield  {author} {\bibinfo {author} {\bibfnamefont {M.}~\bibnamefont
  {Liu}}, \bibinfo {author} {\bibfnamefont {J.}~\bibnamefont {Leveillee}},
  \bibinfo {author} {\bibfnamefont {S.}~\bibnamefont {Lu}}, \bibinfo {author}
  {\bibfnamefont {J.}~\bibnamefont {Yu}}, \bibinfo {author} {\bibfnamefont
  {H.}~\bibnamefont {Kim}}, \bibinfo {author} {\bibfnamefont {C.}~\bibnamefont
  {Tian}}, \bibinfo {author} {\bibfnamefont {Y.}~\bibnamefont {Shi}}, \bibinfo
  {author} {\bibfnamefont {K.}~\bibnamefont {Lai}}, \bibinfo {author}
  {\bibfnamefont {C.}~\bibnamefont {Zhang}}, \bibinfo {author} {\bibfnamefont
  {F.}~\bibnamefont {Giustino}}, \ and\ \bibinfo {author} {\bibfnamefont
  {C.-K.}\ \bibnamefont {Shih}},\ }\href {\doibase 10.1126/sciadv.abi6339}
  {\bibfield  {journal} {\bibinfo  {journal} {Science Advances}\ }\textbf
  {\bibinfo {volume} {7}},\ \bibinfo {pages} {eabi6339} (\bibinfo {year}
  {2021}{\natexlab{c}})},\ \Eprint
  {http://arxiv.org/abs/https://www.science.org/doi/pdf/10.1126/sciadv.abi6339}
  {https://www.science.org/doi/pdf/10.1126/sciadv.abi6339} \BibitemShut
  {NoStop}%
\bibitem [{\citenamefont {Lee}\ \emph {et~al.}(2020{\natexlab{b}})\citenamefont
  {Lee}, \citenamefont {Jin}, \citenamefont {Catuneanu}, \citenamefont {Go},
  \citenamefont {Jung}, \citenamefont {Won}, \citenamefont {Cheong},
  \citenamefont {Kim}, \citenamefont {Liu}, \citenamefont {Kee},\ and\
  \citenamefont {Yeom}}]{lee2020tas2}%
  \BibitemOpen
  \bibfield  {author} {\bibinfo {author} {\bibfnamefont {J.}~\bibnamefont
  {Lee}}, \bibinfo {author} {\bibfnamefont {K.-H.}\ \bibnamefont {Jin}},
  \bibinfo {author} {\bibfnamefont {A.}~\bibnamefont {Catuneanu}}, \bibinfo
  {author} {\bibfnamefont {A.}~\bibnamefont {Go}}, \bibinfo {author}
  {\bibfnamefont {J.}~\bibnamefont {Jung}}, \bibinfo {author} {\bibfnamefont
  {C.}~\bibnamefont {Won}}, \bibinfo {author} {\bibfnamefont {S.-W.}\
  \bibnamefont {Cheong}}, \bibinfo {author} {\bibfnamefont {J.}~\bibnamefont
  {Kim}}, \bibinfo {author} {\bibfnamefont {F.}~\bibnamefont {Liu}}, \bibinfo
  {author} {\bibfnamefont {H.-Y.}\ \bibnamefont {Kee}}, \ and\ \bibinfo
  {author} {\bibfnamefont {H.~W.}\ \bibnamefont {Yeom}},\ }\href {\doibase
  10.1103/PhysRevLett.125.096403} {\bibfield  {journal} {\bibinfo  {journal}
  {Phys. Rev. Lett.}\ }\textbf {\bibinfo {volume} {125}},\ \bibinfo {pages}
  {096403} (\bibinfo {year} {2020}{\natexlab{b}})}\BibitemShut {NoStop}%
\end{thebibliography}%


\begin{thebibliography}{23}%
\makeatletter
\providecommand \@ifxundefined [1]{%
 \@ifx{#1\undefined}
}%
\providecommand \@ifnum [1]{%
 \ifnum #1\expandafter \@firstoftwo
 \else \expandafter \@secondoftwo
 \fi
}%
\providecommand \@ifx [1]{%
 \ifx #1\expandafter \@firstoftwo
 \else \expandafter \@secondoftwo
 \fi
}%
\providecommand \natexlab [1]{#1}%
\providecommand \enquote  [1]{``#1''}%
\providecommand \bibnamefont  [1]{#1}%
\providecommand \bibfnamefont [1]{#1}%
\providecommand \citenamefont [1]{#1}%
\providecommand \href@noop [0]{\@secondoftwo}%
\providecommand \href [0]{\begingroup \@sanitize@url \@href}%
\providecommand \@href[1]{\@@startlink{#1}\@@href}%
\providecommand \@@href[1]{\endgroup#1\@@endlink}%
\providecommand \@sanitize@url [0]{\catcode `\\12\catcode `\$12\catcode
  `\&12\catcode `\#12\catcode `\^12\catcode `\_12\catcode `\%12\relax}%
\providecommand \@@startlink[1]{}%
\providecommand \@@endlink[0]{}%
\providecommand \url  [0]{\begingroup\@sanitize@url \@url }%
\providecommand \@url [1]{\endgroup\@href {#1}{\urlprefix }}%
\providecommand \urlprefix  [0]{URL }%
\providecommand \Eprint [0]{\href }%
\providecommand \doibase [0]{http://dx.doi.org/}%
\providecommand \selectlanguage [0]{\@gobble}%
\providecommand \bibinfo  [0]{\@secondoftwo}%
\providecommand \bibfield  [0]{\@secondoftwo}%
\providecommand \translation [1]{[#1]}%
\providecommand \BibitemOpen [0]{}%
\providecommand \bibitemStop [0]{}%
\providecommand \bibitemNoStop [0]{.\EOS\space}%
\providecommand \EOS [0]{\spacefactor3000\relax}%
\providecommand \BibitemShut  [1]{\csname bibitem#1\endcsname}%
\let\auto@bib@innerbib\@empty
\bibitem [{\citenamefont {Qiao}\ \emph {et~al.}(2017)\citenamefont {Qiao},
  \citenamefont {Li}, \citenamefont {Wang}, \citenamefont {Ruan}, \citenamefont
  {Ye}, \citenamefont {Cai}, \citenamefont {Hao}, \citenamefont {Yao},
  \citenamefont {Chen}, \citenamefont {Wu}, \citenamefont {Wang},\ and\
  \citenamefont {Liu}}]{qiao2017mottness}%
  \BibitemOpen
  \bibfield  {author} {\bibinfo {author} {\bibfnamefont {S.}~\bibnamefont
  {Qiao}}, \bibinfo {author} {\bibfnamefont {X.}~\bibnamefont {Li}}, \bibinfo
  {author} {\bibfnamefont {N.}~\bibnamefont {Wang}}, \bibinfo {author}
  {\bibfnamefont {W.}~\bibnamefont {Ruan}}, \bibinfo {author} {\bibfnamefont
  {C.}~\bibnamefont {Ye}}, \bibinfo {author} {\bibfnamefont {P.}~\bibnamefont
  {Cai}}, \bibinfo {author} {\bibfnamefont {Z.}~\bibnamefont {Hao}}, \bibinfo
  {author} {\bibfnamefont {H.}~\bibnamefont {Yao}}, \bibinfo {author}
  {\bibfnamefont {X.}~\bibnamefont {Chen}}, \bibinfo {author} {\bibfnamefont
  {J.}~\bibnamefont {Wu}}, \bibinfo {author} {\bibfnamefont {Y.}~\bibnamefont
  {Wang}}, \ and\ \bibinfo {author} {\bibfnamefont {Z.}~\bibnamefont {Liu}},\
  }\href {\doibase 10.1103/PhysRevX.7.041054} {\bibfield  {journal} {\bibinfo
  {journal} {Phys. Rev. X}\ }\textbf {\bibinfo {volume} {7}},\ \bibinfo {pages}
  {041054} (\bibinfo {year} {2017})}\BibitemShut {NoStop}%
\bibitem [{\citenamefont {Lin}\ \emph {et~al.}(2020)\citenamefont {Lin},
  \citenamefont {Huang}, \citenamefont {Zhao}, \citenamefont {Qiao},
  \citenamefont {Liu}, \citenamefont {Wu}, \citenamefont {Chen},\ and\
  \citenamefont {Ji}}]{lin2020tase2}%
  \BibitemOpen
  \bibfield  {author} {\bibinfo {author} {\bibfnamefont {H.}~\bibnamefont
  {Lin}}, \bibinfo {author} {\bibfnamefont {W.}~\bibnamefont {Huang}}, \bibinfo
  {author} {\bibfnamefont {K.}~\bibnamefont {Zhao}}, \bibinfo {author}
  {\bibfnamefont {S.}~\bibnamefont {Qiao}}, \bibinfo {author} {\bibfnamefont
  {Z.}~\bibnamefont {Liu}}, \bibinfo {author} {\bibfnamefont {J.}~\bibnamefont
  {Wu}}, \bibinfo {author} {\bibfnamefont {X.}~\bibnamefont {Chen}}, \ and\
  \bibinfo {author} {\bibfnamefont {S.-H.}\ \bibnamefont {Ji}},\ }\href
  {\doibase 10.1007/s12274-019-2584-4} {\bibfield  {journal} {\bibinfo
  {journal} {Nano Res.}\ }\textbf {\bibinfo {volume} {13}},\ \bibinfo {pages}
  {133} (\bibinfo {year} {2020})}\BibitemShut {NoStop}%
\bibitem [{\citenamefont {Liu}\ \emph {et~al.}(2021{\natexlab{a}})\citenamefont
  {Liu}, \citenamefont {Qiao}, \citenamefont {Huang}, \citenamefont {Tang},
  \citenamefont {Ling}, \citenamefont {Zhang}, \citenamefont {Xia},
  \citenamefont {Liao}, \citenamefont {Shi}, \citenamefont {Mao}, \citenamefont
  {Zhu}, \citenamefont {Lu},\ and\ \citenamefont {Fu}}]{liu2021nbse2}%
  \BibitemOpen
  \bibfield  {author} {\bibinfo {author} {\bibfnamefont {Z.-Y.}\ \bibnamefont
  {Liu}}, \bibinfo {author} {\bibfnamefont {S.}~\bibnamefont {Qiao}}, \bibinfo
  {author} {\bibfnamefont {B.}~\bibnamefont {Huang}}, \bibinfo {author}
  {\bibfnamefont {Q.-Y.}\ \bibnamefont {Tang}}, \bibinfo {author}
  {\bibfnamefont {Z.-H.}\ \bibnamefont {Ling}}, \bibinfo {author}
  {\bibfnamefont {W.-H.}\ \bibnamefont {Zhang}}, \bibinfo {author}
  {\bibfnamefont {H.-N.}\ \bibnamefont {Xia}}, \bibinfo {author} {\bibfnamefont
  {X.}~\bibnamefont {Liao}}, \bibinfo {author} {\bibfnamefont {H.}~\bibnamefont
  {Shi}}, \bibinfo {author} {\bibfnamefont {W.-H.}\ \bibnamefont {Mao}},
  \bibinfo {author} {\bibfnamefont {G.-L.}\ \bibnamefont {Zhu}}, \bibinfo
  {author} {\bibfnamefont {J.-T.}\ \bibnamefont {Lu}}, \ and\ \bibinfo {author}
  {\bibfnamefont {Y.-S.}\ \bibnamefont {Fu}},\ }\href {\doibase
  10.1021/acs.nanolett.1c02348} {\bibfield  {journal} {\bibinfo  {journal}
  {Nano Lett.}\ }\textbf {\bibinfo {volume} {21}},\ \bibinfo {pages} {7005}
  (\bibinfo {year} {2021}{\natexlab{a}})}\BibitemShut {NoStop}%
\bibitem [{\citenamefont {Luican-Mayer}\ \emph {et~al.}(2019)\citenamefont
  {Luican-Mayer}, \citenamefont {Zhang}, \citenamefont {DiLullo}, \citenamefont
  {Li}, \citenamefont {Fisher}, \citenamefont {Ulloa},\ and\ \citenamefont
  {Hla}}]{luic2019tas2}%
  \BibitemOpen
  \bibfield  {author} {\bibinfo {author} {\bibfnamefont {A.}~\bibnamefont
  {Luican-Mayer}}, \bibinfo {author} {\bibfnamefont {Y.}~\bibnamefont {Zhang}},
  \bibinfo {author} {\bibfnamefont {A.}~\bibnamefont {DiLullo}}, \bibinfo
  {author} {\bibfnamefont {Y.}~\bibnamefont {Li}}, \bibinfo {author}
  {\bibfnamefont {B.}~\bibnamefont {Fisher}}, \bibinfo {author} {\bibfnamefont
  {S.~E.}\ \bibnamefont {Ulloa}}, \ and\ \bibinfo {author} {\bibfnamefont
  {S.-W.}\ \bibnamefont {Hla}},\ }\href {\doibase 10.1039/c9nr07857f}
  {\bibfield  {journal} {\bibinfo  {journal} {Nanoscale}\ }\textbf {\bibinfo
  {volume} {11}},\ \bibinfo {pages} {22351} (\bibinfo {year}
  {2019})}\BibitemShut {NoStop}%
\bibitem [{\citenamefont {Kresse}\ and\ \citenamefont
  {Furthm\"uller}(1996)}]{kres1996dft}%
  \BibitemOpen
  \bibfield  {author} {\bibinfo {author} {\bibfnamefont {G.}~\bibnamefont
  {Kresse}}\ and\ \bibinfo {author} {\bibfnamefont {J.}~\bibnamefont
  {Furthm\"uller}},\ }\href {\doibase 10.1103/PhysRevB.54.11169} {\bibfield
  {journal} {\bibinfo  {journal} {Phys. Rev. B}\ }\textbf {\bibinfo {volume}
  {54}},\ \bibinfo {pages} {11169} (\bibinfo {year} {1996})}\BibitemShut
  {NoStop}%
\bibitem [{\citenamefont {Perdew}\ \emph {et~al.}(1996)\citenamefont {Perdew},
  \citenamefont {Burke},\ and\ \citenamefont {Ernzerhof}}]{perd1996pbe}%
  \BibitemOpen
  \bibfield  {author} {\bibinfo {author} {\bibfnamefont {J.~P.}\ \bibnamefont
  {Perdew}}, \bibinfo {author} {\bibfnamefont {K.}~\bibnamefont {Burke}}, \
  and\ \bibinfo {author} {\bibfnamefont {M.}~\bibnamefont {Ernzerhof}},\ }\href
  {\doibase 10.1103/PhysRevLett.77.3865} {\bibfield  {journal} {\bibinfo
  {journal} {Phys. Rev. Lett.}\ }\textbf {\bibinfo {volume} {77}},\ \bibinfo
  {pages} {3865} (\bibinfo {year} {1996})}\BibitemShut {NoStop}%
\bibitem [{\citenamefont {Dudarev}\ \emph {et~al.}(1998)\citenamefont
  {Dudarev}, \citenamefont {Botton}, \citenamefont {Savrasov}, \citenamefont
  {Humphreys},\ and\ \citenamefont {Sutton}}]{duda1998plu}%
  \BibitemOpen
  \bibfield  {author} {\bibinfo {author} {\bibfnamefont {S.~L.}\ \bibnamefont
  {Dudarev}}, \bibinfo {author} {\bibfnamefont {G.~A.}\ \bibnamefont {Botton}},
  \bibinfo {author} {\bibfnamefont {S.~Y.}\ \bibnamefont {Savrasov}}, \bibinfo
  {author} {\bibfnamefont {C.~J.}\ \bibnamefont {Humphreys}}, \ and\ \bibinfo
  {author} {\bibfnamefont {A.~P.}\ \bibnamefont {Sutton}},\ }\href {\doibase
  10.1103/PhysRevB.57.1505} {\bibfield  {journal} {\bibinfo  {journal} {Phys.
  Rev. B}\ }\textbf {\bibinfo {volume} {57}},\ \bibinfo {pages} {1505}
  (\bibinfo {year} {1998})}\BibitemShut {NoStop}%
\bibitem [{\citenamefont {Mostofi}\ \emph {et~al.}(2008)\citenamefont
  {Mostofi}, \citenamefont {Yates}, \citenamefont {Lee}, \citenamefont {Souza},
  \citenamefont {Vanderbilt},\ and\ \citenamefont {Marzari}}]{most2008wann}%
  \BibitemOpen
  \bibfield  {author} {\bibinfo {author} {\bibfnamefont {A.~A.}\ \bibnamefont
  {Mostofi}}, \bibinfo {author} {\bibfnamefont {J.~R.}\ \bibnamefont {Yates}},
  \bibinfo {author} {\bibfnamefont {Y.-S.}\ \bibnamefont {Lee}}, \bibinfo
  {author} {\bibfnamefont {I.}~\bibnamefont {Souza}}, \bibinfo {author}
  {\bibfnamefont {D.}~\bibnamefont {Vanderbilt}}, \ and\ \bibinfo {author}
  {\bibfnamefont {N.}~\bibnamefont {Marzari}},\ }\href {\doibase
  10.1016/j.cpc.2007.11.016} {\bibfield  {journal} {\bibinfo  {journal}
  {Comput. Phys. Commun.}\ }\textbf {\bibinfo {volume} {178}},\ \bibinfo
  {pages} {685} (\bibinfo {year} {2008})}\BibitemShut {NoStop}%
\bibitem [{\citenamefont {Gradhand}\ \emph {et~al.}(2012)\citenamefont
  {Gradhand}, \citenamefont {Fedorov}, \citenamefont {Pientka}, \citenamefont
  {Zahn}, \citenamefont {Mertig},\ and\ \citenamefont
  {Gyoerffy}}]{grad2012she}%
  \BibitemOpen
  \bibfield  {author} {\bibinfo {author} {\bibfnamefont {M.}~\bibnamefont
  {Gradhand}}, \bibinfo {author} {\bibfnamefont {D.~V.}\ \bibnamefont
  {Fedorov}}, \bibinfo {author} {\bibfnamefont {F.}~\bibnamefont {Pientka}},
  \bibinfo {author} {\bibfnamefont {P.}~\bibnamefont {Zahn}}, \bibinfo {author}
  {\bibfnamefont {I.}~\bibnamefont {Mertig}}, \ and\ \bibinfo {author}
  {\bibfnamefont {B.~L.}\ \bibnamefont {Gyoerffy}},\ }\href {\doibase
  10.1088/0953-8984/24/21/213202} {\bibfield  {journal} {\bibinfo  {journal}
  {J. Phys.: Condens. Mater.}\ }\textbf {\bibinfo {volume} {24}},\ \bibinfo
  {pages} {213202} (\bibinfo {year} {2012})}\BibitemShut {NoStop}%
\bibitem [{\citenamefont {Guo}\ \emph {et~al.}(2005)\citenamefont {Guo},
  \citenamefont {Yao},\ and\ \citenamefont {Niu}}]{guo2005she}%
  \BibitemOpen
  \bibfield  {author} {\bibinfo {author} {\bibfnamefont {G.~Y.}\ \bibnamefont
  {Guo}}, \bibinfo {author} {\bibfnamefont {Y.}~\bibnamefont {Yao}}, \ and\
  \bibinfo {author} {\bibfnamefont {Q.}~\bibnamefont {Niu}},\ }\href {\doibase
  10.1103/PhysRevLett.94.226601} {\bibfield  {journal} {\bibinfo  {journal}
  {Phys. Rev. Lett.}\ }\textbf {\bibinfo {volume} {94}},\ \bibinfo {pages}
  {226601} (\bibinfo {year} {2005})}\BibitemShut {NoStop}%
\bibitem [{\citenamefont {Zhang}\ \emph {et~al.}(2020)\citenamefont {Zhang},
  \citenamefont {Si}, \citenamefont {Lian}, \citenamefont {Zhou},\ and\
  \citenamefont {Sun}}]{zhang2020mottness}%
  \BibitemOpen
  \bibfield  {author} {\bibinfo {author} {\bibfnamefont {K.}~\bibnamefont
  {Zhang}}, \bibinfo {author} {\bibfnamefont {C.}~\bibnamefont {Si}}, \bibinfo
  {author} {\bibfnamefont {C.-S.}\ \bibnamefont {Lian}}, \bibinfo {author}
  {\bibfnamefont {J.}~\bibnamefont {Zhou}}, \ and\ \bibinfo {author}
  {\bibfnamefont {Z.}~\bibnamefont {Sun}},\ }\href {\doibase
  10.1039/D0TC01719A} {\bibfield  {journal} {\bibinfo  {journal} {J. Mater.
  Chem. C}\ }\textbf {\bibinfo {volume} {8}},\ \bibinfo {pages} {9742}
  (\bibinfo {year} {2020})}\BibitemShut {NoStop}%
\bibitem [{\citenamefont {MacDonald}\ \emph {et~al.}(1988)\citenamefont
  {MacDonald}, \citenamefont {Girvin},\ and\ \citenamefont
  {Yoshioka}}]{macdonald1988t}%
  \BibitemOpen
  \bibfield  {author} {\bibinfo {author} {\bibfnamefont {A.~H.}\ \bibnamefont
  {MacDonald}}, \bibinfo {author} {\bibfnamefont {S.}~\bibnamefont {Girvin}}, \
  and\ \bibinfo {author} {\bibfnamefont {D.}~\bibnamefont {Yoshioka}},\ }\href
  {\doibase 10.1103/PhysRevB.37.9753} {\bibfield  {journal} {\bibinfo
  {journal} {Phys. Rev. B}\ }\textbf {\bibinfo {volume} {37}},\ \bibinfo
  {pages} {9753} (\bibinfo {year} {1988})}\BibitemShut {NoStop}%
\bibitem [{\citenamefont {Weinberg}\ and\ \citenamefont
  {Bukov}(2017)}]{weinberg2017quspin}%
  \BibitemOpen
  \bibfield  {author} {\bibinfo {author} {\bibfnamefont {P.}~\bibnamefont
  {Weinberg}}\ and\ \bibinfo {author} {\bibfnamefont {M.}~\bibnamefont
  {Bukov}},\ }\href {\doibase 10.21468/SciPostPhys.2.1.003} {\bibfield
  {journal} {\bibinfo  {journal} {SciPost Phys.}\ }\textbf {\bibinfo {volume}
  {2}},\ \bibinfo {pages} {003} (\bibinfo {year} {2017})}\BibitemShut {NoStop}%
\bibitem [{\citenamefont {Wen}(2002)}]{wen2002quantum}%
  \BibitemOpen
  \bibfield  {author} {\bibinfo {author} {\bibfnamefont {X.-G.}\ \bibnamefont
  {Wen}},\ }\href {\doibase 10.1103/PhysRevB.65.165113} {\bibfield  {journal}
  {\bibinfo  {journal} {Phys. Rev. B}\ }\textbf {\bibinfo {volume} {65}},\
  \bibinfo {pages} {165113} (\bibinfo {year} {2002})}\BibitemShut {NoStop}%
\bibitem [{\citenamefont {Dodds}\ \emph {et~al.}(2013)\citenamefont {Dodds},
  \citenamefont {Bhattacharjee},\ and\ \citenamefont {Kim}}]{dodds2013quantum}%
  \BibitemOpen
  \bibfield  {author} {\bibinfo {author} {\bibfnamefont {T.}~\bibnamefont
  {Dodds}}, \bibinfo {author} {\bibfnamefont {S.}~\bibnamefont
  {Bhattacharjee}}, \ and\ \bibinfo {author} {\bibfnamefont {Y.~B.}\
  \bibnamefont {Kim}},\ }\href {\doibase 10.1103/PhysRevB.88.224413} {\bibfield
   {journal} {\bibinfo  {journal} {Phys. Rev. B}\ }\textbf {\bibinfo {volume}
  {88}},\ \bibinfo {pages} {224413} (\bibinfo {year} {2013})}\BibitemShut
  {NoStop}%
\bibitem [{\citenamefont {Bieri}\ \emph {et~al.}(2016)\citenamefont {Bieri},
  \citenamefont {Lhuillier},\ and\ \citenamefont
  {Messio}}]{bieri2016projective}%
  \BibitemOpen
  \bibfield  {author} {\bibinfo {author} {\bibfnamefont {S.}~\bibnamefont
  {Bieri}}, \bibinfo {author} {\bibfnamefont {C.}~\bibnamefont {Lhuillier}}, \
  and\ \bibinfo {author} {\bibfnamefont {L.}~\bibnamefont {Messio}},\ }\href
  {\doibase 10.1103/PhysRevB.93.094437} {\bibfield  {journal} {\bibinfo
  {journal} {Phys. Rev. B}\ }\textbf {\bibinfo {volume} {93}},\ \bibinfo
  {pages} {094437} (\bibinfo {year} {2016})}\BibitemShut {NoStop}%
\bibitem [{\citenamefont {Fukui}\ \emph {et~al.}(2005)\citenamefont {Fukui},
  \citenamefont {Hatsugai},\ and\ \citenamefont {Suzuki}}]{fukui2005chern}%
  \BibitemOpen
  \bibfield  {author} {\bibinfo {author} {\bibfnamefont {T.}~\bibnamefont
  {Fukui}}, \bibinfo {author} {\bibfnamefont {Y.}~\bibnamefont {Hatsugai}}, \
  and\ \bibinfo {author} {\bibfnamefont {H.}~\bibnamefont {Suzuki}},\ }\href
  {\doibase 10.1143/JPSJ.74.1674} {\bibfield  {journal} {\bibinfo  {journal}
  {Journal of the Physical Society of Japan}\ }\textbf {\bibinfo {volume}
  {74}},\ \bibinfo {pages} {1674} (\bibinfo {year} {2005})}\BibitemShut
  {NoStop}%
\bibitem [{\citenamefont {Iaconis}\ \emph {et~al.}(2018)\citenamefont
  {Iaconis}, \citenamefont {Liu}, \citenamefont {Hal{\'a}sz},\ and\
  \citenamefont {Balents}}]{iaconis2018spin}%
  \BibitemOpen
  \bibfield  {author} {\bibinfo {author} {\bibfnamefont {J.}~\bibnamefont
  {Iaconis}}, \bibinfo {author} {\bibfnamefont {C.-X.}\ \bibnamefont {Liu}},
  \bibinfo {author} {\bibfnamefont {G.}~\bibnamefont {Hal{\'a}sz}}, \ and\
  \bibinfo {author} {\bibfnamefont {L.}~\bibnamefont {Balents}},\ }\href
  {\doibase 10.21468/SciPostPhys.4.1.003} {\bibfield  {journal} {\bibinfo
  {journal} {SciPost Phys.}\ }\textbf {\bibinfo {volume} {4}},\ \bibinfo
  {pages} {003} (\bibinfo {year} {2018})}\BibitemShut {NoStop}%
\bibitem [{\citenamefont {Tang}\ \emph {et~al.}(2013)\citenamefont {Tang},
  \citenamefont {Fisher},\ and\ \citenamefont {Lee}}]{tang2013low}%
  \BibitemOpen
  \bibfield  {author} {\bibinfo {author} {\bibfnamefont {E.}~\bibnamefont
  {Tang}}, \bibinfo {author} {\bibfnamefont {M.~P.}\ \bibnamefont {Fisher}}, \
  and\ \bibinfo {author} {\bibfnamefont {P.~A.}\ \bibnamefont {Lee}},\ }\href
  {\doibase 10.1103/PhysRevB.87.045119} {\bibfield  {journal} {\bibinfo
  {journal} {Phys. Rev. B}\ }\textbf {\bibinfo {volume} {87}},\ \bibinfo
  {pages} {045119} (\bibinfo {year} {2013})}\BibitemShut {NoStop}%
\bibitem [{\citenamefont {Ruan}\ \emph {et~al.}(2021)\citenamefont {Ruan},
  \citenamefont {Chen}, \citenamefont {Tang}, \citenamefont {Hwang},
  \citenamefont {Tsai}, \citenamefont {Lee}, \citenamefont {Wu}, \citenamefont
  {Ryu}, \citenamefont {Kahn}, \citenamefont {Liou}, \citenamefont {Jia},
  \citenamefont {Aikawa}, \citenamefont {Hwang}, \citenamefont {Wang},
  \citenamefont {Choi}, \citenamefont {Louie}, \citenamefont {Lee},
  \citenamefont {Shen}, \citenamefont {Mo},\ and\ \citenamefont
  {Crommie}}]{ruan2021tase2}%
  \BibitemOpen
  \bibfield  {author} {\bibinfo {author} {\bibfnamefont {W.}~\bibnamefont
  {Ruan}}, \bibinfo {author} {\bibfnamefont {Y.}~\bibnamefont {Chen}}, \bibinfo
  {author} {\bibfnamefont {S.}~\bibnamefont {Tang}}, \bibinfo {author}
  {\bibfnamefont {J.}~\bibnamefont {Hwang}}, \bibinfo {author} {\bibfnamefont
  {H.-Z.}\ \bibnamefont {Tsai}}, \bibinfo {author} {\bibfnamefont {R.~L.}\
  \bibnamefont {Lee}}, \bibinfo {author} {\bibfnamefont {M.}~\bibnamefont
  {Wu}}, \bibinfo {author} {\bibfnamefont {H.}~\bibnamefont {Ryu}}, \bibinfo
  {author} {\bibfnamefont {S.}~\bibnamefont {Kahn}}, \bibinfo {author}
  {\bibfnamefont {F.}~\bibnamefont {Liou}}, \bibinfo {author} {\bibfnamefont
  {C.}~\bibnamefont {Jia}}, \bibinfo {author} {\bibfnamefont {A.}~\bibnamefont
  {Aikawa}}, \bibinfo {author} {\bibfnamefont {C.}~\bibnamefont {Hwang}},
  \bibinfo {author} {\bibfnamefont {F.}~\bibnamefont {Wang}}, \bibinfo {author}
  {\bibfnamefont {Y.}~\bibnamefont {Choi}}, \bibinfo {author} {\bibfnamefont
  {S.~G.}\ \bibnamefont {Louie}}, \bibinfo {author} {\bibfnamefont {P.~A.}\
  \bibnamefont {Lee}}, \bibinfo {author} {\bibfnamefont {Z.-X.}\ \bibnamefont
  {Shen}}, \bibinfo {author} {\bibfnamefont {S.-K.}\ \bibnamefont {Mo}}, \ and\
  \bibinfo {author} {\bibfnamefont {M.~F.}\ \bibnamefont {Crommie}},\ }\href
  {\doibase 10.1038/s41567-021-01321-0} {\bibfield  {journal} {\bibinfo
  {journal} {Nat. Phys.}\ }\textbf {\bibinfo {volume} {17}},\ \bibinfo {pages}
  {1154} (\bibinfo {year} {2021})}\BibitemShut {NoStop}%
\bibitem [{\citenamefont {Chen}\ \emph {et~al.}(2020)\citenamefont {Chen},
  \citenamefont {Ruan}, \citenamefont {Wu}, \citenamefont {Tang}, \citenamefont
  {Ryu}, \citenamefont {Tsai}, \citenamefont {Lee}, \citenamefont {Kahn},
  \citenamefont {Liou}, \citenamefont {Jia}, \citenamefont {Albertini},
  \citenamefont {Xiong}, \citenamefont {Jia}, \citenamefont {Liu},
  \citenamefont {Sobota}, \citenamefont {Liu}, \citenamefont {Moore},
  \citenamefont {Shen}, \citenamefont {Louie}, \citenamefont {Mo},\ and\
  \citenamefont {Crommie}}]{chen2020tase2}%
  \BibitemOpen
  \bibfield  {author} {\bibinfo {author} {\bibfnamefont {Y.}~\bibnamefont
  {Chen}}, \bibinfo {author} {\bibfnamefont {W.}~\bibnamefont {Ruan}}, \bibinfo
  {author} {\bibfnamefont {M.}~\bibnamefont {Wu}}, \bibinfo {author}
  {\bibfnamefont {S.}~\bibnamefont {Tang}}, \bibinfo {author} {\bibfnamefont
  {H.}~\bibnamefont {Ryu}}, \bibinfo {author} {\bibfnamefont {H.-Z.}\
  \bibnamefont {Tsai}}, \bibinfo {author} {\bibfnamefont {R.}~\bibnamefont
  {Lee}}, \bibinfo {author} {\bibfnamefont {S.}~\bibnamefont {Kahn}}, \bibinfo
  {author} {\bibfnamefont {F.}~\bibnamefont {Liou}}, \bibinfo {author}
  {\bibfnamefont {C.}~\bibnamefont {Jia}}, \bibinfo {author} {\bibfnamefont
  {O.~R.}\ \bibnamefont {Albertini}}, \bibinfo {author} {\bibfnamefont
  {H.}~\bibnamefont {Xiong}}, \bibinfo {author} {\bibfnamefont
  {T.}~\bibnamefont {Jia}}, \bibinfo {author} {\bibfnamefont {Z.}~\bibnamefont
  {Liu}}, \bibinfo {author} {\bibfnamefont {J.~A.}\ \bibnamefont {Sobota}},
  \bibinfo {author} {\bibfnamefont {A.~Y.}\ \bibnamefont {Liu}}, \bibinfo
  {author} {\bibfnamefont {J.~E.}\ \bibnamefont {Moore}}, \bibinfo {author}
  {\bibfnamefont {Z.-X.}\ \bibnamefont {Shen}}, \bibinfo {author}
  {\bibfnamefont {S.~G.}\ \bibnamefont {Louie}}, \bibinfo {author}
  {\bibfnamefont {S.-K.}\ \bibnamefont {Mo}}, \ and\ \bibinfo {author}
  {\bibfnamefont {M.~F.}\ \bibnamefont {Crommie}},\ }\href {\doibase
  10.1038/s41567-019-0744-9} {\bibfield  {journal} {\bibinfo  {journal} {Nat.
  Phys.}\ }\textbf {\bibinfo {volume} {16}},\ \bibinfo {pages} {218} (\bibinfo
  {year} {2020})}\BibitemShut {NoStop}%
\bibitem [{\citenamefont {Liu}\ \emph {et~al.}(2021{\natexlab{b}})\citenamefont
  {Liu}, \citenamefont {Leveillee}, \citenamefont {Lu}, \citenamefont {Yu},
  \citenamefont {Kim}, \citenamefont {Tian}, \citenamefont {Shi}, \citenamefont
  {Lai}, \citenamefont {Zhang}, \citenamefont {Giustino},\ and\ \citenamefont
  {Shih}}]{mengke2021nbs2}%
  \BibitemOpen
  \bibfield  {author} {\bibinfo {author} {\bibfnamefont {M.}~\bibnamefont
  {Liu}}, \bibinfo {author} {\bibfnamefont {J.}~\bibnamefont {Leveillee}},
  \bibinfo {author} {\bibfnamefont {S.}~\bibnamefont {Lu}}, \bibinfo {author}
  {\bibfnamefont {J.}~\bibnamefont {Yu}}, \bibinfo {author} {\bibfnamefont
  {H.}~\bibnamefont {Kim}}, \bibinfo {author} {\bibfnamefont {C.}~\bibnamefont
  {Tian}}, \bibinfo {author} {\bibfnamefont {Y.}~\bibnamefont {Shi}}, \bibinfo
  {author} {\bibfnamefont {K.}~\bibnamefont {Lai}}, \bibinfo {author}
  {\bibfnamefont {C.}~\bibnamefont {Zhang}}, \bibinfo {author} {\bibfnamefont
  {F.}~\bibnamefont {Giustino}}, \ and\ \bibinfo {author} {\bibfnamefont
  {C.-K.}\ \bibnamefont {Shih}},\ }\href {\doibase 10.1126/sciadv.abi6339}
  {\bibfield  {journal} {\bibinfo  {journal} {Science Advances}\ }\textbf
  {\bibinfo {volume} {7}},\ \bibinfo {pages} {eabi6339} (\bibinfo {year}
  {2021}{\natexlab{b}})},\ \Eprint
  {http://arxiv.org/abs/https://www.science.org/doi/pdf/10.1126/sciadv.abi6339}
  {https://www.science.org/doi/pdf/10.1126/sciadv.abi6339} \BibitemShut
  {NoStop}%
\bibitem [{\citenamefont {Lee}\ \emph {et~al.}(2020)\citenamefont {Lee},
  \citenamefont {Jin}, \citenamefont {Catuneanu}, \citenamefont {Go},
  \citenamefont {Jung}, \citenamefont {Won}, \citenamefont {Cheong},
  \citenamefont {Kim}, \citenamefont {Liu}, \citenamefont {Kee},\ and\
  \citenamefont {Yeom}}]{lee2020tas2}%
  \BibitemOpen
  \bibfield  {author} {\bibinfo {author} {\bibfnamefont {J.}~\bibnamefont
  {Lee}}, \bibinfo {author} {\bibfnamefont {K.-H.}\ \bibnamefont {Jin}},
  \bibinfo {author} {\bibfnamefont {A.}~\bibnamefont {Catuneanu}}, \bibinfo
  {author} {\bibfnamefont {A.}~\bibnamefont {Go}}, \bibinfo {author}
  {\bibfnamefont {J.}~\bibnamefont {Jung}}, \bibinfo {author} {\bibfnamefont
  {C.}~\bibnamefont {Won}}, \bibinfo {author} {\bibfnamefont {S.-W.}\
  \bibnamefont {Cheong}}, \bibinfo {author} {\bibfnamefont {J.}~\bibnamefont
  {Kim}}, \bibinfo {author} {\bibfnamefont {F.}~\bibnamefont {Liu}}, \bibinfo
  {author} {\bibfnamefont {H.-Y.}\ \bibnamefont {Kee}}, \ and\ \bibinfo
  {author} {\bibfnamefont {H.~W.}\ \bibnamefont {Yeom}},\ }\href {\doibase
  10.1103/PhysRevLett.125.096403} {\bibfield  {journal} {\bibinfo  {journal}
  {Phys. Rev. Lett.}\ }\textbf {\bibinfo {volume} {125}},\ \bibinfo {pages}
  {096403} (\bibinfo {year} {2020})}\BibitemShut {NoStop}%
\end{thebibliography}%

\end{document}


\title{Supplementary Information for "Emergent Quantum Phenomena of Noncentrosymmetric Charge-Density Wave in 1T-Transition Metal Dichalcogenides"}

\author{Cheong-Eung Ahn}
\thanks{These authors contributed equally.}
\affiliation{Department of Physics, Pohang University of Science and Technology, Pohang, 37673, Republic of Korea}
\affiliation{Center for Artificial Low Dimensional Electronic Systems, Institute for Basic Science, Pohang 37673, Korea}

\author{Kyung-Hwan Jin}
\thanks{These authors contributed equally.}
\affiliation{Department of Physics, Jeonbuk National University, Jeonju, 54896, Republic of Korea}
\affiliation{Center for Artificial Low Dimensional Electronic Systems, Institute for Basic Science, Pohang 37673, Korea}

\author{Young-Jae Choi}
\affiliation{Center for Artificial Low Dimensional Electronic Systems, Institute for Basic Science, Pohang 37673, Korea}

\author{Jae Whan Park}
\affiliation{Center for Artificial Low Dimensional Electronic Systems, Institute for Basic Science, Pohang 37673, Korea}

\author{Han Woong Yeom}
\affiliation{Department of Physics, Pohang University of Science and Technology, Pohang, 37673, Republic of Korea}
\affiliation{Center for Artificial Low Dimensional Electronic Systems, Institute for Basic Science, Pohang 37673, Korea}

\author{Ara Go}
\email{arago@jnu.ac.kr}
\affiliation{Department of Physics, Chonnam National University, Gwangju 61186, Korea} 

\author{Yong Baek Kim}
\email{yongbaek.kim@utoronto.ca}
\affiliation{Department of Physics, University of Toronto, Toronto, Ontario M5S 1A7, Canada} 

\author{Gil Young Cho}
\email{gilyoungcho@postech.ac.kr}
\affiliation{Department of Physics, Pohang University of Science and Technology, Pohang, 37673, Republic of Korea}
\affiliation{Center for Artificial Low Dimensional Electronic Systems, Institute for Basic Science, Pohang 37673, Korea}
\affiliation{Asia-Pacific Center for Theoretical Physics, Pohang, Gyeongbuk, 37673, Korea}

\maketitle

\tableofcontents

\section{STM Images of ACDS CDW}
\begin{figure*}[!ht]
\includegraphics{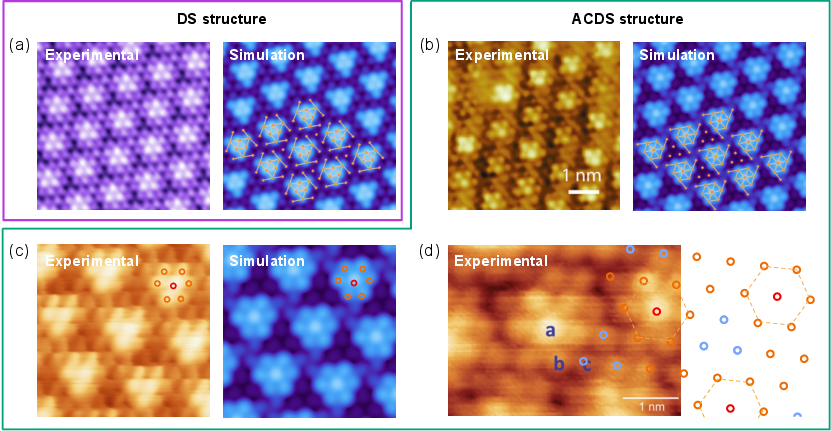}
\caption{\textbf{(a)} Experimental \protect\cite{qiao2017mottness} (left) and simulated (right) STM image of the DS TaS${}_2$. Experimental STM images of \textbf{(b)} TaSe${}_2$ (V$_b$ = -0.1 V and I$_t$ = 150 pA), \textbf{(c)} NbSe${}_2$ (V$_b$ = -1.0 V and I$_t$ = 40 pA), and \textbf{(d)} TaS${}_2$ (V$_b$ = 1.0 V and I$_t$ = 1 nA), proposed to exhibit the ACDS structure, along with the corresponding simulated STM images. The experimental images are taken from \protect\cite{lin2020tase2,liu2021nbse2,luic2019tas2}, respectively. The positions of the top-layer X atoms are indicated in the simulated STM images. {We note that the DS structure \textbf{(a)} has triangular-shaped, three bright protrusions, which are markedly different from the ACDS structure \textbf{(b-d)}.}}
\label{fig:STM_exp}
\end{figure*}

For the DS CDW structure of 1T-MX$_2$, STM shows three bright protrusions around the DS center in the top $X$ layer. For example, see the STM image of 1T-TaS${}_2$, which was reported previously in \cite{qiao2017mottness}, in Figure \ref{fig:STM_exp}(a). For the ACDS CDW structure, we predict a single bright protrusion at the center which is surrounded by six smaller protrusions (Figure \ref{fig:STM_exp}(b)-(d)). Such features have been observed in 1T-TaSe${}_2$ \cite{lin2020tase2}, 1T-NbSe${}_2$ \cite{liu2021nbse2}, and 1T-TaS${}_2$ \cite{luic2019tas2}. Although this aspect was overlooked in the mentioned papers, it highlights the difficulty in reproducing the observed STM image using the DS CDW structure, while the ACDS CDW structure provides a more consistent interpretation.

\section{Density Functional Theory}
We performed first-principles calculations within the framework of density-functional theory using the Perdew-Burke-Ernzerhof-type generalized gradient approximation for the exchange-correlation functional, as implemented in the Vienna ab initio simulation package \cite{kres1996dft,perd1996pbe}. All the calculations are carried out using the kinetic energy cutoff of 520 eV on a $9 \times 9 \times 1$ Monkhorst-Pack k-point mesh for the $\sqrt{13} \times \sqrt{13}$ superlattice. All structures are fully optimized until the residual forces are less than 0.01 eV/\AA. The electronic self-consistent iteration is converged to $10^{-5}$ eV precision of the total energy. The spin-orbit coupling is included in the self-consistent electronic structure calculation. A vacuum layer of 18\AA${}$ is used to ensure decoupling between neighboring layers.

To accurately represent electronic correlations in 1T-MX${}_2$, the DFT+$U$ correction is employed, which captures the Coulomb interaction of Nb 4d and Ta 5d orbitals on the mean-field level, following the simplified (rotational invariant) approach introduced by Dudarev \cite{duda1998plu}. The effective Hubbard $U$ ($U_\text{eff}=U-J$) parameters are applied to the 4d or 5d electrons of M atom. We construct Wannier representations by projecting the Bloch states from the first-principles calculation of ACDS MX${}_2$ systems \cite{most2008wann}.

\begin{figure*}[!ht]
\centering
\includegraphics{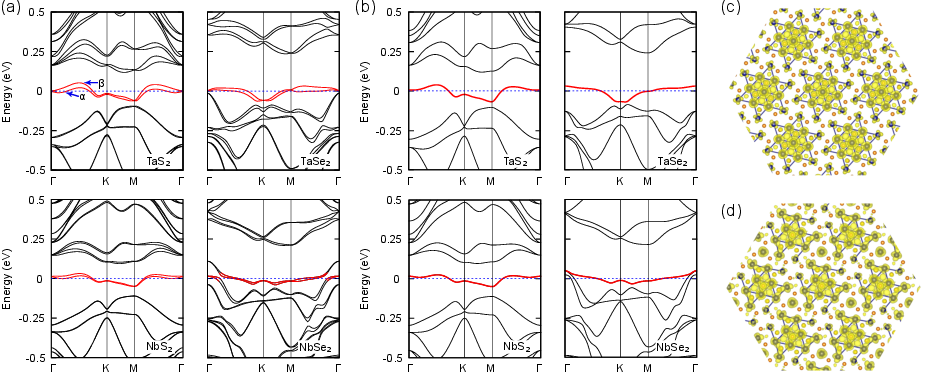}
\caption{Calculated band structure of ACDS CDW TaS${}_2$, TaSe${}_2$, NbS${}_2$, and NbSe${}_2$ \textbf{(a)} with and \textbf{(b)} without spin-orbit coupling, respectively. The flat bands of ACDS CDW are marked by the red line. Calculated real-space spreads of the flat bands of \textbf{(c)} DS and \textbf{(d)} ACDS CDW TaS${}_2$, respectively. In reference to the CDW cluster center, electrons are distributed locally, and in the case of the ACDS CDW, the electron delocalization is enhanced due to the additional charge distribution at the isolated single transition metal site.}
\label{fig:DFT_band}
\end{figure*}

\begin{figure}[!ht]
\centering
\includegraphics{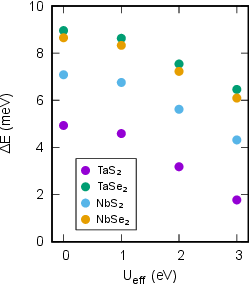}
\caption{The energy difference ($\Delta E$=$E_{DS}$-$E_{ACDS}$) between the DS and ACDS as a function of Hubbard $U$.}
\label{fig:Edff_U}
\end{figure}

The DFT results are summarized in Table \ref{tab:DFT_energy_ACDSvDS} and Figure \ref{fig:DFT_band}. {The inclusion of the Hubbard $U$ effect indeed changes the energy differences between the DS and ACDS structures. Our finding shows that with increasing $U$, the energy difference ($\Delta E$=$E_{DS}$-$E_{ACDS}$) between the DS and ACDS structures decreases Figure \ref{fig:Edff_U}. Notably, both systems manifest flat bands, implying the potential occurrence of spin splitting within these bands or the transition to a Mott phase owing to correlation effects. Consequently, such gap opening (due to the correlation) in flat bands contribute to the relative stability of the structures. These observations suggest a greater feasibility for experimental realization of the ACDS structure.}

\begin{table}[!h]
\centering
\begin{tabular}{c|ccc}
\hline
material & $a$ (\AA) & $\Delta E$ (meV) & $W_\text{FB}$ (meV)\\
\hline
$\mathrm{NbS_2}$  & 12.22 & 7.08 & 81 \\
$\mathrm{NbSe_2}$ & 12.59 & 8.65 & 63 \\
$\mathrm{TaS_2}$  & 12.17 & 4.93 & 114 \\
$\mathrm{TaSe_2}$ & 12.59 & 8.96 & 94 \\
\hline
\end{tabular}
\caption{(a) Calculated lattice constant of 1T-MX$_2$ with the ACDS structure, (b) energy difference per atom ($\Delta E=E_\text{DS}-E_\text{ACDS}$) between the DS and ACDS structure, and (c) band width ($W_\text{FB}$) of flat bands.}
\label{tab:DFT_energy_ACDSvDS}
\end{table}

\section{Spin Hall Conductivity}
The intrinsic spin Hall conductivities were calculated by using the wannier90 code \cite{most2008wann} for evaluating the Kubo formula \cite{grad2012she,guo2005she}:
\begin{equation}
\sigma_{xy}^{s_{z}}=\frac{e^{2}}{\hbar}\sum_{\bm{k}} f_{n}(\bm{k})\Omega_{n}^{s_{z}}(\bm{k}),\nonumber
\end{equation}
with 
\begin{equation}
\Omega _{n}^{s_{z}}(\bm{k})=2 \text{Im} \sum_{m\neq n}\frac{\bra{\psi _{n,\bm{k}}}\widehat{j}_{x}^{s_{z}}\ket{\psi_{m,\bm{k}}} \bra{\psi _{m,\bm{k}}}\nu_{y}\ket{\psi_{n,\bm{k}}}}{(E_{m}-E_{n})^{2}},\nonumber
\end{equation}
where $\widehat{j}_{x}^{s_z}$ is the spin current operator defined as $\frac{1}{2}(\widehat{s}_{z}\nu _{x}+\nu _{x}\widehat{s}_{z})$. Here $\nu _{x}$ is the velocity operator (along $x$), $f_{n}(\bm{k})$ is a Fermi distribution function at momentum $\bm{k}$. $\Omega _{n}^{s_z}(\bm{k})$ is the spin Berry curvature. By integrating $\Omega _{n}^{s_z}(\bm{k})$ over the occupied states, we obtain $\sigma_{xy}^{s_z}$ as a function of Fermi level.

\begin{figure*}[!ht]
\centering
\includegraphics{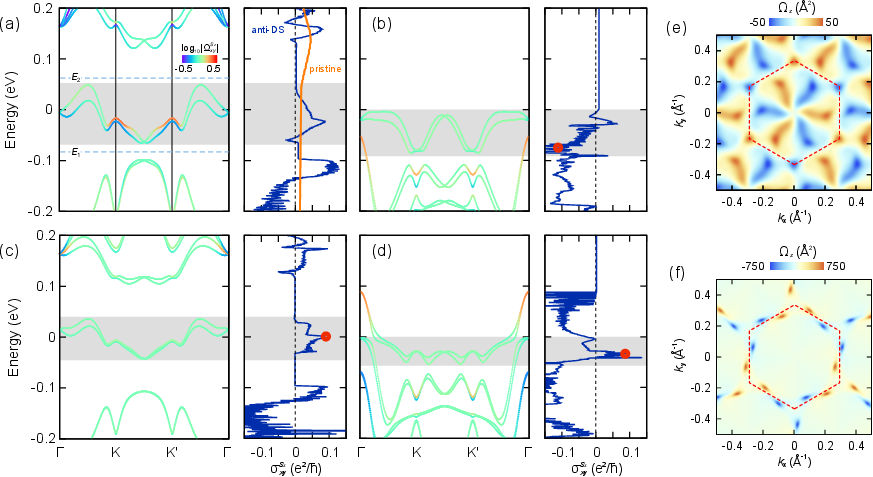}
\caption{Calculated spin Berry curvature ($\Omega_{xy}^{S_z}$) projected band and corresponding SHC ($\sigma_{xy}^{S_z}$) for ACDS \textbf{(a)} TaS${}_2$, \textbf{(b)} TaSe${}_2$, \textbf{(c)} NbS${}_2$, and \textbf{(d)} NbSe${}_2$, respectively. The shaded area represents regions attributed to the flat band. \textbf{(e,f)} Berry curvature distributions when the Fermi level is located at $E_1$ and $E_2$ in Figure \ref{fig:SHC_other}(a), respectively.}
\label{fig:SHC_other}
\end{figure*}

For the units of the SHC, the unit of $\sigma_{xy}^{s_z}$ for 2D monolayer systems is $e^{2}/{\hbar}$ ($\cong 2.43 \times 10^{-4}\Omega ^{-1}$), whereas $(\Omega m)^{-1}$ for 3D bulk systems. To convert the values of the $\sigma_{xy}^{s_{z}}$ in monolayer limits to those in bulk, one needs to divide the values of $\sigma_{xy}^{s_{z}}$ in the monolayer by its thickness (TaS${}_2$: 3.21\AA; TaSe${}_2$: 3.44\AA; NbS${}_2$: 3.20\AA; NbSe${}_2$: 3.48\AA).

Bands that create a crossing but then also form a hybridization gap with the inclusion of SOC will give rise to a large spin Berry curvature. In here, the gapped crossings at $K$ and $K'$ point generate large SHC (Figure \ref{fig:SHC_other}(a-d)). Furthermore, the strong density of states influenced by the flatness of the bands also plays a role. Figure \ref{fig:SHC_other}(e) and \ref{fig:SHC_other}(f) depict the Berry curvature distributions at the Fermi level positions of $E_1$ and $E_2$, respectively, where $E_1$ excludes the flat bands and $E_2$ includes the flat bands. When located at the $E_1$ level without the inclusion of the Rashba flat band, the maximum value of the Berry curvature is around 50 \AA${}^2$. However, in the case of the $E_2$ level, which includes the flat band, the maximum value of the Berry curvature significantly increases to 750 \AA${}^2$. Therefore, due to the localized states resulting from ACDS formation and the influence of the SOC gap, an enhanced spin Hall effect can be induced.

\section{Strain on ACDS CDW Structure}
Strain serves as an effective way for manipulating the electronic structure of 2D materials. In the context of TaSe$_2$, there has been evidence that altering strain can adjust the energy level of the flat band. Specifically, this material exhibits a Mott insulating state under tensile strain, while the collapse of "Mottness" occurs under compressive strain \cite{zhang2020mottness}. Notably, strain can also effectively tune the hopping parameter of localized orbitals that constitute the flat band. This adjustment holds the potential to regulate the flat band's band width and control correlated interactions.

\begin{figure}[!h]
\centering
\includegraphics{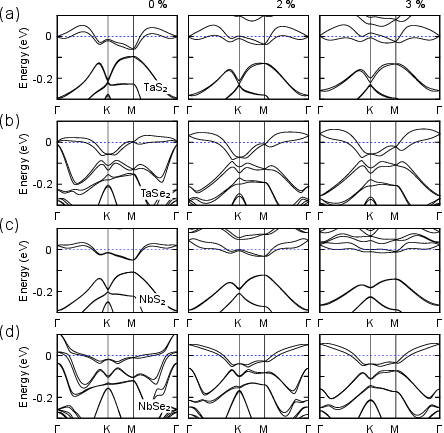}
\caption{Calculated band structure of ACDS CDW (column) TaS${}_2$, TaSe${}_2$, NbS${}_2$, and NbSe${}_2$ under (row) 0\%, 2\%, 3\% tensile strain. Under tensile strain, the band width of the Rashba flat band reduces by up to $\sim$30\%.}
\label{fig:DFT_strain}
\end{figure}

Figure \ref{fig:DFT_strain} showcases the strain-dependent band structure of the ACDS structure in 1T-MX$_2$. As tensile strain increases, the Rashba flat band's position progressively shifts upward, leading to a reduction in the band width of the flat band by up to 30\%. This is attributed to the decreasing hopping overlap between localized orbitals with increasing strain.

{It is also interesting to note that the energy difference between the two structures decreases under the strain, making the formation of the ACDS more likely in the presence of the strain. Once a metastable ACDS structure is formed, it is unlikely to transit to the DS structure due to the strain. This is due to the huge energy barrier between the DS and ACDS structures. To investigate the energy barrier for transition between the DS and ACDS structures, the Nudged Elastic Band (NEB) method was employed. The results suggest that an energy of 12 meV/atom is required for the ACDS structure to transit to the DS structure. Therefore, it is anticipated that the ACDS structure will stable even under applied strain.}

\section{Chern Insulator in TaSe${}_2$}
\begin{figure}[!h]
\centering
\includegraphics{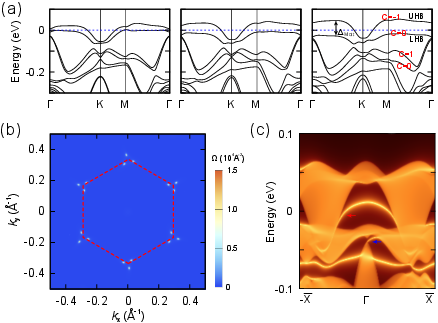}
\caption{\textbf{(a)} Calculated band structures of ACDS CDW TaSe${}_2$, with the on-site Coulomb repulsion $U$ of 0.6, 0.8, and 1.0 eV, respectively. \textbf{(b)} Calculated Berry curvature of TaSe${}_2$ (1.0 eV) when considering the occupied bands up to the LHB. \textbf{(c)} Calculated edge states of TaSe${}_2$ (1.0 eV) with a zigzag boundary. The chiral edge modes are indicated by red and blue arrows.}
\label{fig:Qah_tase2}
\end{figure}

In the ACDS CDW structure, we can expect the realization of Mott states through the correlation effects induced by the Rashba flat band. To investigate this, we performed calculations using the DFT+U formalism to explore the phase transitions under various values of Hubbard U. Figure \ref{fig:Qah_tase2}(a) show the band structures of ACDS CDW TaSe$_2$ as a function of on-site Coulomb repulsion values. By considering the U value, we can effectively describe the splitting of the lower and upper Hubbard bands (LHB and UHB) originating flat band as well as the Mott state. As the U value increases, we observe that the Mott gap gradually becomes larger. For TaSe$_2$, the hybridization with neighboring states leads to a topological Mott phase.

Figure \ref{fig:Qah_tase2}(b) displays the distribution of Berry curvature of TaSe$_2$ ($U=$1 eV) when considering the occupied bands up to the lower Hubbard band (LHB). The non-zero Berry curvature is strongly localized near the $K$ ($K'$) points, and integrating the Berry curvature over the first BZ gives the Chern number of 1, confirming a quantum anomalous Hall (QAH) phase based on Rashba flat band. A more explicit demonstration of QAH phase is the presence of the chiral states at the edges. Figure \ref{fig:Qah_tase2}(c) shows the gapless chiral edge states. Despite the absence of a global gap in the band structure due to the narrowness of the bands, the presence of a 1D chiral edge mode along the zigzag edge can be observed. This edge mode connects the LHB and upper Hubbard band (UHB) and exhibits a chiral nature.

\section{Effective Spin Model}
We built a Wannier tight-binding model which contains a set of parameters $\{ t_1, t_2, t_3, \cdots, \lambda_R, \lambda_{\text{SOC}}\}$. Here $t_i$, $\lambda_R$, and $\lambda_{\text{SOC}}$ denote the $i$-th neighbor hopping parameters, and strength of Rashba, and intrinsic SOC, respectively. We find $t_1,t_2,t_3,\lambda_R$ are of similar order of magnitude, while $t_i$ ($i>3$) and $\lambda_\text{SOC}$ are an order-of-magnitude smaller, and thus ignored. The parameters found from DFT are reported in Table \ref{tab:parameter_values} and \ref{tab:parameter_values_strained}.

To study the strongly-correlated phases in the Mott insulator limit, we include a large on-site repulsion, $V=U\sum_in_{i\uparrow}n_{i\downarrow}$. Performing $t/U$ perturbation theory \cite{macdonald1988t}, we obtain an effective spin model at second order. Note that to obtain Eq. (1), we additionally apply a $\pi/2$-rotation along the $z$-axis to the spin coordinates relative to the lattice coordinates for convenience. In the absence of spin-orbit coupling ($\lambda_R=0$), we only have the Heisenerg interactions $J_1^{(0)}=4t_1^2/U$ and
\begin{equation}
\frac{J_2}{J_1^{(0)}}=\left(\frac{t_2}{t_1}\right)^2,\qquad\frac{J_3}{J_1^{(0)}}=\left(\frac{t_3}{t_1}\right)^2.\nonumber
\end{equation}
Including spin-orbit coupling, the nearest-neighbor Heisenberg interaction is weakened to $J_1/J_1^{(0)}=1-\left(\lambda_R/t_1\right)^2$, and the DM and Ising interactions are introduced,
\begin{equation}
\frac{J_\parallel}{J_1^{(0)}}=2\left(\frac{\lambda_R}{t_1}\right)^2,\qquad\frac{D}{J_1^{(0)}}=2\left(\frac{\lambda_R}{t_1}\right).\nonumber
\end{equation}
The parameter values for the materials considered in our paper are listed in Table \ref{tab:parameter_values} and \ref{tab:parameter_values_strained}.
\begin{table}[!h]
\centering
\begin{tabular}{c|cccc|ccc}
\hline
material & $t_1$ & $t_2$ & $t_3$ & $\lambda_R$ & $\lambda_R/t_1$ & $J_2/J_1^{(0)}$ & $J_3/J_1^{(0)}$\\
\hline
$\mathrm{NbS_2}$  & 5.20 & -3.19 & -4.29 & 2.0 & 0.38 & 0.376 & 0.681\\
$\mathrm{NbSe_2}$ & 3.62 &  3.46 & -1.64 & 2 & 0.55 & 0.915 & 0.206\\
$\mathrm{TaS_2}$  & 6.61 & -5.32 & -5.91 & 4 & 0.61 & 0.648 & 0.799\\
$\mathrm{TaSe_2}$ & 4.62 & -5.46 &  1.64 & 4 & 0.87 & 1.397 & 0.126\\
\hline
\end{tabular}
\caption{Hopping and Rashba SOC parameters (in units of meV) of flat bands and corresponding spin-spin interaction strengths of the effective spin model of ACDS CDWs. For $\mathrm{NbSe_2}$, we present the parameters ignoring the hybridizaiton of the flat bands.}
\label{tab:parameter_values}
\end{table}

\begin{table}[!h]
\centering
\begin{tabular}{c|cccc|ccc}
\hline
strain & $t_1$ & $t_2$ & $t_3$ & $\lambda_R$ & $\lambda_R/t_1$ & $J_2/J_1^{(0)}$ & $J_3/J_1^{(0)}$\\
\hline
0.00\% & 5.20 & -3.19 & -4.29 & 2.0 & 0.38 & 0.376 & 0.681\\
0.25\% & 5.30 & -2.94 & -4.04 & 2.1 & 0.40 & 0.308 & 0.581\\
0.50\% & 5.40 & -2.69 & -3.79 & 2.3 & 0.42 & 0.248 & 0.493\\
0.75\% & 5.50 & -2.44 & -3.54 & 2.4 & 0.43 & 0.197 & 0.414\\
1.00\% & 5.60 & -2.19 & -3.29 & 2.5 & 0.45 & 0.153 & 0.345\\
\hline
\end{tabular}
\caption{Hopping and Rashba SOC parameters (in units of meV) of strained $\mathrm{NbS_2}$ flat bands and resulting spin-spin interaction strengths of the effective spin model of ACDS CDWs.}
\label{tab:parameter_values_strained}
\end{table}


\section{Exact Diagonalization}
Exact diagonalization (ED) was performed using the \texttt{QuSpin} Python package \cite{weinberg2017quspin}. The spin Hamiltonian was placed on a $3\times 6$ triangular lattice with periodic boundary conditions. The phase diagrams are colored according to the momentum which has the largest momentum-space spin-spin correlations for the ground state,
\begin{equation}
S\left(\mathbf{k}\right)=\frac{1}{N}\sum_{\mathbf{r},\mathbf{r}'}e^{i\mathbf{k}\cdot\left(\mathbf{r}'-\mathbf{r}\right)}\ev{\mathbf{S}_{\mathbf{r}'}\cdot\mathbf{S}_{\mathbf{r}}}_0.
\label{eq:ED_SScorr}
\end{equation}
We also find points where the second order derivative of energy with respect to the parameters $J_2,J_3$, diverges. The derivatives were approximately evaluated using the finite difference method.

\begin{figure}[!h]
\includegraphics[width=\linewidth]{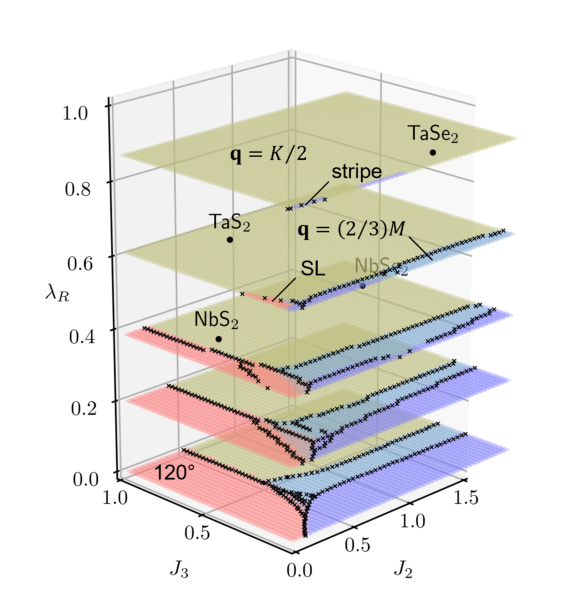}
\caption{Exact diagonalization phase diagram. Colors indicate the momentum which has the largest momentum-space spin-spin correlations for the ground state, Eq. (\ref{eq:ED_SScorr}), while the points indicate diverging energy derivatives with respect to $J_2,J_3$.}
\label{fig:ED_phase_diagram}
\end{figure}

The ED phase diagram is shown in Figure \ref{fig:ED_phase_diagram}. We find a non-trivial region whose phase boundaries are not characterized by changes in the spin-spin correlation peak momentum. Compared to the variational Monte Carlo phase diagram (Figure 4(a)), we expect this region to correspond to a spin liquid. While both phase diagrams show similar spin liquid regions, and both show an overall trend of incommensurate orders ($\Gamma$KI, $\Gamma$MI, IS) replacing the commensurate orders (120${}^\circ$, stripe) as $\lambda_R$ is increased, the phase boundaries are fairly different, especially between the incommensurate orders. We attribute this to finite size effects, namely, the ED calculations are unable to properly differentiate the incommensurate orders, resulting in greatly shifted phase boundaries.

\section{Classical Magnetic Order}
The Luttinger-Tisza method relaxes the constraint of spin size from $\norm{\mathbf{s}_{\mathbf{r}}}=1/2$ for each site $\mathbf{r}$, to the global constraint $\sum_{\mathbf{r}}\norm{\mathbf{s}_{\mathbf{r}}}^2=N/4$. The Luttinger-Tisza ground state can then be analytically found as
\begin{equation}
\mathbf{s}_{\mathbf{r}}=\frac{1}{2}\left[\cos\left(\mathbf{q}\cdot\mathbf{r}\right)\mathbf{u}_1+\sin\left(\mathbf{q}\cdot\mathbf{r}\right)\mathbf{u}_2\right],
\label{eq:classical_ansatz}
\end{equation}
where $\mathbf{q}=\arg\min_{\mathbf{k}}\left\{\lambda_0\left(\mathcal{J}_{\mathbf{k}}\right)\right\}$ is the momentum with the smallest eigenvalue of the momentum-space spin interaction matrix $\mathcal{J}_{\mathbf{k}}$, and $\mathbf{u}_1$ and $\mathbf{u}_2$ are real vectors determined from the corresponding eigenvector of $\mathcal{J}_{\mathbf{k}}$ and the time-reversal conjugate. If the Luttinger-Tisza ground state additionally satisfies $\norm{\mathbf{s}_{\mathbf{r}}}=1/2$, we can be certain that the resulting state is the classical ground state. This constraint requires $\mathbf{u}_1$ and $\mathbf{u}_2$ being orthonormal, i.e. the Luttinger-Tisza method fails when $\mathbf{u}_1$ and $\mathbf{u}_2$ fails to be orthonormal.

In order to fill parameter values where the Luttinger-Tisza ground state fails to satisfy the spin size constraint, we treat Eq. (\ref{eq:classical_ansatz}) as an ansatz where $\mathbf{q}$, $\mathbf{u}_1$, $\mathbf{u}_2$ are variational parameters, also known as a single-$Q$ ansatz. To restrict $\mathbf{u}_1$ and $\mathbf{u}_2$ to orthonormal vectors, we used $\mathbf{u}_1=R_{\bm{\theta}}\hat{\mathbf{x}}$ and $\mathbf{u}_2=R_{\bm{\theta}}\hat{\mathbf{y}}$ where $R_{\bm{\theta}}$ is a $\mathrm{SO}(3)$ rotation matrix parametrized by Euler's rotation vector $\bm{\theta}$. The optimization was carried out using a combination of simulated annealing and gradient descent. Phase boundaries between magnetic orders in Figure 4(a,b) were drawn based on divergences of the second or third order derivative of energy with respect to the parameters $J_2,J_3$.

\begin{figure}[!h]
\includegraphics[width=0.48\textwidth]{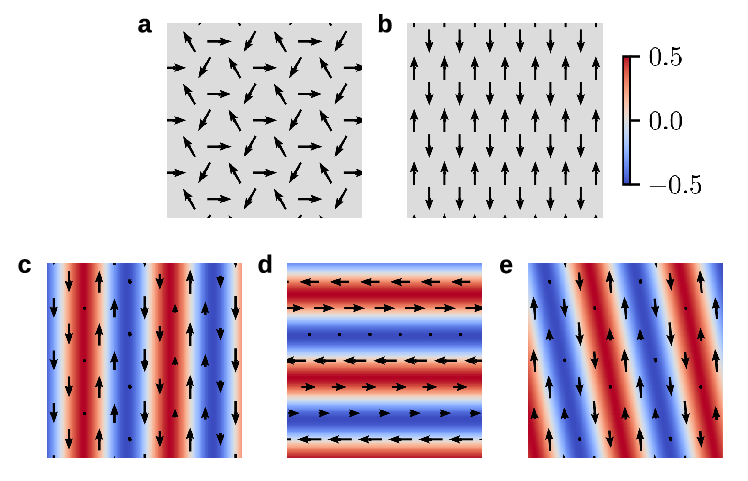}
\caption{Real-space spin configurations of magnetic orders. \textbf{(a)} 120${}^\circ$, \textbf{(b)} stripe, \textbf{(c)} $\Gamma$MI, \textbf{(d)} $\Gamma$KI, \textbf{(e)} IS orders. Only the in-plane component $(S_x, S_y)$ of spins is shown, while the background colors indicate $S_z$.}
\label{fig:magnetic_orders}
\end{figure}

We find the following five classical ground states, depicted in Figure \ref{fig:magnetic_orders}. First, the 120${}^\circ$ antiferromagnetic order has an ordering momentum of $\mathbf{q}=K$, and $\hat{\mathbf{u}},\hat{\mathbf{v}}$ lie on the xy-plane with a SO(2) degeneracy. The stripe order has an ordering momentum of $\mathbf{q}=M$, and $\hat{\mathbf{u}}$ is parallel to $\mathbf{q}$. The orientation vector $\hat{\mathbf{v}}$ does not affect the single-$Q$ ansatz for $\mathbf{q}=M$. $\Gamma$KI has an incommensurate ordering momentum $\mathbf{q}=\eta K$ with $\eta \in (0,1)$, which varies continuously with the system parameters. $\hat{\mathbf{u}},\hat{\mathbf{v}}$ are perpendicular to $\mathbf{q}$. Similarly, $\Gamma$MI has an incommensurate ordering momentum $\mathbf{q}=\eta M$ with $\eta \in (0,1)$ which varies continuously with the system parameters, and $\hat{\mathbf{u}},\hat{\mathbf{v}}$ are perpendicular to $\mathbf{q}$. Finally, the incommensurate spiral order has an ordering momentum which varies continuously with the system parameters, and does not remain fixed on a high-symmetry line. $\hat{\mathbf{u}}$ and $\hat{\mathbf{v}}$ are perpendicular to a spiral axis $\hat{\mathbf{w}}$, which also varies continuously with the system parameters and is not aligned from $\mathbf{q}$.

\section{Candidate Spin Liquids}
\paragraph{Spinon Mean Field Theory.} We use the standard fermionic partons to generate the candidate spin liquid states \cite{wen2002quantum}. Spin operators are given by
\begin{equation}
S_i^a=\frac{1}{2} \sum_{\alpha,\beta}f_{i,\alpha}^\dagger\sigma_{\alpha\beta}^a f_{i,\beta},
\label{eq:spinon_definition}
\end{equation}
where $f_{i,\alpha}$, $\alpha\in\{\uparrow,\downarrow\}$ is a canonical fermion operator. To project out the spurious components of spinon states, we further impose the constraint $n_i=\sum_{\alpha}f_{i,\alpha}^\dagger f_{i,\alpha}=1$ on states. Formally, it can be implemented by the Gutzwiller projector $\mathcal{P}_G=\prod_in_i(2-n_i)$. Then, our candidate spin liquid wavefunctions can be written as  
\begin{equation}
\ket{\Psi_\text{SL}\left(\mathbf{x}\right)}=\mathcal{P}_G\ket{\overline{\mathrm{GS}}_\text{kin}\left(\mathbf{x}\right)}, \nonumber
\end{equation}
where $\ket{\overline{\mathrm{GS}}_\text{kin}\left(\mathbf{x}\right)}$ is a fermionic ground state of a non-interacting mean-field Hamiltonian $\overline{H}_\text{kin}\left(\mathbf{x}\right)$ of spinons with a set of mean-field parameters $\mathbf{x}$. Essentially, $\mathbf{x}$ acts as the variational parameters for the spin liquid wavefunction $\ket{\Psi_\text{SL}\left(\mathbf{x}\right)}$, whose energy is optimized against the spin Hamiltonian Eq. (1).

The mean-field Hamiltonian is constructed by condensing the \textit{spin liquid decoupling channels} \cite{dodds2013quantum} of the quartic spinon Hamiltonian, obtained by direct substitution of Eq. (\ref{eq:spinon_definition}) to Eq. (1). The mean-field Hamiltonian can be written as
\begin{equation}
\begin{split}
\overline{H}_\text{kin}&=-\sum_{\ev{i,j}}\left(\mathbf{f}_i^\dagger\mathfrak{t}_{ij}^\dagger\mathbf{f}_j+\tilde{\mathbf{f}}_i^\dagger\mathfrak{d}_{ij}^\dagger\mathbf{f}_j\right)\\
&\qquad-\frac{J_2}{4}\sum_{\ev{\ev{i,j}}}\left(\chi_{ij}^*\mathbf{f}_i^\dagger\mathbf{f}_j+\delta_{ij}^*\tilde{\mathbf{f}}_i^\dagger\mathbf{f}_j\right)\\
&\qquad-\frac{J_3}{4}\sum_{\ev{\ev{\ev{i,j}}}}\left(\chi_{ij}^*\mathbf{f}_i^\dagger\mathbf{f}_j+\delta_{ij}^*\tilde{\mathbf{f}}_i^\dagger\mathbf{f}_j\right)+\text{h.c.},
\end{split}
\label{eq:MF_Hamiltonian}
\end{equation}
where $\chi_{ij},e_{ij}^\alpha,\delta_{ij},d_{ij}^\alpha$ are the mean field parameters, corresponding to the expectation value of the singlet hopping channel $X_{ij}$, the triplet hopping channels $E_{ij}^\alpha$, the singlet pairing channel $\Delta_{ij}$, and the triplet pairing channels $D_{ij}^\alpha$,
\begin{equation}
\begin{cases*}
X_{ij}=\mathbf{f}_i^\dagger\mathbf{f}_j\\
E_{ij}^\alpha=\mathbf{f}_i^\dagger\sigma^\alpha\mathbf{f}_j
\end{cases*},\qquad\begin{cases*}
\Delta_{ij}=\tilde{\mathbf{f}}_i^\dagger\mathbf{f}_j\\
D_{ij}^\alpha=\tilde{\mathbf{f}}_i^\dagger\sigma^\alpha\mathbf{f}_j
\end{cases*}.\nonumber
\end{equation}
The nearest neighbor hopping matrix $\mathfrak{t}_{ij}$ is given by
\begin{equation}
\begin{split}
\mathfrak{t}_{ij}&=\left[\left(\frac{J_1}{4}+\frac{J_\parallel}{8}+\frac{D}{8}\right)\chi_{ij}-i\frac{D}{8}e_{ij}^\parallel\right]\mathbb{I}\\
&\qquad+\left[i\frac{D}{8}\chi_{ij}+\left(\frac{J_\parallel}{8}+\frac{D}{8}\right)e_{ij}^\parallel\right]\left(\mathbf{r}_{ij}\cdot\bm{\sigma}\right)\\
&\qquad+\frac{D}{8}e_{ij}^\perp\left[\left(\hat{\mathbf{z}}\times\mathbf{r}_{ij}\right)\cdot\bm{\sigma}\right]+\frac{D}{8}e_{ij}^z\sigma^z,
\end{split}
\label{eq:MF_hopping_matrix}
\end{equation}
where $e_{ij}^\parallel=\mathbf{r}_{ij}\cdot\mathbf{e}_{ij}$ and $e_{ij}^\perp=\left(\hat{\mathbf{z}}\times\mathbf{r}_{ij}\right)\cdot\mathbf{e}_{ij}$. An equivalent definition applies to the nearest-neighbor pairing matrix $\mathfrak{d}_{ij}$ in terms of $\delta_{ij}$ and $d_{ij}^\alpha$. The mean field parameters serve as the ansatz parameters for variational Monte Carlo.

\paragraph{Fully Symmetric U(1) States.} For now, we enforce the full lattice symmetry group assuming a $\mathrm{U}(1)$ invariant gauge group. Using the projective symmetry group classification \cite{wen2002quantum}, we identify six non-trivial mean field ansatz, each with 1-4 independent real mean-field parameters. For simplicity, we express the different mean-field solutions using a larger set of 6 parameters: $\chi,\tilde{e}^\parallel,\tilde{e}^\perp,\tilde{e}^z,\chi^{(2)},\chi^{(3)}$, some of which are trivial (Table \ref{tab:symmetric_MF_PSGs}). The nearest-neighbor hopping matrices, Eq. (\ref{eq:MF_hopping_matrix}), are given by
\begin{equation}
\begin{cases*}
\mathfrak{t}^\text{U(1)}_{\mathbf{r}+\mathbf{l}_0,\mathbf{r}}=\mathfrak{t}^\text{U(1)}_{\left(n_1-1,n_2-1\right),\left(n_1,n_2\right)}=\eta_0^{n_2}\mathfrak{t}^\text{U(1)}_0\\
\mathfrak{t}^\text{U(1)}_{\mathbf{r}+\mathbf{l}_1,\mathbf{r}}=\mathfrak{t}^\text{U(1)}_{\left(n_1+1,n_2\right),\left(n_1,n_2\right)}=\eta_0^{n_2}\mathfrak{t}^\text{U(1)}_1\\
\mathfrak{t}^\text{U(1)}_{\mathbf{r}+\mathbf{l}_2,\mathbf{r}}=\mathfrak{t}^\text{U(1)}_{\left(n_1,n_2+1\right),\left(n_1,n_2\right)}=\eta_0\mathfrak{t}^\text{U(1)}_2
\end{cases*},
\label{eq:ansatz_symU1}
\end{equation}
where
\begin{equation}
\mathfrak{t}^\text{U(1)}_a=t_0\mathbb{I}+it_\parallel\sigma_a^\parallel+i\frac{D}{8}\tilde{e}^\perp\sigma_a^\perp+i\frac{D}{8}\tilde{e}^z\sigma^z,\nonumber
\end{equation}
where
\begin{equation}
\begin{cases*}
t_0=\left(\frac{J_1}{4}+\frac{J_\parallel}{8}+\frac{D}{8}\right)\chi+\frac{D}{8}\tilde{e}^\parallel\\
t_\parallel=\frac{D}{8}\chi+\left(\frac{J_\parallel}{8}+\frac{D}{8}\right)\tilde{e}^\parallel
\end{cases*},\nonumber
\end{equation}
and $\sigma_a^\parallel=\mathbf{l}_a\cdot\bm{\sigma}$, $\sigma_a^\perp=\left(\hat{\mathbf{z}}\times\mathbf{l}_a\right)\cdot\bm{\sigma}$. The second-neighbor hopping terms are given by
\begin{equation}
\begin{cases*}
\chi_{\mathbf{r}+\mathbf{l}^{(2)}_0,\mathbf{r}}=\chi_{\left(n_1-1,n_2-2\right),\left(n_1,n_2\right)}=\eta_0^{n_2}\chi^{(2)}\\
\chi_{\mathbf{r}+\mathbf{l}^{(2)}_1,\mathbf{r}}=\chi_{\left(n_1+2,n_2+1\right),\left(n_1,n_2\right)}=\chi^{(2)}\\
\chi_{\mathbf{r}+\mathbf{l}^{(2)}_2,\mathbf{r}}=\chi_{\left(n_1-1,n_2+1\right),\left(n_1,n_2\right)}=\eta_0^{n_2}\chi^{(2)}
\end{cases*},\nonumber
\end{equation}
and the third-neighbor hopping terms are given by
\begin{equation}
\begin{cases*}
\chi_{\mathbf{r}+2\mathbf{l}_0,\mathbf{r}}=\chi_{(n_1-2,n_2-2),(n_1,n_2)}=\chi^{(3)}\\
\chi_{\mathbf{r}+2\mathbf{l}_1,\mathbf{r}}=\chi_{(n_1+2,n_2),(n_1,n_2)}=\chi^{(3)}\\
\chi_{\mathbf{r}+2\mathbf{l}_2,\mathbf{r}}=\chi_{(n_1,n_2+2),(n_1,n_2)}=\chi^{(3)}
\end{cases*}.\nonumber
\end{equation}

\begin{table}[!h]
\begin{tabular}{c|c|cccccc}
\hline
$\mathrm{U}(1)$ PSG & $\eta_0$ & $\chi$ & $\tilde{e}^{\parallel}$ & $\tilde{e}^{\perp}$ & $\tilde{e}^{z}$ & $\chi^{(2)}$ & $\chi^{(3)}$\\
\hline
U1A00 & $+1$ & O & O & X & X & O & O\\
U1A10 & $+1$ & X & X & O & X & X & X\\
U1A11 & $+1$ & X & X & X & O & X & X\\
U1B00 & $-1$ & X & X & X & O & O & O\\
U1B01 & $-1$ & X & X & O & X & X & X\\
U1B11 & $-1$ & O & O & X & X & X & X\\
\hline
\end{tabular}
\caption{Table of allowed mean field parameters for the six distinct non-trivial $\mathrm{U}(1)$ PSGs. Only the parameters marked by O can take non-zero values. The other PSGs are either trivial or produce equivalent ansatze for the mean-field Hamiltonian, Eq. (\ref{eq:MF_Hamiltonian}).}
\label{tab:symmetric_MF_PSGs}
\end{table}

\begin{figure*}[!ht]
\includegraphics[width=\linewidth]{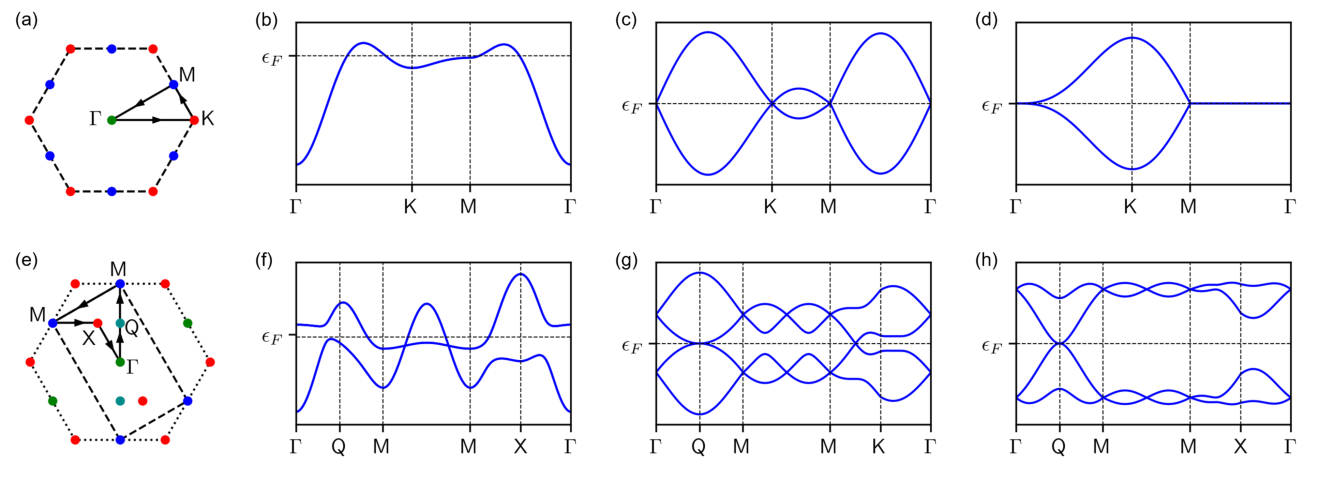}
\caption{Example spinon band structures of the six symmetric spinon mean field states. Spinon band structures of \textbf{(b)} U1A00, \textbf{(c)} U1A10, \textbf{(d)} U1A11 states are depicted with the Brillouin zone path specified in \textbf{(a)}. Spinon band structures of \textbf{(f)} U1B00, \textbf{(g)} U1B01, \textbf{(h)} U1B11 states are depicted with the Brillouin zone path specified in \textbf{(e)}. Note that the bands depicted in (b,f) are doubly degenerate.}
\label{fig:MF_bands_symm}
\end{figure*}

The three U1A states have two energy bands over the full Brillouin zone (Figure \ref{fig:MF_bands_symm}(a)). The U1A00 states has four mean field parameters, and as such, can take on various band structures. It typically describes a Fermi surface state (Figure \ref{fig:MF_bands_symm}(b)). The U1A10 and U1A11 states have only one mean field parameter. This parameter only controls the overall energy scale without affecting the topological properties of the band. The U1A10 state is a semi-metal with anisotropic Dirac cones at the six $\Gamma$, $K$, and $M$ points (Figure \ref{fig:MF_bands_symm}(c)). The U1A11 state is a semi-metal where the entire $\overline{\Gamma M}$ lines lie on the Fermi energy, and the band dispersion is quadratic at the $\Gamma$ point (Figure \ref{fig:MF_bands_symm}(d)).

Due to the insertion of $\pi$-flux, the translational symmetry of the U1B states is halved and the six-fold rotational symmetry is lowered to two-fold (only for the spinons, spins retain the full symmetry). As such, we consider the four energy bands over the halved Brillouin zone (Figure \ref{fig:MF_bands_symm}(e)). The U1B00 state has three mean field parameters, and as such, can take on various band structures. Specifically, it can correspond to a Fermi surface state (Figure \ref{fig:MF_bands_symm}(f)), a semi-metal with six Dirac cones, or a trivial band insulator. Note that the latter does not correspond to a spin liquid. The U1B01 state has a single mean field parameter. It describes a semi-metal with quadratic dispersion at the $Q$ points, and a Dirac cone at the $\overline{MX}$ midpoints (Figure \ref{fig:MF_bands_symm}(g)). The U1B11 states has two mean field parameters, and typically describes a quadratic band touching (QBT) at the $Q$ points. In the absence of SOC, the U1B11 state reduces to one mean field parameter, and becomes fine tuned to instead have doubly degenerate Dirac cones at the $Q$ points.

\paragraph{Chiral Descendants.} As discussed in the main text, we also consider the minimally symmetry broken and $\mathbb{Z}_2$ descendants of the U1B11 states. For spontaneous time-reversal symmetry breaking, we consider the four chiral subgroups, generated by
\begin{equation}
\mathrm{SG}^{c\tau\tau'}=\ev{T_1,T_2,C_2^x\mathcal{T}^{\tau},C_6^z\mathcal{T}^{\tau'}}\subsetneq\mathrm{SG},\nonumber
\end{equation}
where $\tau,\tau'\in\{0,1\}$. We denote the four descendants of the U1B11 ansatz as the c$\tau\tau'$U1B11 ansastze. We find that the c00U1B11 ansatz is unchanged from the original U1B11 ansatz. The c01U1B11 ansatz can evolve the QBT to a chiral spinon Fermi surface ansatz. The c10U1B11 and c11U1B11 ansatze, which are Kalmeyer-Laughlin-type chiral spin liquids \cite{bieri2016projective}, can gap out the QBT. Numerically evaluating the Chern number \cite{fukui2005chern}, we find that both have a Chern number of $C=2$ ($C=1$ for each of the $Q$ points).

\paragraph{Nematic Descendants.} Next, we consider nematic spin liquids, which have reduced rotational symmetry. Specifically, we consider spin liquids satisfying the reduced subgroups,
\begin{equation}
\begin{cases*}
\mathrm{SG}^{n3}=\ev{T_1,T_2,\mathcal{T},C_2^x,C_3^z}\subsetneq\mathrm{SG}\\
\mathrm{SG}^{n2}=\ev{T_1,T_2,\mathcal{T},C_2^x,C_2^z}\subsetneq\mathrm{SG}
\end{cases*},\nonumber
\end{equation}
where $C_3^z=\left(C_2^z\right)^2$ and $C_2^z=\left(C_6^z\right)^3$. We find that the n3U1B11 ansatz is equivalent to the U1B11 ansatz. The n2U1B11 ansatz has four additional parameters from the U1B11 ansatz, which can allow each QBT to evolve into two Dirac cones (Figure 4(g)). We note that the n2U1B11 ansatz can also topologically-trivially gap out the QBT, realizing magnetic order.

\paragraph{Z2 Descendants.} Finally, we consider the $\mathbb{Z}_2$ descendants, also known as \textit{Higgs} descendants. We identify four $\mathbb{Z}_2$ descendants of the U1B11 ansatz. These can be written so that in addition to the six hopping mean-field parameters of Eq. (\ref{eq:ansatz_symU1}), we can also have non-trivial pairing mean field parameters $\delta$ and $\tilde{d}^\parallel$ which gives the nearest-neighbor pairing matrices as
\begin{equation}
\begin{cases*}
\mathfrak{d}_{\mathbf{r}+\mathbf{l}_0,\mathbf{r}}=\mathfrak{d}_{\left(n_1-1,n_2-1\right),\left(n_1,n_2\right)}=\eta_0^{n_2}\mathfrak{d}_0\\
\mathfrak{d}_{\mathbf{r}+\mathbf{l}_1,\mathbf{r}}=\mathfrak{d}_{\left(n_1+1,n_2\right),\left(n_1,n_2\right)}=\eta_0^{n_2}\mathfrak{d}_1\\
\mathfrak{d}_{\mathbf{r}+\mathbf{l}_2,\mathbf{r}}=\mathfrak{d}_{\left(n_1,n_2+1\right),\left(n_1,n_2\right)}=\eta_0\mathfrak{d}_2
\end{cases*},\nonumber
\end{equation}
where
\begin{equation}
\begin{split}
\mathfrak{d}_a&=\left[\left(\frac{J_1}{4}+\frac{J_\parallel}{8}+\frac{D}{8}\right)\delta+\frac{D}{8}\tilde{d}^\parallel\right]\mathbb{I}\\
&\qquad+i\left[\frac{D}{8}\delta+\left(\frac{J_\parallel}{8}+\frac{D}{8}\right)\tilde{d}^\parallel\right]\left(\mathbf{l}_a\cdot\bm{\sigma}\right).
\end{split}\nonumber
\end{equation}
Among the 8 mean field parameters, only 2-5 can take non-trivial values (Table \ref{tab:symmetric_MF_Z2PSGs}). We find that the Z2B01111 ansatz is unchanged from the original U1B11 ansatz (after a particle-hole transformation), and Z2B00110 and Z2B10111 ansatze remain QBTs despite the additional parameters. The Z2B11110 ansatz can either split the QBTs into four Dirac cones, or topologically-trivially gap out the QBT, similar to the n2U1B11 ansatz.

\begin{table}[!h]
\begin{tabular}{c|cccccc|cc}
\hline
$\mathbb{Z}_2$ PSG label & $\chi$ & $\tilde{e}^\parallel$ & $\tilde{e}^\perp$ & $\tilde{e}^z$ & $\chi^{(2)}$ & $\chi^{(3)}$ & $\delta$ & $\tilde{d}^\parallel$\\
\hline
Z2B00110 & O & O & X & X & X & X & O & O\\
Z2B01111 & X & X & X & X & X & X & O & O\\
Z2B10111 & X & X & O & X & X & X & O & O\\
Z2B11110 & X & X & X & O & O & O & O & O\\
\hline
\end{tabular}
\caption{Table of allowed mean field parameters among the four $\mathbb{Z}_2$ algebraic PSGs descendant from the U1B11 $\mathrm{U}(1)$ PSG. Only the parameters marked by O can take non-zero values.}
\label{tab:symmetric_MF_Z2PSGs}
\end{table}

\paragraph{Summary.} In summary, we consider the six non-trivial fully symmetric $\mathrm{U}(1)$ spin liquids (U1A00, U1A10, U1A11, U1B00, U1B01, U1B11) and the five minimal-descendants of the U1B11 state which can stabilize the QBT (c01U1B11, c10U1B11, c11U1B11, n2U1B11, Z2B11110) as candidate spin liquid wavefunctions.

\section{Variational Monte Carlo}
\begin{figure}[!h]
\includegraphics[width=0.9\linewidth]{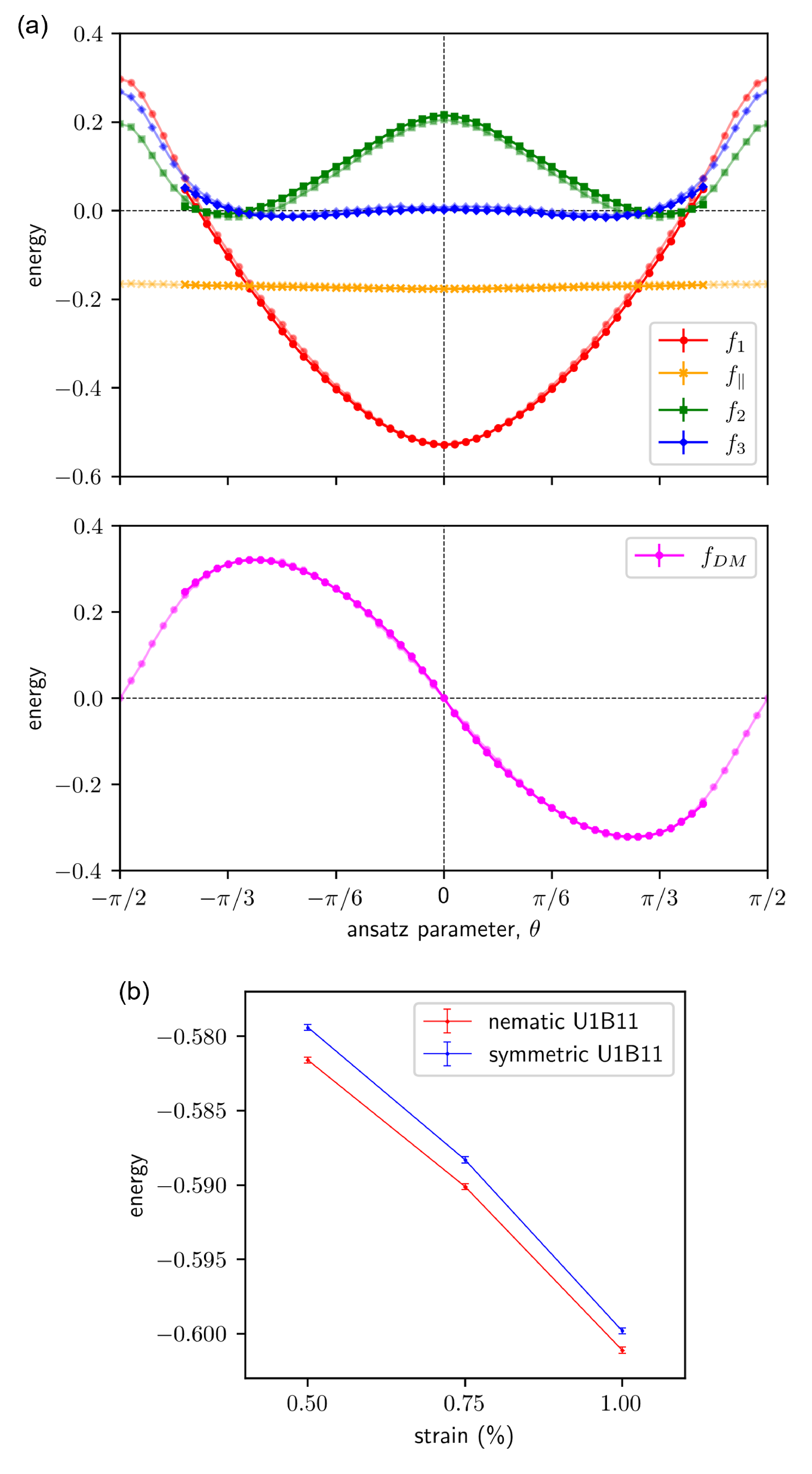}
\caption{(a) VMC evaluation of $f_1,f_\parallel,f_\text{DM},f_2,f_3$ for the fully symmetric U1B11 state, Eq. (\ref{eq:VMC_U1B11_energy}), (b) VMC evalauation of the nematic and fully symmetric U1B11 state energies of ACDS NbS${}_2$ as a function of the strain. Energy is in units of $J_1^{(0)}$.}
\label{fig:VMC_U1B11}
\end{figure}

\paragraph{Destabilization of the Dirac Spin Liquid.} We provide additional details on the destabilization of the Dirac spin liquid, described by the U1B11 ansatz, from the inclusion of Rashba spin orbit coupling. The U1B11 ansatz has two mean field parameters, which in terms of the Gutzwiller projected state, reduces to just one parameter. Specifically, we optimized $\theta\in[-\pi/2,\pi/2)$ which maps to the mean field parameters as
\begin{equation}
\mathfrak{t}_a\left(\theta\right)=\cos\theta\:\mathbb{I}+i\sin\theta\:\left(\mathbf{l}_a\cdot\bm{\sigma}\right).\nonumber
\end{equation}
For $\theta=0$, we have a Dirac spin liquid, while for $\theta\neq 0$, we have a QBT. The energy can generally be written as
\begin{equation}
\begin{split}
\frac{E\left(\theta\right)}{N}&=f_1\left(\theta\right)J_1+f_\parallel\left(\theta\right)J_\parallel+f_\text{DM}\left(\theta\right)D\\
&\qquad+f_2\left(\theta\right)J_2+f_3\left(\theta\right)J_3,
\end{split}
\label{eq:VMC_U1B11_energy}
\end{equation}
where the functions $f_1,f_\parallel,f_\text{DM},f_2,f_3$ are shown in Figure \ref{fig:VMC_U1B11}(a). While $f_1(\theta)$, $f_\parallel(\theta)$, $f_2(\theta)$, $f_3(\theta)$ are symmetric with respect to $\theta\mapsto-\theta$, $f_\text{DM}(\theta)$ is antisymmetric. Thus, for $D\neq 0$, the optimal $\theta$ is displaced from $0$. Therefore, within the U1B11 ansatz, QBT is preferred to the DSL for $\lambda_R>0$. As discussed in the main text, this motivates the study of the descendants of U1B11.

\paragraph{Comparison Between Spin Liquids.} To obtain the VMC phase diagram (Figure 4(a,b)), we first optimized the eleven candidate wavefunctions and compared them with one another. Although gradient descent was used when ever possible, the U1B00, U1A00, and c01U1B11 ansatze were optimized with grid search. This is because these ansatze describe Fermi surface states, for which the energy landscape is composed of disconnected and discontinuous plateaus, which cannot be optimized using local optimization methods, such as gradient descent. We found the n2U1B11 ansatz has lowest energy for all relevant parameter values. See for example Figure \ref{fig:VMC_U1B11}(b).

\paragraph{Magnetic Order.} To identify the transition from NDSL to magnetic order, we modify the mean-field spinon Hamiltonian to include both kinetic and magnetic terms \cite{iaconis2018spin},
\begin{equation}
\overline{H}\left(\mathbf{x},h\right)=\overline{H}_\text{kin}\left(\mathbf{x}\right)+h\overline{H}_\text{mag}, \nonumber
\end{equation}
where
\begin{equation}
\overline{H}_\text{mag}=-\sum_{\mathbf{r}}\left[\cos\left(\mathbf{q}_c\cdot\mathbf{r}\right)\hat{\mathbf{u}}_c+\sin\left(\mathbf{q}_c\cdot\mathbf{r}\right)\hat{\mathbf{v}}_c\right]\cdot\mathbf{S}_{\mathbf{r}}, \nonumber 
\end{equation}
and $\mathbf{q}_c$, $\hat{\mathbf{u}}_c$, $\hat{\mathbf{v}}_c$ correspond to the classical magnetic order. Note that as VMC is performed on a finite sized system, the closest commensurate $\mathbf{q}_c$, and corresponding corrections to $\hat{\mathbf{u}}_c$, $\hat{\mathbf{v}}_c$, were used. This ansatz was optimized for system sizes $N=6\times 6$, $12\times 12$, $18\times 18$, and when ambiguous, we performed finite size scaling of the parameter $h$ to determine whether or not $h=0$ in the thermodynamic limit. When $h\neq 0$ is extrapolated under the scaling, we conclude that the corresponding state is a trivial magnetic ordered state. Otherwise, it is a spin liquid state.

\section{STM Images of Spin Liquids}
STM probes the electron spectral function, $\diff I/\diff V\propto A_e\left(\mathbf{k},\omega\right)$, and can serve as an experimental signature for spin liquids. To describe the electron correlations, we require not only the spinons, but also the chargons. We use a slave-rotor formalism, $c_{i,s}=f_{i,s}b_i$ where $c_{i,s}$ and $f_{i,s}$ are the electron and fermionic spinon annihilation operators as used previously, and $b_i$ is the bosonic chargon annihilation operator. Taking the mean field approximation \cite{tang2013low}, electron correlators factorize locally (in space and time) into the spinon and chargon correlators. At the zero-temperature, a lowest-order approximation for the electron spectral function is given by
\begin{equation}
\begin{split}
&A_e\left(\mathbf{k}_e,\omega\right)\\
&=\sum_{\mathbf{k}_f+\mathbf{k}_b=\mathbf{k}_e}\sum_\nu\frac{M_\nu\left(\mathbf{k}_f,\mathbf{k}_b\right)}{2\Omega_{\mathbf{k}_b}}\left[\delta\left(\omega-\xi_{\mathbf{k}_f}^\nu-\Omega_{\mathbf{k}_b}\right)\right.\\
&\qquad\qquad\qquad\qquad\left.+\delta\left(\omega-\xi_{\mathbf{k}_f}^\nu+\Omega_{\mathbf{k}_b}\right)\right],
\end{split}
\label{eq:sMF_STM}
\end{equation}
where $\nu$ indexes the relevant spinon bands, $\xi_{\mathbf{k}_f}^\nu$ is the $\nu$-th spinon band, $\Omega_{\mathbf{k}_b}$ is the chargon band. $M_\nu\left(\mathbf{k}_f,\mathbf{k}_b\right)$ is the form factor, which we approximate as constant. Overall, the electron spectral function is peaked at $\mathbf{Q}_e=\mathbf{Q}_f+\mathbf{Q}_b$ where $\mathbf{Q}_f$ are the momentum of energy minima/maxima closest to the Fermi level in the spinon bands such as the position of QBTs and Dirac cones, and $\mathbf{Q}_b$ are the energy minima of the chargon bands. The spinon dispersion, chargon effective mass, and form factor determine secondary features of the STM signal.

Let us consider the QBT states given by the U1B11 ansatz. The spinons touch the Fermi level at the two $Q$ points, while the chargons have four band minima located at
\begin{equation}
\mathbf{Q}_b=\left(\pm\frac{\pi}{3},0\right),\left(\pm\frac{2\pi}{3},\mp\frac{\pi}{\sqrt{3}}\right).\nonumber
\end{equation}
We can identify the peak positions $\mathbf{Q}_e=\mathbf{Q}_f+\mathbf{Q}_b$ as the two $K$ points and the six $X$ points. The electron spectral function in the vicinity of each of these points are given by
\begin{equation}
A_e\left(\mathbf{Q}_e+\mathbf{k},\omega\right)\underset{\sim}{\propto}\Theta\left(\sqrt{2\left(m_f+\chi^{-1}\right)\left(-\omega-\Delta\right)}-k\right),\nonumber
\end{equation}
where $\Delta$ is the insulating gap, $m_f$ is the spinon effective mass, and $\Theta(x)$ is the Heaviside step function.

In real experiments, finite temperature and interactions with the environment result in a finite lifetime of the quasi-particles, and the STM tip has a finite resolution. These have the overall effect of broadening out the $\delta$-functions in Eq. (\ref{eq:sMF_STM}). As such, the STM image of Figure 4(c) was drawn replacing the step function with a logistic function to make it more realistic. As the steepness depends on many fine details of the experiment and is likely obfuscated by noise, we have chosen an arbitrary steepness.

Let us consider the NDSL states given by the n2U1B11 ansatz. The QBT is split into two anisotropic Dirac cones. The position of the chargon minima are also slightly shifted, and have anisotropic curvature. The momenta and dispersions were tracked numerically. The electron spectral function in the vicnity of the peak positions can be approximated as
\begin{equation}
A_e\left(\mathbf{Q}_e+\mathbf{k},\omega\right)\underset{\sim}{\propto}R\left(\left(-\omega-\Delta\right)-\frac{1}{2}\mathbf{k}^TX_b\mathbf{k}\right),\nonumber
\end{equation}
where $X_b$ is the $2\times 2$ Hessian matrix of the chargon dispersion, and $R(x)=x\Theta(x)$ is the ramp function. We note the anisotropicity of the spinon Dirac cones only affect the overall strength of correlations, not the dissipation of the peak. This function was used to draw the STM image in Figure 4(d).

\begin{figure}[t!]
\centering
\includegraphics[width=0.56\linewidth]{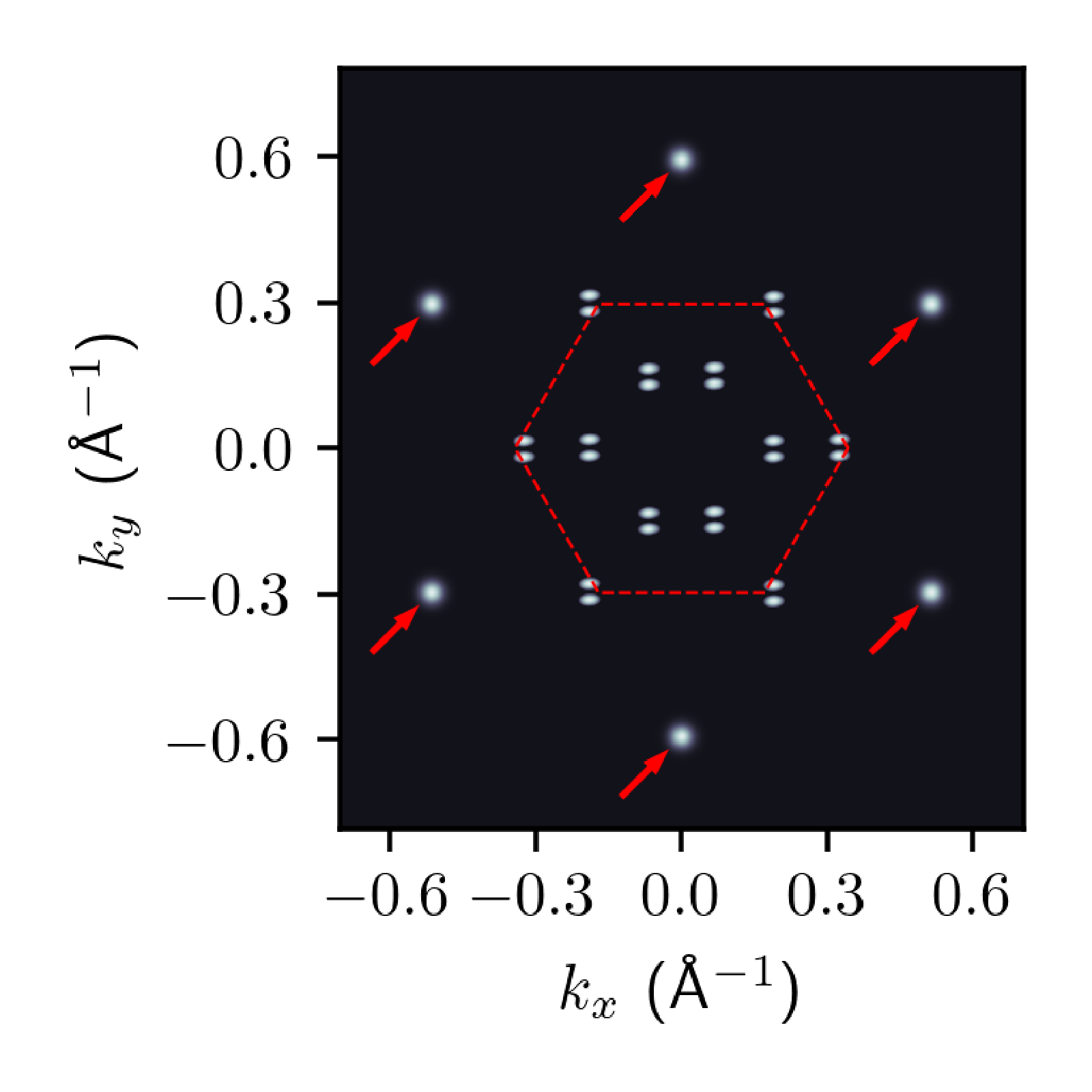}
\caption{Simulated STM image of the nematic spin liquid in real momentum space. The red dashed lines represent the reduced Brillouin zone of the ACDW CDW unit cell. The six peaks outside the reduced Brillouin zone (indicated by the red arrows) correspond to the periodicity of the ACDS CDW (CDW vectors). Within and along the boundaries of the reduced Brillouin zone, we have additional peaks due to the nematic spin liquid.}
\label{fig:MF_STM_extended}
\end{figure}

\begin{figure}[t!]
\centering
\includegraphics[width=0.92\linewidth]{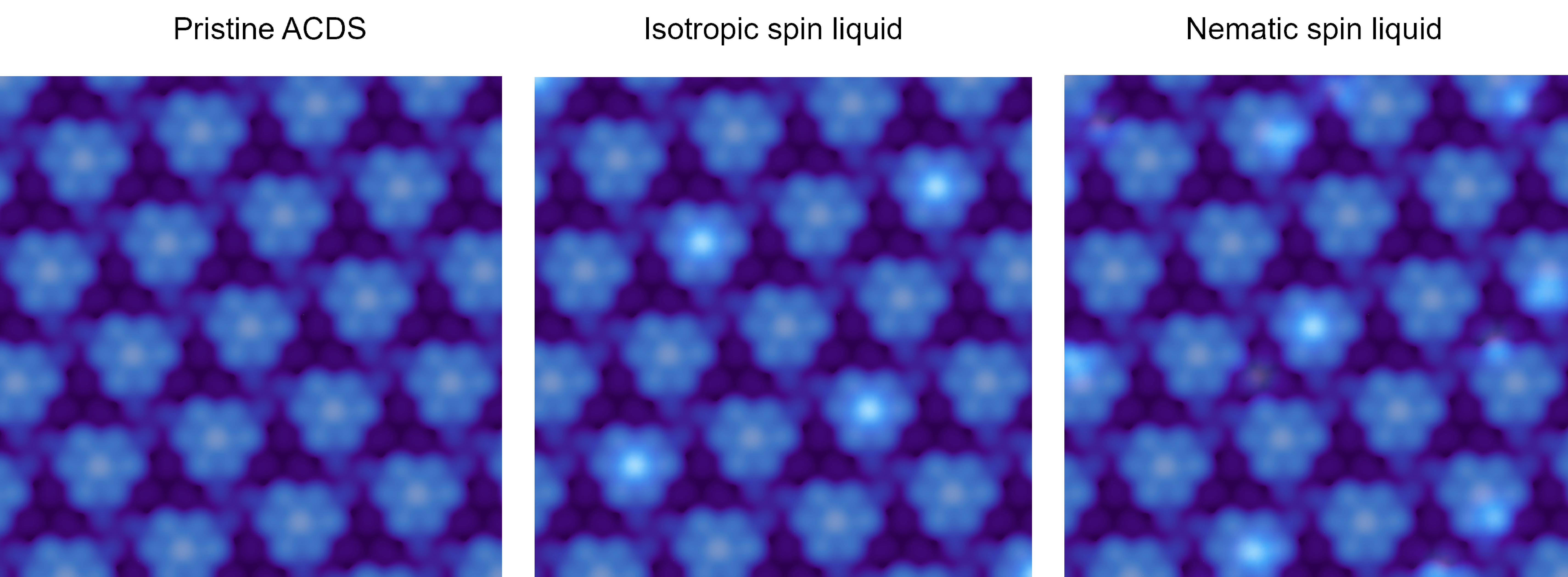}
\caption{Simulated STM image of the ACDS CDW without spin liquid features, with the isotropic (fully symmetric) spin liquid, and with the nematic spin liquid, in real space. The additional electron-electron correlations corresponding to the spin liquid produce super-modulations. The isotropic spin liquid produces commensurate super-modulations with 3-fold rotational symmetry, while the nematic spin liquid produces incommensurate super-modulations with broken 3-fold rotational symmetry.}
\label{fig:MF_STM_realspace}
\end{figure}

{We note that the STM images of Figure 4(c,d) are drawn in the reduced Brillouin zones of ACDS CDWs, and the pattern repeats with the periodicity of the reciprocal lattice. When taking a full 2D Fourier trasformation of the STM images obtained in experiments, the momentum is not reduced by the CDW periodicity, and the spin liquid peaks (when the spin liquid is stabilized) coexist with peaks we can attribute to the ACDS CDW, as observed in Figure 3 of \cite{ruan2021tase2}. The peaks due to the ACDS CDW can be distinguished from those due to the spin liquids because they occur at different momentum. First, we know where the peaks due to the CDW are, because we already know that the ACDS is a $\sqrt{13}\times\sqrt{13}$ CDW. Essentially, these peaks define the reduced BZ of the ACDS CDW in the STM images and live outside of the reduced BZ. On the other hand, the peaks due to the spin liquid will occur inside the reduced BZ. A full STM image including both features in real momentum space is shown in Figure \ref{fig:MF_STM_extended}. In real space, this corresponds to a super-modulation of the intensity [Fig.\ref{fig:MF_STM_realspace}]. While this super-modulation can be seen in our simulated images, they can easily become obscured in real-experiments due to noise, and taking the Fourier transform will allow us to see the features more distinctly. This is especially true for the nematic spin liquid because it produces incommensurate super-modulations which may be difficult to distinguish from noise.}

\section{Substrate influence on surface orientation}
{In the ACDS structure, the noncentrosymmetric distortions occur in the top and bottom chalcogen layers, leading to the possibility of observing different STM images for these two surfaces. However, in experimental situations, the TMD monolayer is grown on a specific substrate like graphene, leading to the two different structural models (upside and downside orientation). Note that we refer to a configuration where six anions gather around the selected anion atom to form a chalcogen layer network pointing upwards as the "upside orientation" and downwards as the "downside orientation". Having this in mind, we find that the substrate can give a favor on one orientation over the other, explaining naturally why only one type of the two surfaces has been observed in experiments. To be specific, we chose graphene as the substrate \cite{chen2020tase2,ruan2021tase2,liu2021nbse2} and constructed a heterostructure with a 5$\times$5 graphene supercell and TaSe${}_2$ ACDS structure to minimize the mismatch Figure \ref{fig:TaSe2_gra}.

\begin{figure*}[!ht]
\includegraphics[scale=1.0]{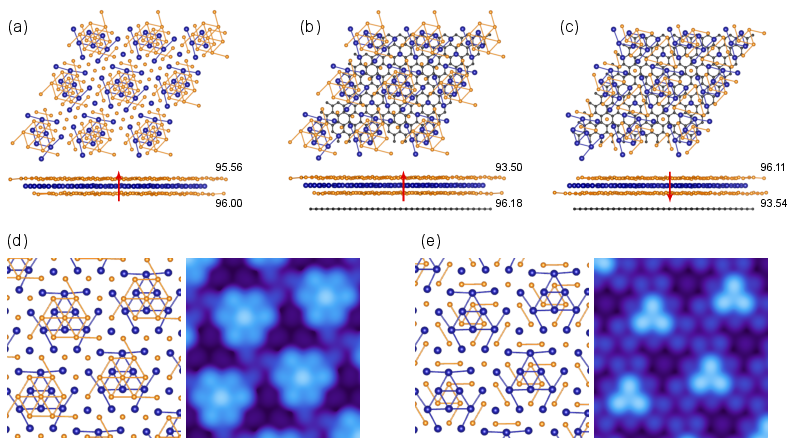}
\caption{(a) Top and side view of ACDS TaSe${}_2$ monolayer. (b),(c) Top and side views of the heterostructure of the ACDS TaSe${}_2$ monolayer and graphene with upside and downside orientations, respectively. The numbers written in the side view represent the charge accumulated on the upper and lower Se layers. The red arrows indicate the dipole field between the top and bottom Se layers. To enhance the visibility, only the top Se layer is depicted in the tow view. (d),(e) The atomic structures of ACDS and simulated STM images in upside and downside orientations, respectively.}
\label{fig:TaSe2_gra}
\end{figure*}

We next examined the energy difference between the DS and ACDS structures on graphene. For the case without the substrate, the energy difference between the DS and ACDS structures is 8.96 meV/atom. However, when placed on graphene, electrons are transferred from graphene, reducing the energy difference between the DS and ACDS structure. In particular, for the upside orientation, this difference was calculated to be 3.56 meV/atom, and for the downside orientation, it was 4.70 meV/atom. This indicates that energetically, TaSe${}_2$ grown on graphene prefers the upside configuration. It is likely due to the different charge distributions between the top and bottom Se layers. In the ACDS structure (upside orientation) without the substrate, charge is relatively more distributed in the bottom Se layer, leading to the formation of a dipole field between the two Se layers (indicated by the red arrows in Fig. \ref{fig:TaSe2_gra}). Considering the interaction with graphene, electron transfer from graphene to the ACDS structure would preferentially increase the dipole field when the ACDS structure is in contact with the upside configuration. Consequently, we believe that the experimental STM images predominantly represent the surface corresponding to the upside orientation. We also found the similar results for the other ACDS TMDs like TaS${}_2$, NbSe${}_2$, and NbS${}_2$. As a side note, the STM simulation images for the downside orientation surface show three bright protrusions, which have not been experimentally confirmed Fig. \ref{fig:TaSe2_gra}(d),(e).}

\section{Appropriate U values and emerging states}
{We can estimate the suitable $U$ values by comparing theoretically-calculated DOS and the spectral gaps in experimentally-measured dI/dV for the ACDS systems. The reported gaps for TaS${}_2$, TaSe${}_2$, and NbSe${}_2$ were found to be approximately 0.25 eV, 0.2 eV, and 0.15 eV, respectively \cite{luic2019tas2,lin2020tase2,mengke2021nbs2}. We extracted $U$ values that could describe these spectral gaps for the ACDS structures [Fig. \ref{fig:PDOS_varU}, Table \ref{tab:Estimated_U} ]. Since the experimental dI/dV for NbS${}_2$ is not available at this moment, we extrapolated U values based on the spectral gap observed in TaS${}_2$, by noting that Hubbard $U$ is expected to be inversely proportional to the square of the band widths \cite{lee2020tas2}. The NbS${}_2$ ACDS structure has the band width, which is approximately 1.4 times larger than that of the TaS${}_2$ ACDS structure, and this leads to $U_{eff}$=2.3 eV for the NbS${}_2$ ACDS structure.

\begin{figure*}[!ht]
\includegraphics[scale=1.0]{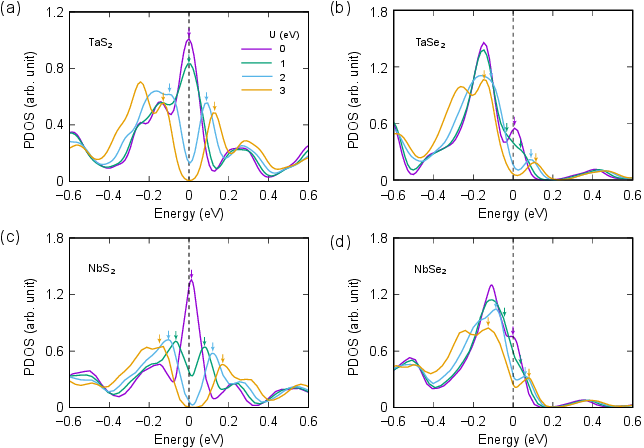}
\caption{The partial density of states (PDOS) of the transition metal atom’s dz2 orbital as a function of the Hubbard $U$ for (a) TaS${}_2$, (b) TaSe${}_2$, (c) NbS${}_2$ and (d) NbSe${}_2$. The arrows indicate the position of the flat bands for the given $U$ values, whose energy splitting is the main source of the spectral gaps. They are then compared with the experimentally observed dI/dV in Ref. \cite{luic2019tas2,lin2020tase2,mengke2021nbs2}.}
\label{fig:PDOS_varU}
\end{figure*}

For those extracted $U$ values, we can determine emerging quantum phenomena in each material. Our main findings suggest that various quantum phenomena can emerge depending on the varying strengths of correlations (varying $U$ values). For TaS${}_2$, the well-defined flat band leads to a Mott insulating state for the estimated $U_{eff}$=2.8 eV, which will support the incommensurate magnetic order (See the phase diagram [Fig.4a] of the main text). In the case of TaSe${}_2$, the estimated $U_{eff}$ is approximately 1.9 eV. For this $U$, the proximity of the flat bands to valence states and mixing with those states result in the quantum anomalous Hall state. NbS${}_2$ is inside a Mott insulator ($U_{eff}$=2.3 eV), which will exhibit the strain-induced nematic spin liquid phase. For NbSe${}_2$, it was observed that the estimated $U_{eff}$ is relatively small ($U_{eff}$=1.4 eV), which leads to a giant spin Hall conductivity, approximately 0.14 $e^{2}/{\hbar}$. Hence, we find that various emerging quantum phenomena that we discovered in our manuscript will appear in appropriate materials.}

\begin{table}[!h]
\centering
\begin{tabular}{c|c|c}
\hline
material & $U_{eff}$ (eV) & Emerging Phenomena \\
\hline
$\mathrm{NbS_2}$  & 2.8 & \makecell{Strain-enabled spin liquid \\ (Mott insulator)}  \\
$\mathrm{NbSe_2}$ & 1.9 & \makecell{Quantum anomalous Hall effect\\ (Interaction-enabled topological state)}  \\
$\mathrm{TaS_2}$  & 2.3 & \makecell{Incommensurate magnetic order\\ (Mott insulator)}  \\
$\mathrm{TaSe_2}$ & 1.4 & Giant spin Hall conductivity \\
\hline
\end{tabular}
\caption{Estimated $U$ values and corresponding emerging phenomena.}
\label{tab:Estimated_U}
\end{table}

\bibliographystyle{apsrev4-1}
\bibliography{refs}